\documentclass[journal,12pt,draftclsnofoot,onecolumn,english]{IEEEtran}

\usepackage{epsfig}
\usepackage{times}
\usepackage{float}
\usepackage{afterpage}
\usepackage{amsmath}
\usepackage{amstext,cite}
\usepackage{amssymb,bm}
\usepackage{latexsym}
\usepackage{color}
\usepackage{graphicx}
\usepackage{amsmath}
\usepackage{amsthm}
\usepackage{graphicx}
\usepackage[center]{caption}
\usepackage{pstricks}
\usepackage{caption}
\usepackage{subcaption}
\usepackage{booktabs}
\usepackage{multicol}
\usepackage{lipsum}
\usepackage[T1]{fontenc}
\usepackage{hyperref}
 \usepackage{aecompl}
 \usepackage{mathrsfs}

 \usepackage{arydshln}

\usepackage{soul}
\usepackage{cancel}

\usepackage{changes}
\definechangesauthor[color=red,name={Mingyue Ji}]{MJ}

\allowdisplaybreaks

\long\def\comment#1{}

\newcommand{\fv}{{\mathbf f}}

\newcommand{\sv}{{\mathbf s}}

\newcommand{\vv}{{\mathbf v}}

\newcommand{\Ac}{{\mathcal A}}
\newcommand{\Bc}{{\mathcal B}}

\newcommand{\Dc}{{\mathcal D}}

\newcommand{\Gc}{{\mathcal G}}
\newcommand{\Hc}{{\mathcal H}}

\newcommand{\Kc}{{\mathcal K}}
\newcommand{\Lc}{{\mathcal L}}

\newcommand{\Pc}{{\mathcal P}}
\newcommand{\Qc}{{\mathcal Q}}

\newcommand{\Sc}{{\mathcal S}}

\newcommand{\Uc}{{\mathcal U}}

\newcommand{\Vc}{{\mathcal V}}

\newcommand{\Zc}{{\mathcal Z}}

\newcommand{\asf}{{\mathsf a}}
\newcommand{\bsf}{{\mathsf b}}

\newcommand{\msf}{{\mathsf m}}
\newcommand{\nsf}{{\mathsf n}}

\newcommand{\qsf}{{\mathsf q}}

\newcommand{\ysf}{{\mathsf y}}

\newcommand{\Ksf}{{\mathsf K}}
\newcommand{\Lsf}{{\mathsf L}}
\newcommand{\Msf}{{\mathsf M}}
\newcommand{\Nsf}{{\mathsf N}}

\newcommand{\Rsf}{{\mathsf R}}

\newcommand{\Tsf}{{\mathsf T}}

\newtheorem{thm}{Theorem}
\newtheorem{cor}{Corollary}
\newtheorem{lem}{Lemma}

\newtheorem{rem}{Remark}
\newtheorem{example}{Example}

\providecommand{\definitionname}{Definition}

\usepackage{amsmath}
\usepackage{tikz}
\usetikzlibrary{calc}

\usepackage[nodisplayskipstretch]{setspace} 

\newcommand{\tikzmark}[1]{\tikz[overlay,remember picture] \node (#1) {};}
\newcommand{\DrawBox}[4][]{%
    \tikz[overlay,remember picture]{%
        \coordinate (TopLeft)     at ($(#2)+(-0.2em,0.9em)$);
        \coordinate (BottomRight) at ($(#3)+(0.2em,-0.3em)$);
        \path (TopLeft); \pgfgetlastxy{\XCoord}{\IgnoreCoord};
        \path (BottomRight); \pgfgetlastxy{\IgnoreCoord}{\YCoord};
        \coordinate (LabelPoint) at ($(\XCoord,\YCoord)!0.5!(BottomRight)$);
        \draw [red,#1] (TopLeft) rectangle (BottomRight);
        \node [below, #1, fill=none, fill opacity=1] at (LabelPoint) {#4};
    }
}

\begin{document}

\title{On   Secure Distributed Linearly Separable  Computation}  
\author{
Kai~Wan,~\IEEEmembership{Member,~IEEE,} 
Hua~Sun,~\IEEEmembership{Member,~IEEE,}
Mingyue~Ji,~\IEEEmembership{Member,~IEEE,}  
and~Giuseppe Caire,~\IEEEmembership{Fellow,~IEEE}

\thanks{
K.~Wan and G.~Caire are with the Electrical Engineering and Computer Science Department, Technische Universit\"at Berlin, 10587 Berlin, Germany (e-mail:  kai.wan@tu-berlin.de; caire@tu-berlin.de). The work of K.~Wan and G.~Caire was partially funded by the European Research Council under the ERC Advanced Grant N. 789190, CARENET.}
\thanks{
H.~Sun is with the Department of Electrical Engineering, University of North Texas, Denton, TX 76203, USA (email: hua.sun@unt.edu). The work of H.~Sun was supported in part by NSF Award 2007108.
}
\thanks{
M.~Ji is with the Electrical and Computer Engineering Department, University of Utah, Salt Lake City, UT 84112, USA (e-mail: mingyue.ji@utah.edu). The work of M.~Ji was supported in part by NSF Awards 1817154 and 1824558.}
}
\maketitle

\begin{abstract}
Distributed linearly separable  computation, where a user asks some distributed servers to compute a linearly separable function,  was recently formulated by the same authors and aims to alleviate the bottlenecks of stragglers and communication cost in   distributed computation. For this purpose, the data center assigns a subset of input datasets to each server, and each server   computes some coded packets  on the assigned datasets, which are then sent to the user. The user should recover the task function from the answers of a subset of servers, such the   effect of stragglers could be tolerated.

In this paper, we formulate     a novel secure  framework for this distributed linearly separable computation, where we aim to let the user only 
  retrieve the  desired task function without obtaining any other information about the input datasets, even if it receives the answers of all servers.  In order to preserve the security of the input datasets, some common randomness variable independent of the datasets should be introduced into the transmission. 

We   show that   any   non-secure linear-coding based computing scheme   for the original distributed linearly separable computation problem, can be made secure without increasing the communication cost (number of symbols the user should receive). Then we focus on the case where the computation cost of each server  (number of datasets assigned to each server) is minimum and aim to minimize the size of the randomness variable (i.e., randomness size) introduced in the system
  while achieving the optimal communication cost. We first propose an information theoretic   converse bound on the  randomness size. 
We then propose secure computing schemes based on two well-known data assignments, namely {\it fractional repetition assignment} and {\it cyclic assignment}. These schemes are optimal subject to using these assignments. Motivated by the observation of  the general limitation of these two schemes on the randomness size, we propose a computing scheme with novel assignment, which strictly outperforms the above two schemes. Some additional optimality results  are also obtained.
\end{abstract}


\section{Introduction}
\label{sec:intro}
 Distributed linearly separable computation, which is a generalization of many existing distributed computing problems such as    distributed gradient coding~\cite{gradiencoding} and  distributed linear transform~\cite{shortdot2016dutta}, was originally proposed in~\cite{linearcomput2020wan} considering two important bottlenecks in the distributed computation systems:  communication cost and stragglers. 
 In this computation scenario,  a user aims to compute  a    function of $\Ksf$ datasets ($D_1,\ldots,D_{\Ksf}$) on a finite field $\mathbb{F}_{\qsf}$ through $\Nsf$ distributed servers. 
The task function can be seen as    $\Ksf_{\rm c}$ linear combinations of $\Ksf$ intermediate messages (the $n^{\text{th}}$ intermediate message $W_n$ is a function of   dataset $D_n$ and contains $\Lsf$ symbols).  
 The problem contains three phases, {\it assignment, computing, decoding}. During the assignment phase, the data center with access to the $\Ksf$ datasets assigns $\Msf$ datasets to each server, where $\Msf$ represents the computation cost of each server. During the computing phase, each server first computes the intermediate message of each dataset assigned to it, and then transmits a coded packet of the computed intermediate messages to the user. During the decoding phase, from the answers of any $\Nsf_{\rm r}$ servers, the user should recover the task function such that the system can tolerate $\Nsf-\Nsf_{\rm r}$ stragglers. The worst-case   number of symbols (normalized by $\Lsf$) needed to be received is defined as the communication cost. 
  The objective is to minimize the communication cost for each given computation cost. 
  The optimality  results for some  cases have been founded in the literature and are summarized below: 
\begin{itemize}
\item  {\it $\Ksf_{\rm c}=1$.} The computation problem reduces to the  distributed gradient coding problem in~\cite{gradiencoding}.  When the computation cost is minimum (i.e., $\Msf=\frac{\Ksf}{\Nsf}(\Nsf-\Nsf_{\rm r}+1)$), the gradient coding scheme in~\cite{gradiencoding} achieves the optimal communication cost (equal to $\Nsf_{\rm r}$) as proved in~\cite{linearcomput2020wan}. Then some extended gradient coding schemes were proposed in~\cite{efficientgradientcoding,adaptiveGC2020} which characterize the 
    optimal communication cost under the constraint of linear coding, for each possible computation cost.  
  \item  {\it Minimum comptation cost $\Msf=\frac{\Ksf}{\Nsf}(\Nsf-\Nsf_{\rm r}+1)$.} The optimal communication cost with the cyclic assignment (an  assignment widely used in the related distributed computing problems)   was characterized in~\cite{linearcomput2020wan}, when the computation cost is minimum.
  \item   For the general case,~\cite{cctradeoff2020wan} proposed a computing scheme under some
parameter regimes, which is  order
optimal within a factor of $2$ under the constraint of the cyclic assignment.  
\end{itemize}

 In this paper, we consider  a novel secure  framework for this distributed linearly separable computation problem, where we aim to let the user only 
retrieve   the   desired task function  without obtaining any other information about the $\Ksf$ datasets.  We notice that this security model has been widely used in the literature in the context of
secure multiparty computation~\cite{BGW1988,CCD1988} and secure aggregation  for federated learning~\cite{practicalsecure2016Bonawitz,privacy2017Bonawitz}.

\begin{figure}
\centerline{\includegraphics[scale=0.25]{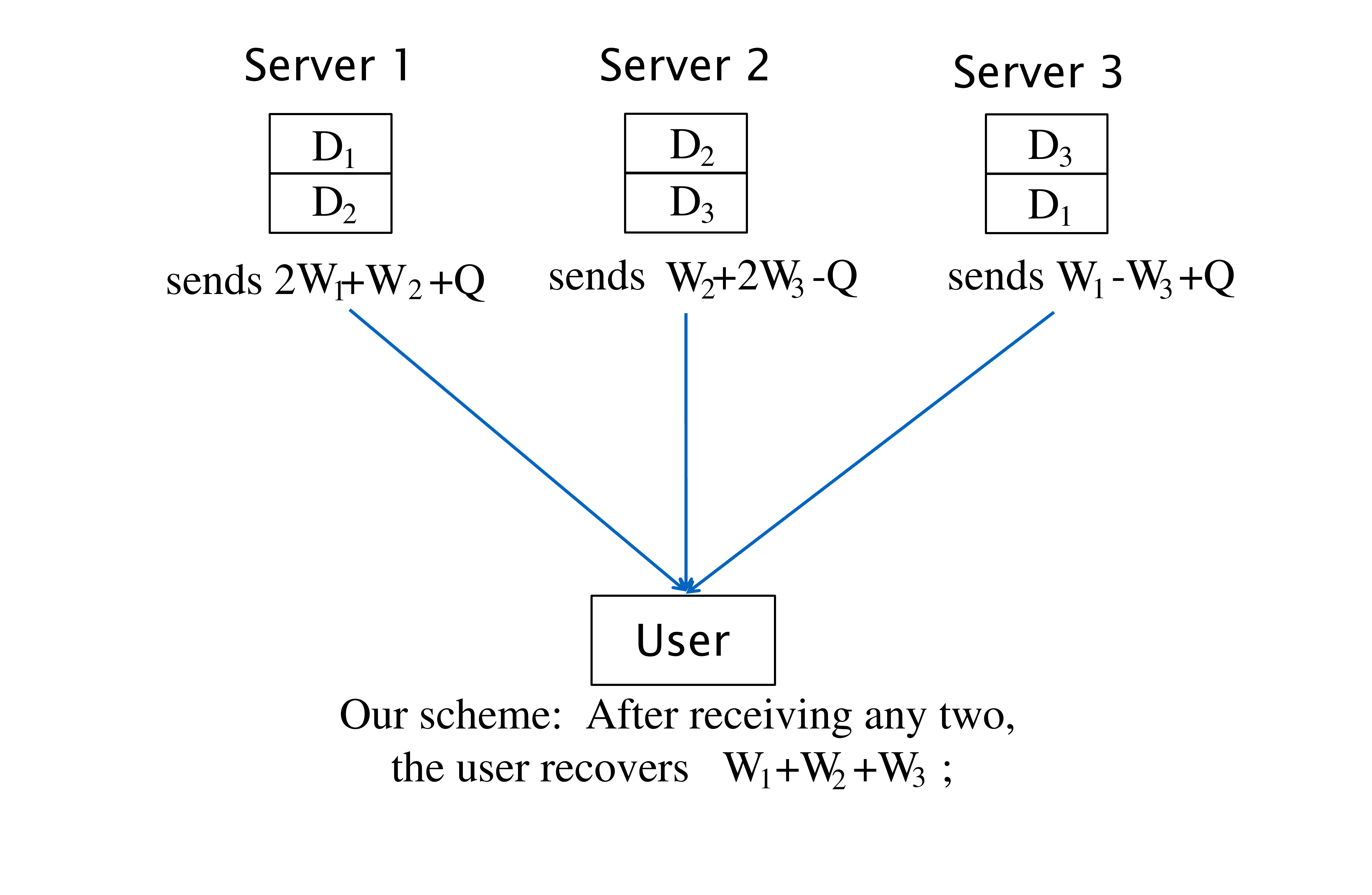}}
\caption{\small  Secure distributed linearly separable  computation with $\Ksf=\Nsf=3$, $\Nsf_{\rm r}=2$, $\Ksf_{\rm c}=1$, and  $\Msf=2$.}
\label{fig: system_model}
\end{figure}

Let us focus on a small but instructive example in Fig.~\ref{fig: system_model}, where  $\Ksf=\Nsf=3$, $\Nsf_{\rm r}=2$, $\Ksf_{\rm c}=1$, $\Msf=2$, and the task function is $W_1+W_2+W_3$. Assume the field is $\mathbb{F}_{3}$. 
We use the cyclic assignment in~\cite{gradiencoding} to assign $D_1$ to servers $1,3$; assign $D_2$ to servers $1,2$; assign $D_3$ to servers $2,3$.  In addition, for the sake of secure computation,   the data center generates a randomness variable $Q$  uniformly   over $[\mathbb{F}_{\qsf}]^{\Lsf}$, which is independent of 
the datasets, and assigns    $Q $ to each server.
In the computing phase of the novel proposed scheme, server $1$ computes $2W_1 + W_2 + Q$;   server $2$ computes $W_2 + 2W_3 - Q$;   server $3$ computes $W_1 - W_3 + Q$. It can be seen that from the answers of any two servers, the user can recover the task function $W_1+W_2+W_3$. Moreover,  even if the user receives the answers of all servers, 
it cannot get any other information about the messages (nor the datasets) because   $Q$ is unknown to it. Notice that the communication cost in this example is $2$, which is the 
same as the gradient coding scheme in~\cite{gradiencoding}.

 The above example shows that it is possible to preserve the security of the datasets (except the task function) from the user.
  The main questions we ask in this paper are (i) do we need additional communication cost to satisfy this   security constraint? (ii) how much randomness is required to guarantee security? 
 
Compared to the existing works on coded distributed secure computation on matrix multiplication in~\cite{lagrange2019yu,chang2019privatesecure,aliasgari2019private,secure2019jia,updowlink2019kakar,yu2020entangle,GCSAsecure2020chen,securezhu2021},
the main differences of the consider secure problem are as follows:
 (i) in the above existing works, the data center assigns the coded version of all input datasets   to the  distributed servers, 
 while in the considered problem the assignment phase is uncoded; 
 (ii) the above existing works  aim to preserve the security of the input datasets from the servers,  where each distributed server can only access the datasets assigned to it 
 while in the considered problem we aim to preserve the security of the input datasets (except the task function) from the user who may receive all answers of the servers.

\subsection*{Contributions}
In this paper, we formulate the secure distributed linearly separable computation problem. We first show that  any   non-secure linear-coding based computing scheme   for the original distributed linearly separable computation problem, can be made secure without increasing the communication cost. Then we focus on the secure distributed linearly separable computation problem where $\Ksf_{\rm c}=1$ and $\Msf=\frac{\Ksf}{\Nsf}(\Nsf-\Nsf_{\rm r}+1)$ (i.e., the computation cost is minimum), and aim to minimize the randomness size\footnote{\label{foot:reason} This randomness should be broadcasted from the data center to the servers and   stored at the servers; thus reducing the randomness size   can reduce the communication cost from the data center and  the the storage cost at the servers.} while achieving the optimal communication cost $\Nsf_{\rm r}$. Our contributions on this objective  are as follows:
\begin{itemize}
\item For each possible assignment, we propose an information theoretic   converse bound on the randomness size, which is also a converse bound on the   randomness size while achieving the optimal communication cost.
\item When $\Nsf-\Nsf_{\rm r}+1$ divides $\Nsf$, we propose a secure computing scheme with the fractional repetition assignment in~\cite{gradiencoding}, which coincides with the proposed converse bound on the randomness size.
\item Under the constraint of the widely used cyclic assignment~\cite{gradiencoding,improvedGC2017halbawi,MDSGC2018raviv,efficientgradientcoding,adaptiveGC2020,linearcomput2020wan,cctradeoff2020wan}, we propose an optimal  secure computing scheme in the sense that minimum randomness size is achieved.
\item Motivated by the observation that the computing scheme with   the fractional repetition assignment can only work for the case where $\Nsf-\Nsf_{\rm r}+1$ divides $\Nsf$ and that the computing scheme with the cyclic assignment is highly sub-optimal in terms of the randomness size, we propose a new computing scheme with novel assignment strategies.
The novel computing scheme can cover the optimality results of the computing scheme with the fractional repetition assignment; in general it needs a   lower randomness size while achieving the optimal communication cost than that of the computing scheme with the cyclic assignment. 
We also prove that it is optimal when $\frac{\Nsf-\Nsf_{\rm r}+1}{\text{GCD}(\Nsf,\Nsf-\Nsf_{\rm r}+1)}\leq 4 $. 
\end{itemize}

 

\subsection*{Paper Organization}
The rest of this paper is organized as follows.
 Section~\ref{sec:model} introduces the secure distributed linearly separable computation  problem. 
Section~\ref{sec:main} provides   the main results in this paper and   some numerical evaluations. 
Section~\ref{sec:achievable}  presents the proposed   distributed computing schemes. 
Section~\ref{sec:con} concludes the paper and some of the proofs are given in the Appendices.

\subsection*{Notation Convention}
Calligraphic symbols denote sets,  
bold lower-case letters denote vector, bold
upper-case letters denote matrices,
and sans-serif symbols denote system parameters.
We use $|\cdot|$ to represent the cardinality of a set or the length of a vector;
$[a:b]:=\left\{ a,a+1,\ldots,b\right\}$; 
  $[n] := [1:n]$;
$\mathbb{F}_{\qsf}$ represents a  finite field with order $\qsf$;         
$\mathbf{M}^{\text{T}}$  and $\mathbf{M}^{-1}$ represent the transpose  and the inverse of matrix $\mathbf{M}$, respectively;
$\mathbf{I}_n$ represents the identity matrix with dimension $n \times n$;
${\bf 0}_{m \times n}$ represents the zero  matrix with dimension $m\times n$; 
 the matrix $[a;b]$ is written in a Matlab form, representing $[a,b]^{\text{T}}$;
$(\mathbf{M})_{m \times n}$ represents the dimension of matrix $\mathbf{M}$ is $m \times n$;
$\mathbf{M}^{(\Sc)_{\rm r}}$ represents the sub-matrix of $\mathbf{M}$ which is composed of the rows  of $\mathbf{M}$ with indices in $\Sc$ (here $\rm r$ represents `rows'); 
$\mathbf{M}^{(\Sc)_{\rm c}}$ represents the sub-matrix of $\mathbf{M}$ which is composed of the columns  of $\mathbf{M}$ with indices in $\Sc$ (here $\rm c$ represents `columns'); 
 $\text{Mod} (b,a)$ represents the modulo operation on $b$ with  integer divisor $a$ and in this paper we let $\text{Mod}(b,a)\in \{1,\ldots,a \}$ (i.e., we let $ \text{Mod}(b,a)=a$ if $a$ divides $b$);
$\text{GCD}(b,a)$ represents the Greatest Common Divisor of integers $b$ and $a$;   
we let $\binom{x}{y}=0$ if $x<0$ or $y<0$ or $x<y$.
 In this paper, for each set  of integers  $\Sc$, we sort the elements in $\Sc$ in an increasing order and denote the $i^{\text{th}}$ smallest element by $\Sc(i)$, i.e., $\Sc(1)<\ldots<\Sc(|\Sc|)$.

\section{System Model}
\label{sec:model}
We formulate a $(\Ksf,\Nsf,\Nsf_{\rm r},\Ksf_{\rm c}, \Msf)$ secure linearly separable  computation problem  over the canonical user-server distributed system, as illustrated in Fig.~\ref{fig: system_model}.  
Compared to the distributed computing framework in~\cite{linearcomput2020wan}, an additional security constraint  will be added. The detailed system model is as follows.

The user wants to compute a function 
$$
f(D_1,   \ldots, D_{\Ksf})
$$
on  $\Ksf$ independent datasets $D_1,  \ldots, D_{\Ksf}$.   As the data sizes are large, 
the computing task function is distributed over a group of $\Nsf$ servers. 
 For distributed computation to be possible, we assume that the function is \emph{linearly separable} 
   with respect to the datasets, i.e., that there exist  functions $f_1, \ldots, f_{\Ksf}$ such that $f(\cdot)$ can be written as 
      \begin{subequations}
\begin{align}
   f(D_1, \ldots, D_{\Ksf}) &= g\big(f_1(D_1), \ldots, f_{\Ksf} (D_{\Ksf}) \big) \\
&= {\bf G} \ [W_1;\ldots;W_{\Ksf}],\label{eq:separated objective}
\end{align}
     \end{subequations} 
where  ${\bf G}$ is a $\Ksf_{\rm c} \times \Ksf$ matrix    and we model $f_k(D_k)$,  $k \in [\Ksf]$ as the $k$-th message  $W_k$ and $f_k(\cdot)$ is an arbitrary function. Notice that when $\Ksf_{\rm c}=1$, without loss of generality, we assume that 
\begin{align}
   f(D_1, \ldots, D_{\Ksf}) ={\bf G} \ [W_1;\ldots;W_{\Ksf}] = W_1 +\cdots+W_{\Ksf} .\label{eq:separated objective Kc=1}
\end{align}
In this paper, we assume that $f(D_1, \ldots, D_{\Ksf})$ contains one linear combination of the $\Ksf$ messages, and that 
 the $\Ksf$ messages are independent. Each message is composed of $\Lsf$ uniformly i.i.d.   symbols  over a finite field $\mathbb{F}_{\qsf}$ for some large enough prime-power $\qsf$. 
 As in~\cite{linearcomput2020wan} we assume that  $\frac{\Ksf}{\Nsf} $ is an integer.

A computation scheme for our problem contains three phases, {\it data assignment}, {\it computing}, and {\it decoding}. 
\paragraph*{Data assignment phase}
The data center/global server  assigns each dataset $D_k$ where $k \in [\Ksf]$ to a subset of the $\Nsf$ servers in an uncoded manner. The set of datasets assigned to server $n \in [\Nsf]$ is denoted by $\Zc_n$, where $\Zc_n \subseteq [\Ksf]$. 
The   assignment  constraint is that 
\begin{align}
|\Zc_n| \leq \Msf. 
\end{align}
 The assignment for all servers is denoted by $\mathbf{Z}=(\Zc_1,\ldots,\Zc_{\Nsf})$.

As an additional problem  constraint, we impose that   the user learns no further information about  $(D_1,\ldots,D_{\Ksf})$ other than  the task  function $ f(D_1, \ldots, D_{\Ksf})$. To this purpose, the data center also generates  a randomness variable  $Q \in \Qc$,  and assign $Q $ to each server  $k\in [\Ksf]$. Notice that   
\begin{align}
I(Q ; D_1, \ldots, D_{\Ksf} )=I(Q ; W_1, \ldots, W_{\Ksf} )=0. \label{eq:independent key}
\end{align}

 The   randomness
size $\eta$ measures the   amount of   randomness, i.e.,
\begin{align}
\eta= \frac{H(Q )}{\Lsf}. \label{eq:def of eta}
\end{align}

\paragraph*{Computing phase}
Each server $n \in[\Nsf]$ first computes the message $W_k = f_k (D_k)$ for each $k \in \Zc_n$.   Then it generates
\begin{align}
 X_n = \psi_n(\{W_k:  k \in \Zc_n\},  Q )
 \end{align}
  where the encoding function $\psi_n$ is such that
\begin{align} 
\psi_n &:  [\mathbb{F}_{\qsf}]^{ |\Zc_n| \Lsf} \times   | \Qc |  \to [\mathbb{F}_{\qsf}]^{ \Tsf_n },  
\label{eq: encoding function def}
\end{align}
and  $ \Tsf_n$ represents the length of $ X_n $. Finally, server $n$ sends $X_n$ to the user.

 
\paragraph*{Decoding phase}
The computation scheme should tolerate   $\Nsf - \Nsf_{\rm r}$ stragglers.
As the user does not know a priori which servers are stragglers, the computation scheme should be designed so that from the answers of any $\Nsf_{\rm r}$   servers, the user can recover ${\bf G}  [W_1;\ldots;W_{\Ksf}]$.  
Hence, for any subset of servers $\Ac \subseteq [\Nsf]$ where $|\Ac|=\Nsf_{\rm r}$, with the definition
\begin{align}
X_{\Sc}:=\{X_n: n\in \Sc \},
\end{align}
for any set $\Sc\subseteq [\Nsf]$, 
  there exists a decoding function $\phi_{\Ac}$ such that
        \begin{subequations}
\begin{align}
& \phi_{\Ac}\big( X_{\Ac}  \big) = {\bf G} \ [W_1;\ldots;W_{\Ksf}]  , \\
 &\phi_{\Ac} :  [\mathbb{F}_{\qsf}]^{\sum_{n \in \Ac} \Tsf_n }   \to [\mathbb{F}_{\qsf}]^{  \Ksf_{\rm c} \Lsf}.
\end{align} 
\label{eq:decoding model}
      \end{subequations}
      Notice that $Q$  is unknown to the user, and thus $Q$ cannot be used in the decoding procedure in~\eqref{eq:decoding model}. 
 In order to protect the security, 
even if receiving the answers of all servers in $[\Nsf]$,  
 the user cannot learn any information about the messages except the desired task function; 
 it should satisfy that\footnote{\label{foot:markov chain}Notice $X_{[\Nsf]}$ is a function of $(W_{1},\ldots,W_{\Nsf})$ and $Q$. By the data processing inequality, the security constraint in~\eqref{eq:security} is equivalent to $I\big(D_{1},\ldots, D_{\Ksf};   X_{[\Nsf]} | {\bf G}   [W_1;\ldots;W_{\Ksf}]  \big)=0$.} 
\begin{align}
I\big(W_{1},\ldots, W_{\Ksf};   X_{[\Nsf]} | {\bf G}   [W_1;\ldots;W_{\Ksf}]  \big)=0. \label{eq:security}
\end{align}

We denote the communication cost by,  
\begin{align}
\Rsf  :=  \max_{\Ac  \subseteq [\Nsf]: |\Ac|= \Nsf_{\rm r}} \frac{ \sum_{n \in \Ac} \Tsf_n}{   \Lsf  }, \label{eq:communicaton rate}
\end{align}
representing 
 the maximum normalized   number of symbols received by the user from any $\Nsf_{\rm r}$ responding servers.

When the computation cost is minimum,  it was proved in~\cite[Lemma 1]{linearcomput2020wan} that {\bf each  dataset is assigned to $\Nsf - \Nsf_{\rm r}+1$ servers and    each server obtains $\Msf$ datasets}, where
$$
\Msf=|\Zc_1|=\cdots=|\Zc_{\Nsf}|= \frac{\Ksf}{\Nsf} (\Nsf-\Nsf_{\rm r} +1).
$$

In this paper,  we mainly focus on the case where $\Ksf_{\rm c}=1$ and   the computation cost is minimum and search for the minimum communication cost $\Rsf^{\star}$. In addition, with the optimal communication cost, we aim to search the minimum  
 randomness size $\eta^{\star}$ necessary to achieve the security constraint~\eqref{eq:security}.

 \section{Main Results}
\label{sec:main}
 In this section, we present our main results.
 
In the following, we  show that compared to the distributed linearly separable computation problem in~\cite{linearcomput2020wan},   
 the optimal communication cost does not change when the  security constraint in~\eqref{eq:security} is added.
 \begin{thm}
\label{thm:optimal communication cost}
 For the $(\Ksf,\Nsf,\Nsf_{\rm r},\Ksf_{\rm c}, \Msf)$ secure distributed linearly separable computation problem with  $\Msf= \frac{\Ksf}{\Nsf} (\Nsf-\Nsf_{\rm r} +1)$ and $\Ksf_{\rm c}=1$,
 the optimal communication cost is $\Rsf^{\star}=\Nsf_{\rm r}$.
 \hfill $\square$ 
\end{thm}  
   \begin{IEEEproof}
   \paragraph*{Converse}
Obviously,    the converse bound for the   distributed linearly separable computation problem in~\cite{linearcomput2020wan} which is without the  security constraint in~\eqref{eq:security} is also a converse bound for the considered secure distributed linearly separable computation problem. Hence, from~\cite[(16a)]{linearcomput2020wan}  we have 
\begin{align}
\Rsf^{\star}\geq \Nsf_{\rm r}.\label{eq:converse for communication}
\end{align}

\paragraph*{Achievability} 
We can use an extension of the distributed computing scheme in~\cite{linearcomput2020wan} for the case $\Ksf_{\rm c}=1$ and  $\Msf= \frac{\Ksf}{\Nsf} (\Nsf-\Nsf_{\rm r} +1)$. 

{\it Assignment phase.} 
The cyclic assignment is used, which was   widely used in the  existing works on   the distributed computing problems~\cite{gradiencoding,improvedGC2017halbawi,MDSGC2018raviv,efficientgradientcoding,adaptiveGC2020,linearcomput2020wan,cctradeoff2020wan}. More precisely,
we divide all the $\Ksf$ datasets into $\Nsf$ non-overlapping and equal-length groups, 
where the $i^{\text{th}}$ group for each $i\in [\Nsf]$ is $\Gc_i=\{k\in [\Ksf]: \text{Mod}(k,\Nsf) =i\}$ containing $\frac{\Ksf}{\Nsf}$ datasets.\footnote{\label{foot:recall mod}  Recall that by convention, we let  $\text{Mod}(b,a) =a$ if $a$ divides $b$.}
We assign all datasets  in $\Gc_i$ to the servers in  
\begin{align}
\Hc_i= \big\{\text{Mod}(i,\Nsf),  \text{Mod}(i-1,\Nsf),\ldots,  \text{Mod}(i-\Nsf+\Nsf_{\rm r} , \Nsf ) \big\}. \label{eq:Hk in cyclic assignment}
\end{align}  

Thus the set of groups assigned  to  server $n \in [\Nsf]$   is 
\begin{align}
\Zc^{\prime}_n=   \big\{\text{Mod}(n,\Nsf) , \text{Mod}(n+1,\Nsf)  , \ldots, \text{Mod}(n+\Nsf-\Nsf_{\rm r} ,\Nsf)  \big\}  
 \label{eq:cyclic assignment n divides k}
\end{align}
with cardinality $\Nsf-\Nsf_{\rm r} +1$.

{\it Computing phase.}
 For each $i\in [\Nsf]$,  we define a merged message as  $W^{\prime}_i=\sum_{k\in [\Ksf]: \text{Mod}(k,\Nsf)=i} W_k $.
All datasets in $\Gc_i$ are assigned to each server in  $\Hc_i$, which can compute $W^{\prime}_i$. 
We then introduce 
$Q$ as a set of 
$\Nsf_{\rm r}-1$ independent   randomness    variables $Q_{1},\ldots,Q_{\Nsf_{\rm r}-1}$, where
$Q_j, j\in [\Nsf_{\rm r}-1]$ is uniformly i.i.d. over $[\mathbb{F}_{\qsf}]^{\Lsf}$, and    we assign $Q$  to each server. 

In the computing phase, 
we let each server transmit one linear combination of  merged messages and randomness    variables, such that the user can receive  $\Nsf_{\rm r}$ linear combinations of merged messages from any set of $\Nsf_{\rm r}$ responding servers, and then recover ${\bf F^{\prime}} [W^{\prime}_{1}; \ldots; W^{\prime}_{\Nsf}; Q_1;\ldots; Q_{\Nsf_{\rm r}-1}]$ where\footnote{\label{foot:recall Mm} Recall that $(\mathbf{M})_{m \times n}$ represents the dimension of matrix $\mathbf{M}$ is $m \times n$.}
 \begin{equation}\setstretch{1.25}
{\bf F^{\prime}} =\begin{bmatrix}      \  
\tikzmark{left1} \textcolor{white}{0} 1 &\cdots &1 \textcolor{white}{0} & \textcolor{white}{0} 0 &\cdots & 0\textcolor{white}{0} \ \\  \
\textcolor{white}{0} * &\cdots  &  * \textcolor{white}{0} &\tikzmark{left2} \textcolor{white}{0}   +& \cdots & + \textcolor{white}{0} \ \\ \ 
\textcolor{white}{0} \vdots &\ddots & \vdots \textcolor{white}{0} &\textcolor{white}{0} \vdots & \ddots& \vdots \textcolor{white}{0}  \ \\ \
  \textcolor{white}{0}   * &\cdots  &  * \textcolor{white}{0} \tikzmark{right1}  &\textcolor{white}{0} + & \cdots & + \textcolor{white}{0} \tikzmark{right2} \   \end{bmatrix}. \label{eq:F prime cyclic secure}
\end{equation}
 \DrawBox[thick, black,  dashed ]{left1}{right1}{\textcolor{black}{\footnotesize$ ({\bf F} )_{(\Nsf_{\rm r}\times \Nsf)} $   }}
\DrawBox[thick, red, dashed]{left2}{right2}{\textcolor{red}{\footnotesize$({\bf S}^{\prime} )_{(\Nsf_{\rm r}-1) \times (\Nsf_{\rm r}-1)} $}}

Notice that each `$*$' represents a    symbol uniformly i.i.d over $\mathbb{F}_{\qsf}$, and `$+$' represents  the generic element  of the matrix (i.e., ${\bf S}^{\prime}$ can be any full-rank matrix over $[\mathbb{F}_{\qsf}]^{(\Nsf_{\rm r}-1) \times (\Nsf_{\rm r}-1)}$).
    The next step is to determine the transmission vector   of each server $n\in [\Nsf]$, denoted by $\sv_{n}$  where the  transmitted linear combination by server $n$ is 
\begin{align}
X_n= \sv_{n} \ {\bf F^{\prime}}  \ [W^{\prime}_{1}; \ldots; W^{\prime}_{\Nsf};Q_1;\ldots; Q_{\Nsf_{\rm r}-1} ]. \label{eq:exiting linear combination}
\end{align}
Notice that the number of merged messages which server $n$ cannot compute is $\Nsf_{\rm r}-1$ and that $Q_1,\ldots, Q_{\Nsf_{\rm r}-1}$  have been assigned to server $n$.
The sub-matrix of ${\bf F^{\prime}}$ including the   columns with the indices in $[\Nsf]\setminus \Zc^{\prime}_n$ has the dimension $ \Nsf_{\rm r}  \times  (\Nsf_{\rm r}-1)$. 
As each `$*$' is uniformly i.i.d. over $\mathbb{F}_{\qsf}$,   a vector basis for the left-side null space of this sub-matrix contains one linearly  independent vectors with high probability. 
Hence, we let $\sv_{n}$ be this left-side null space vector, such that in the linear combination~\eqref{eq:exiting linear combination}
 the coefficients  of the merged messages which server $n$ cannot compute are $0$. It was   proved in~\cite{linearcomput2020wan} that for each set $\Ac \subseteq [\Nsf]$ where $|\Ac|=\Nsf_{\rm r}$, the vectors $\sv_{n}$ where $n \in \Ac$  are linearly independent with high probability. Hence, the user can recover ${\bf F }^{\prime}   [W^{\prime}_{1}; \ldots; W^{\prime}_{\Nsf};Q_1;\ldots; Q_{\Nsf_{\rm r}-1} ]$ from the answer of workers in $\Ac$, which contains the desired task function.\footnote{\label{foot:pick one}Notice that if we choose the value of each   `$*$' uniformly i.i.d. over  $\mathbb{F}_{\qsf}$, the user can recover the desired task function with high probability. Hence, we can choose the values of `$*$'s such that the scheme is decodable.} 

 For the security, it can be seen that from the answers of all server, the user can only recover ${\bf F }^{\prime}   [W^{\prime}_{1}; \ldots; W^{\prime}_{\Nsf};Q_1;\ldots; Q_{\Nsf_{\rm r}-1} ]$ containing $\Nsf_{\rm r}$ linearly independent combinations. In addition, ${\bf S}^{\prime} $
is full-rank (with rank equal to $\Nsf_{\rm r}-1$). Hence, the user can only recover $W_{1}+\cdots+W_{\Ksf}$
without $Q_1,\ldots, Q_{\Nsf_{\rm r}-1}$, i.e.,  $H(X_{[\Nsf]}|W_{1},\ldots,W_{\Ksf})=H(X_{[\Nsf]}|W_{1}+\cdots+W_{\Ksf})=(\Nsf-1)\Lsf$, and thus the security
constraint in~\eqref{eq:security} holds. 

It can be seen that the communication cost of the proposed scheme is $\Nsf_{\rm r}$ and the size of randomness is  $\Nsf_{\rm r}-1$.
 \end{IEEEproof}

From Theorem~\ref{thm:optimal communication cost}, it can be seen  that the additional security constraint does not increase the communication cost. 
More interestingly, in Appendix~\ref{sec:extension of linear coding} we will show the following theorem.
\begin{thm}
\label{thm:extension of linear coding}
Any    linear-coding based computing scheme   for the   $(\Ksf,\Nsf,\Nsf_{\rm r},\Msf,\Ksf_{\rm c})$ non-secure distributed linearly separable computation problem where $\Ksf_{\rm c}\in [\Ksf]$,  $\Msf=\frac{\Ksf}{\Nsf}(\Nsf-\Nsf_{\rm r}+\msf)$ and $\msf\in [\Nsf_{\rm r}]$, can be made secure without increasing the communication cost.
\hfill $\square$ 
\end{thm}

From Theorem~\ref{thm:extension of linear coding}, we can add the security into the distributed computing schemes in~\cite{efficientgradientcoding,adaptiveGC2020} for the case that   $\Ksf_{\rm c}=1$, and also add the security into the distributed computing scheme  in~\cite{linearcomput2020wan} for the case that $\Msf=\frac{\Ksf}{\Nsf}(\Nsf-\Nsf_{\rm r}+1)$, without increasing the communication cost. 

In the rest of this paper, we focus on the $(\Ksf,\Nsf,\Nsf_{\rm r},\Ksf_{\rm c}, \Msf)$ secure distributed linearly separable computation problem with  $\Msf= \frac{\Ksf}{\Nsf} (\Nsf-\Nsf_{\rm r} +1)$ and $\Ksf_{\rm c}=1$, and aim to   minimize the randomness size $\eta$ while achieving the optimal communication cost $\Rsf^{\star}=\Nsf_{\rm r}$.

  We first introduce a  novel converse bound on $\eta$ for a fixed assignment, whose proof can be found in Appendix~\ref{sec:compound}.
\begin{thm}
\label{thm:converse lemma}
For the $(\Ksf,\Nsf,\Nsf_{\rm r},\Ksf_{\rm c}, \Msf)$ secure distributed linearly separable computation problem with  $\Msf= \frac{\Ksf}{\Nsf} (\Nsf-\Nsf_{\rm r} +1)$ and $\Ksf_{\rm c}=1$,
for a fixed     assignment $\mathbf{Z}=(\Zc_1,\ldots,\Zc_{\Nsf})$, if there  exists an ordered set of servers in $[\Nsf]$ denoted by $\sv =(s_1,\ldots,s_{|\sv| })$, such that 
\begin{align}
\Zc_{s_i} \setminus \big(\Zc_{s_1}\cup \cdots \Zc_{s_{i-1}}\big)\neq \emptyset, \ \forall i\in [|\sv|], \label{eq:vector constraint}
\end{align} 
it must hold that
\begin{align}
\eta \geq |\sv|-1. \label{eq:converse lemma}
\end{align}
\hfill $\square$ 
\end{thm}  
  Notice that while deriving the converse bound  in Theorem~\ref{thm:converse lemma}, we do not use the constraint that communication cost is minimum. Hence, it is a   converse on the  randomness size, which is also a converse bound on the randomness size while  achieving the optimal communication cost.

A general converse bound over all possible assignments can be directly obtained from Theorem~\ref{thm:converse lemma}. 
 \begin{cor}
\label{cor:general cor}
  For the $(\Ksf,\Nsf,\Nsf_{\rm r},\Ksf_{\rm c}, \Msf)$ secure distributed linearly separable computation problem with  $\Msf= \frac{\Ksf}{\Nsf} (\Nsf-\Nsf_{\rm r} +1)$ and $\Ksf_{\rm c}=1$,
 it must hold that
\begin{align}
\eta^{\star} \geq  \min_{\mathbf{Z}} \ \max_{\sv  :   \Zc_{s_i} \setminus \big(\Zc_{s_1}\cup \cdots \Zc_{s_{i-1}}\big)\neq \emptyset, \forall i\in [|\sv|]  } |\sv|-1. \label{eq:converse general cor}
\end{align}
 \hfill $\square$ 
\end{cor}

To solve the min-max optimization problem in~\eqref{eq:converse general cor} is highly combinatorial and becomes a part of on-going works. For some specific cases, this optimization problem has been solved in this paper  (see Theorems~\ref{thm:division} and~\ref{thm:exact optimality Mleq4}).  In the following, we provide a generally loosen version of the converse bound in~\eqref{eq:converse general cor}.
 \begin{cor}
\label{cor:converse cor}
For the $(\Ksf,\Nsf,\Nsf_{\rm r},\Ksf_{\rm c}, \Msf)$ secure distributed linearly separable computation problem with  $\Msf= \frac{\Ksf}{\Nsf} (\Nsf-\Nsf_{\rm r} +1)$ and $\Ksf_{\rm c}=1$,
 it must hold that
\begin{align}
\eta^{\star} \geq \left\lceil  \frac{\Nsf}{\Nsf-\Nsf_{\rm r}+1} \right\rceil -1. \label{eq:converse cor}
\end{align}
\hfill $\square$ 
\end{cor} 
  \begin{IEEEproof}
By definition, there are $\Ksf$ datasets in the library and we assign $\frac{\Ksf}{\Nsf}(\Nsf-\Nsf_{\rm r}+1)$ datasets to
each server.   Hence, for any possible assignment, we can   find 
$$
\left\lceil \frac{\Ksf}{\Msf} \right\rceil= \left\lceil \frac{\Nsf}{\Nsf-\Nsf_{\rm r}+1} \right\rceil
$$ servers, where each server has some dataset  which is not assigned to 
 other $\left\lceil \frac{\Nsf}{\Nsf-\Nsf_{\rm r}+1} \right\rceil -1 $ servers. By Theorem~\ref{thm:converse lemma}, we have 
 $\eta^{\star} \geq \left\lceil \frac{\Nsf}{\Nsf-\Nsf_{\rm r}+1} \right\rceil-1$. 
  \end{IEEEproof}

 We then characterize the optimal randomness size  for the case where $\Nsf-\Nsf_{\rm r}+1$ divides $\Nsf$.
\begin{thm}
\label{thm:division}
For the $(\Ksf,\Nsf,\Nsf_{\rm r},\Ksf_{\rm c}, \Msf)$ secure distributed linearly separable computation problem where  $\Msf= \frac{\Ksf}{\Nsf} (\Nsf-\Nsf_{\rm r} +1)$, $\Ksf_{\rm c}=1$, and $\Nsf-\Nsf_{\rm r}+1$ divides $\Nsf$, 
  to achieve the optimal communication cost, the minimum  randomness size is 
\begin{align}
&\eta^{\star}= \frac{\Nsf}{\Nsf-\Nsf_{\rm r}+1}-1 . \label{eq:gradient eta rho rep}   
\end{align}  
  \hfill $\square$ 
\end{thm}
\begin{IEEEproof}
The converse part of~\eqref{eq:gradient eta rho rep} directly comes from Corollary~\ref{cor:converse cor}. We then describe  the achievable scheme, which is based on the fractional repetition assignment   in~\cite{gradiencoding}.

We define that $\nsf:=\frac{\Nsf}{\Nsf-\Nsf_{\rm r}+1}  $ which is a positive integer. 
We divide the $\Ksf$ datasets into $\nsf $ groups, where the $i^{\text{th}}$ group is $\Dc_i=\left[ (i-1) \frac{\Ksf}{\nsf}+1 : i \frac{\Ksf}{\nsf} \right]$ for each $i\in \left[\nsf \right]$. 
We assign all datasets in $\Dc_i$ to servers in $[(i-1)(\Nsf-\Nsf_{\rm r}+1)+1 : i(\Nsf-\Nsf_{\rm r}+1) ]$.

In the computing phase, we introduce  $\nsf-1$ independent randomness    variables $Q_1,\ldots, Q_{\nsf-1}$, where $Q_j$ is uniformly i.i.d. over $[\mathbb{F}_{\qsf}]^{\Lsf}$. 

We let each server in $[ \Nsf-\Nsf_{\rm r}+1 ]$ compute
\begin{align}
A_1 =  Q_1 + \sum_{k\in \Dc_1} W_k ; \label{eq:A1}
\end{align}
 for   each $i\in \left[2: \nsf -1  \right]$, we let each server in $[(i-1)(\Nsf-\Nsf_{\rm r}+1)+1 : i(\Nsf-\Nsf_{\rm r}+1) ]$ compute
\begin{align}
A_i=   -Q_{i-1} +Q_i +   \sum_{k\in \Dc_i} W_k   ; \label{eq:Ai}
\end{align}
finally, we let each server in $\left[ \Nsf_{\rm r} : \Nsf \right]$ compute
\begin{align}
A_{\nsf} = -Q_{\nsf-1} + \sum_{k\in \Dc_{\nsf}} W_k . \label{eq:Alast}
\end{align}

Recall that each group has $\Nsf-\Nsf_{\rm r}+1$ servers.
 For the decodability, 
 from the answers of any $\Nsf_{\rm r}$ responding servers (i.e., there are $\Nsf_{\rm r}$ stragglers), 
the user always receives $A_1,A_2,\ldots, A_{\nsf}$. By summing $A_1,A_2,\ldots, A_{\nsf}$, the user recovers $W_{1}+\cdots+W_{\Ksf}$.

For the security, from the answers of all servers, the user receives $A_1,A_2,\ldots, A_{\nsf}$, totally $\nsf$ linear combinations. In the linear space of these linear combinations, there is  $W_{1}+\cdots+W_{\Ksf}$. The projection of this $\nsf$-dimensional linear space on $[Q_1;\ldots; Q_{\nsf-1}]$, has the dimension equal to $\nsf-1$. Hence, the user can only recover $W_{1}+\cdots+W_{\Ksf}$ without   $Q_1 , \ldots ,  Q_{\nsf-1}$.   
\end{IEEEproof} 
  
  In the following theorem, we focus on the cyclic assignment.
  \begin{thm}
  \label{thm:cyclic}
For the $(\Ksf,\Nsf,\Nsf_{\rm r},\Ksf_{\rm c}, \Msf)$ secure distributed linearly separable computation problem with  $\Msf= \frac{\Ksf}{\Nsf} (\Nsf-\Nsf_{\rm r} +1)$ and $\Ksf_{\rm c}=1$,  
 to achieve the optimal communication cost, the minimum  randomness size under the constraint of the cyclic assignment is
\begin{align}
\eta^{\star}_{\text{cyc}}=\Nsf_{\rm r}-1. \label{eq:gradient eta cyc} 
\end{align}  
  \hfill $\square$ 
\end{thm}  
\begin{IEEEproof}  
The achievability part   was   described in the proof of Theorem~\ref{thm:optimal communication cost}. In the following, we prove the converse part.

If the cyclic assignment is used, let us focus on an ordered set of  $\Nsf_{\rm r}$ neighbouring servers 
\begin{align}
\sv= \big(\Nsf_{\rm r}, \Nsf_{\rm r}-1,\ldots, 1 \big) . \label{eq:found max cyc}
\end{align} 
 For each $n\in [\Nsf_{\rm r}]$, 
dataset $D_n$ is assigned to servers in $\{n,\text{Mod}(n-1,\Nsf),\ldots, \text{Mod}(n-\Nsf+\Nsf_{\rm r},\Nsf) \}$; thus servers in $\{\Nsf_{\rm r},\Nsf_{\rm r}-1,\ldots, n+1 \}$ do not know $D_n$. 
Hence, the ordered set $\sv$  in~\eqref{eq:found max cyc} satisfies the constraint in~\eqref{eq:vector constraint}, and we have 
$
\eta^{\star}_{\text{cyc}}\geq |\sv|-1= \Nsf_{\rm r}-1 $, which proves~\eqref{eq:gradient eta cyc}.
  \end{IEEEproof}
  
  Comparing Theorems~\ref{thm:division} and~\ref{thm:cyclic}, it can be seen that the computing scheme with the cyclic assignment is highly sub-optimal where the multiplicative gap to the optimality could be unbounded.\footnote{\label{foot:unbounded example} For example, when $\Nsf=2(\Nsf-\Nsf_{\rm r}+1)$ and $\Nsf$ is very large, the optimal randomness size is $1$ as shown in~\eqref{eq:gradient eta rho rep}, while the needed randomness size of the computing scheme with the cyclic assignment is $\Nsf_{\rm r}-1=\frac{\Nsf}{2}$.} 
  However, when   $\Nsf-\Nsf_{\rm r}+1$ does not divide $\Nsf$, the fractional repetition assignment 
  in~\cite{gradiencoding} cannot be used.  
  On the observation that most of existing works on this distributed  linearly separable  computing problem (without security)
are either based on the cyclic assignment (such as~\cite{gradiencoding,MDSGC2018raviv,efficientgradientcoding,adaptiveGC2020,linearcomput2020wan,yangelastic2019,cctradeoff2020wan}) or the fractional repetition assignment (such as~\cite{gradiencoding,replicationcode2020}), we need to design new assignments for the considered secure computation problem.

For the ease of  notation, we define that 
\begin{align}
\Msf^{\prime}:=\Nsf-\Nsf_{\rm r}+1. \label{eq:M prime}
\end{align}

\begin{figure}
\centerline{\includegraphics[scale=0.5]{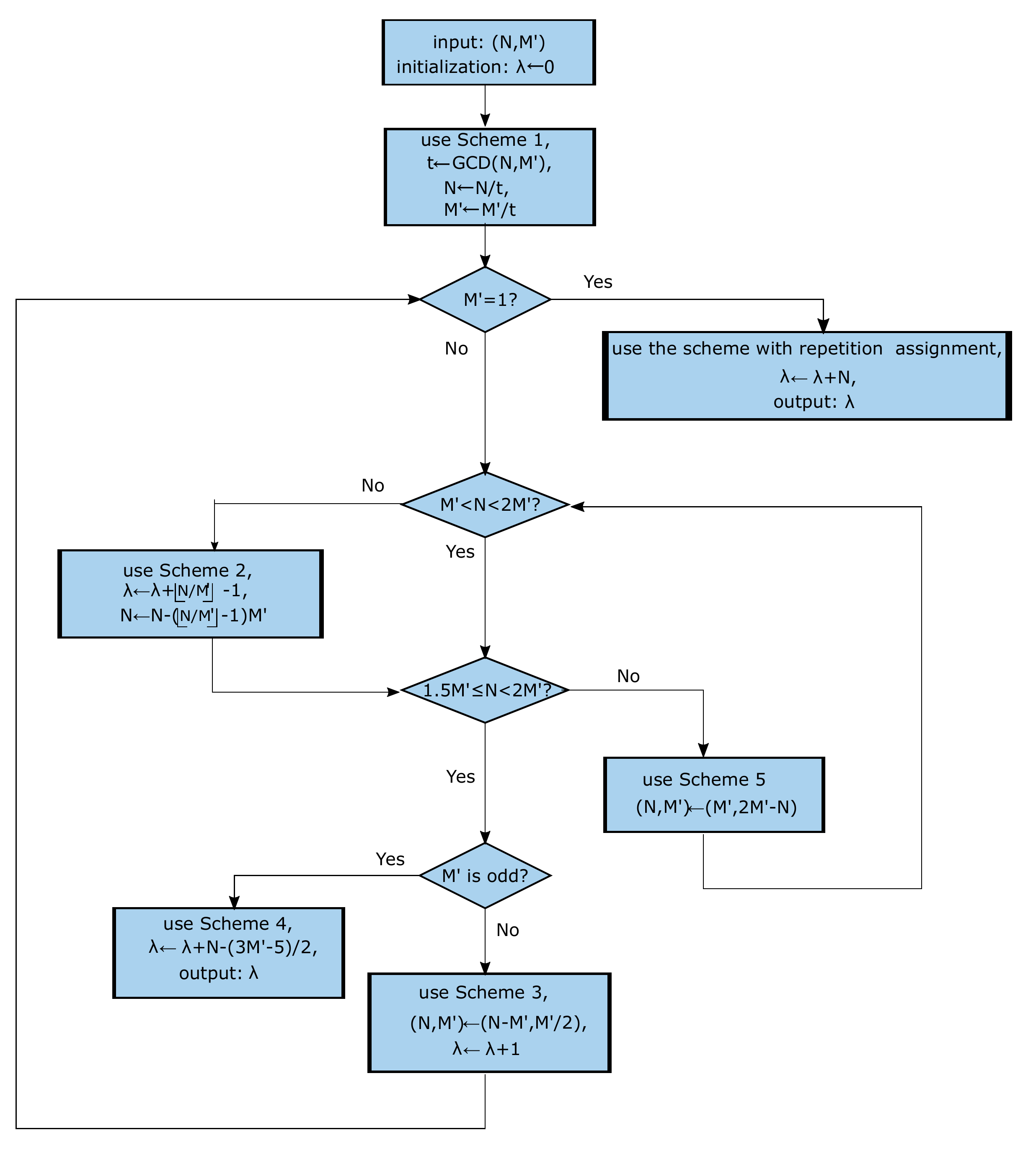}}
\caption{\small Flow diagram of the combined scheme in Theorem~\ref{thm:combined scheme}. Notice that the condition to use Scheme~5 is that $\Msf^{\prime}<\Nsf <1.5\Msf^{\prime}$; in this case, $2\Msf^{\prime}-\Nsf >1$.}
\label{fig: diagram}
\end{figure}

In Section~\ref{sec:achievable}, we will propose five novel achievable schemes for different ranges of system parameters. The performance of the combined scheme  
 given in the following theorem is based on a recursive  algorithm illustrated in Fig.~\ref{fig: diagram}, which will be explained in Remark~\ref{rem:highlight combined scheme}. 
\begin{thm}
\label{thm:combined scheme}
For the $(\Ksf,\Nsf,\Nsf_{\rm r},\Ksf_{\rm c}, \Msf)$ secure distributed linearly separable computation problem with  $\Msf= \frac{\Ksf}{\Nsf} \Msf^{\prime}$ and $\Ksf_{\rm c}=1$, 
 to achieve the optimal  communication cost, 
the    randomness size  $\eta=  h(\Nsf,\Msf^{\prime})-1$ is achievable,    where the function $h(\cdot,\cdot)$ has the following properties:
 \begin{itemize}
 \item By directly using the scheme with the  fractional  repetition assignment for Theorem~\ref{thm:division}, we have
\begin{align}
h(\Nsf,1)=\Nsf.\label{eq:direct rep}
\end{align}  
 \item By   {\it Scheme~1} described in Section~\ref{sub:GCD}  we have  
\begin{align}
h(\Nsf,\Msf^{\prime} )=h\left(\frac{\Nsf}{\text{GCD}(\Nsf, \Msf^{\prime})},\frac{\Msf^{\prime}}{\text{GCD}(\Nsf,\Msf^{\prime})} \right). \label{eq:from GCD}
\end{align} 
 \item For the case where  $\Nsf > 2\Msf^{\prime}$,  by {\it Scheme~2} described in Section~\ref{sub:partial rep} we have  
\begin{align}
h(\Nsf,\Msf^{\prime})=h\big(\Nsf-\left\lfloor \Nsf/\Msf^{\prime}-1\right\rfloor \Msf^{\prime},\Msf^{\prime}\big)+ \left\lfloor \Nsf/\Msf^{\prime} -1\right\rfloor  . \label{eq:from partial rep}
\end{align} 
 \item For the case where $1.5 \Msf^{\prime} \leq  \Nsf < 2 \Msf^{\prime}$ and $\Msf^{\prime}$  is even, by  {\it Scheme~3} described in Section~\ref{sub:M is even}  we have  
\begin{align}
 h(\Nsf,\Msf^{\prime})= h\left(\Nsf-\Msf^{\prime},\frac{\Msf^{\prime}}{2} \right) +1. \label{eq:M is even}
\end{align} 
 \item For the case where $1.5 \Msf^{\prime} \leq  \Nsf < 2 \Msf^{\prime}$ and $\Msf^{\prime}$ is odd, by  {\it Scheme~4} described in Section~\ref{sub:M is odd} we have  
 \begin{align}
 h(\Nsf,\Msf^{\prime})=   \Nsf-\frac{3\Msf^{\prime}-5}{2}; \label{eq:M is odd}
 \end{align}
 \item For the case where $\Msf^{\prime}< \Nsf <  1.5 \Msf^{\prime}$, by  {\it Scheme~5} described in Section~\ref{sub:less than 1.5M}  we have  
\begin{align}
h(\Nsf,\Msf^{\prime})=h(\Msf^{\prime},2\Msf^{\prime}-\Nsf). \label{eq:less than 1.5M}
\end{align} 
  \end{itemize}
    \hfill $\square$ 
 \end{thm} 
 
Notice that it can be seen that, the needed randomness size of the combined scheme for    Theorem~\ref{thm:combined scheme} is  
$$
h(\Nsf,\Msf^{\prime})-1 \leq \frac{\Ksf}{\Nsf}(\Nsf-\Msf^{\prime}+1)-1=\Nsf_{\rm r}-1,
$$
where $\Nsf_{\rm r}-1$ is the needed randomness size of the computing scheme with the cyclic assignment for  Theorem~\ref{thm:cyclic}.
In addition,
 the multiplicative gap 
between the needed randomness sizes of the computing scheme with the cyclic assignment for  Theorem~\ref{thm:cyclic}  and  
 the combined scheme for Theorem~\ref{thm:combined scheme}, could be unbounded. We provide some examples:  (i) let us focus on the example where $\Ksf=\Nsf= n \Msf +1$ and $\Msf$ does not divide $\Nsf$. By  Scheme~2, $h(\Nsf,\Msf^{\prime})  = h(\Nsf,\Msf )= h (\Msf+1,\Msf)+n-1 $; by Scheme~5,  $h(\Msf+1,\Msf)=2$. Hence, the needed randomness size is $h(\Nsf,\Msf^{\prime})-1=n$, while that  of the computing scheme with the cyclic assignment is $\Nsf_{\rm r}-1=\Nsf-\Msf= (n-1)\Msf +1$. (ii) We then focus on the example where $\Ksf=\Nsf=1.5\Msf$. By Scheme~3, $h(\Nsf,\Msf^{\prime}) = h(\Nsf,\Msf ) =h (0.5\Msf,0.5\Msf)+1$; by Scheme~1,  $h (0.5\Msf,0.5\Msf)=1$. Hence, the needed   randomness size is $h(\Nsf,\Msf^{\prime})-1=1$,  while that  of the computing scheme with the cyclic assignment is $\Nsf_{\rm r}-1=\Nsf-\Msf= 0.5 \Msf$.  (iii) Finally, we focus on the example where $\Ksf=\Nsf=\frac{3\Msf+1}{2}$ and $\Msf$ is an odd not dividing $\Nsf$. By Scheme~4, $h(\Nsf,\Msf^{\prime})=  h(\Nsf,\Msf )=3$. Hence, the   needed   randomness size is $h(\Nsf,\Msf^{\prime})-1=2$,  while that  of the computing scheme with the cyclic assignment is $\Nsf_{\rm r}-1=\Nsf-\Msf=\frac{ \Msf+1}{2}$.  

\begin{rem}[High-level ideas for Theorem~\ref{thm:combined scheme}]
\label{rem:highlight combined scheme}
\em
We divide the $\Ksf$ datasets into $\Nsf$ non-overlapping and  equal-length groups, where the $i^{\text{th}}$ group denoted by  $\Gc_i=\{k\in [\Ksf]: \text{Mod}(k,\Nsf) =i\}$ contains $\frac{\Ksf}{\Nsf}$ datasets, for each $i\in [\Nsf]$.
Group $\Gc_i$ is assigned to $\Msf^{\prime}=\Nsf-\Nsf_{\rm r}+1$ servers, each of which can compute the merged message $W^{\prime}_i$.
Hence, we treat the   $(\Ksf,\Nsf,\Nsf_{\rm r},1, \Msf) $ secure distributed linearly separable computation problem as the $(\Nsf,\Nsf,\Nsf_{\rm r},1,\Msf^{\prime})$ secure distributed linearly separable computation problem.

As   in Appendix~\ref{sec:extension of linear coding}, the design on the computing phase contains two stages. 
\begin{itemize}
\item In the first stage, we do not consider the security constraint in~\eqref{eq:security}.  We   let each server send one linear combination of merged messages which it can compute, such that from any set of $\Nsf_{\rm r}$ responding servers, the user can recover $W^{\prime}_1+\cdots +W^{\prime}_{\Nsf}$. 
Assume that from the answers of all servers, the user can recover ${\bf F} [W^{\prime}_1;\ldots;W^{\prime}_{\Nsf}]$ where the  dimension of ${\bf F}$ is   $\lambda \times \Nsf$ and $\lambda$ represents the number of totally transmitted linearly independent combinations of merged messages.
Thus the transmission of  server $n\in [\Nsf]$ can be expressed as $\sv_n {\bf F} [W^{\prime}_1;\ldots;W^{\prime}_{\Nsf}]$, where $\sv_n$ represents the transmission vector of server $n$.
 \item In the second stage, we take the security constraint in~\eqref{eq:security} into consideration.  We introduce $\lambda-1$ independent randomness    variables $Q_{1}, \ldots, Q_{\lambda-1}$, where $Q_i, i\in [\lambda-1]$ is uniformly i.i.d. over $[\mathbb{F}_{\qsf}]^{\Lsf}$. 
We then generate the matrix ${\bf F}^{\prime}=[({\bf F})_{\lambda \times \Nsf}, ({\bf S})_{\lambda \times (\lambda-1)}]$, where $
    \mathbf{S}= [{\bf 0}_{1 \times (\lambda- 1 )}; \mathbf{S}^{\prime} ]
$
    and 
    $
    \mathbf{S}^{\prime}
    $     is full-rank with dimension $(\lambda-1) \times (\lambda-1 )$.

   We let each server $n\in [\Nsf]$ transmit 
${\bf s}_n   {\bf F}^{\prime}  [W_{1,1} ;\ldots;W_{\Ksf};Q_1;\ldots;Q_{\lambda-1}].$
It is proved in Appendix~\ref{sec:extension of linear coding} that the resulting scheme is decodable and secure. The needed  randomness size $\eta$ is equal to $\lambda-1$.
\end{itemize}
The second stage can be immediately obtained once the first stage is fixed. Hence, now we only need to focus on the first stage  where 
 we aim to minimize the number of totally transmitted linearly independent combinations (i.e., $\lambda$) for the $(\Nsf,\Nsf,\Nsf_{\rm r},1,\Msf^{\prime})$ non-secure  distributed linearly separable computation problem (for the sake of simplicity, we will call it $(\Nsf,\Msf^{\prime})$ non-secure problem since $\Nsf_{\rm r}=\Nsf-\Msf^{\prime}+1$).  
   Notice that if in the first stage ${\bf F}$ is chosen as that in~\eqref{eq:F prime cyclic secure} where all elements outside the first line are chosen i.i.d. over $\mathbb{F}_{\qsf}$, the computing scheme becomes the scheme with the cyclic assignment for Theorem~\ref{thm:cyclic}.  To reduce the number of totally transmitted linearly independent combinations, in the combined  scheme  for Theorem~\ref{thm:combined scheme} we design more structured ${\bf F}$.

The flow diagram of the combined scheme for Theorem~\ref{thm:combined scheme}  where we have $\lambda=h(\Nsf,\Msf^{\prime})$,  is given in Fig.~\ref{fig: diagram}.  The procedure in the flow diagram is finished when either $\Msf^{\prime}=1$ or $1.5 \Msf^{\prime} \leq \Nsf< 2\Msf^{\prime}$ and $\Msf^{\prime}$ is odd. There must exist an output for each input case because when none of the above two constraints  are satisfied, $\Msf^{\prime}$ will be further reduced.
   \hfill $\square$ 
  \end{rem}

Comparing the achievable scheme in Theorem~\ref{thm:combined scheme} with the proposed converse bounds in Corollaries~\ref{cor:general cor} and~\ref{cor:converse cor}, we can characterize the following optimality result, whose proof could be found in Appendix~\ref{sec:exact optimality Mleq4}.
\begin{thm}
\label{thm:exact optimality Mleq4}
For the $(\Ksf,\Nsf,\Nsf_{\rm r},\Ksf_{\rm c},\Msf)$ secure distributed linearly separable computation problem with  $\Msf= \frac{\Ksf}{\Nsf} \Msf^{\prime}$, $\Ksf_{\rm c}=1$, and $\frac{\Msf^{\prime}}{\text{GCD}(\Nsf, \Msf^{\prime})} \leq 4$, 
 to achieve  the optimal  communication cost, 
 the minimum randomness size is $h(\Nsf,\Msf^{\prime})-1$, where   $h(\cdot,\cdot)$ is defined in Theorem~\ref{thm:combined scheme}.
    \hfill $\square$ 
\end{thm}
Notice that when $\frac{\Msf^{\prime}}{\text{GCD}(\Nsf, \Msf^{\prime})}=1$, we have that $\Msf^{\prime}$ divides $\Nsf$;  in this case Theorem~\ref{thm:exact optimality Mleq4}
reduces to Theorem~\ref{thm:division}.

\begin{figure}
    \centering
    \begin{subfigure}[t]{0.5\textwidth}
        \centering
        \includegraphics[scale=0.6]{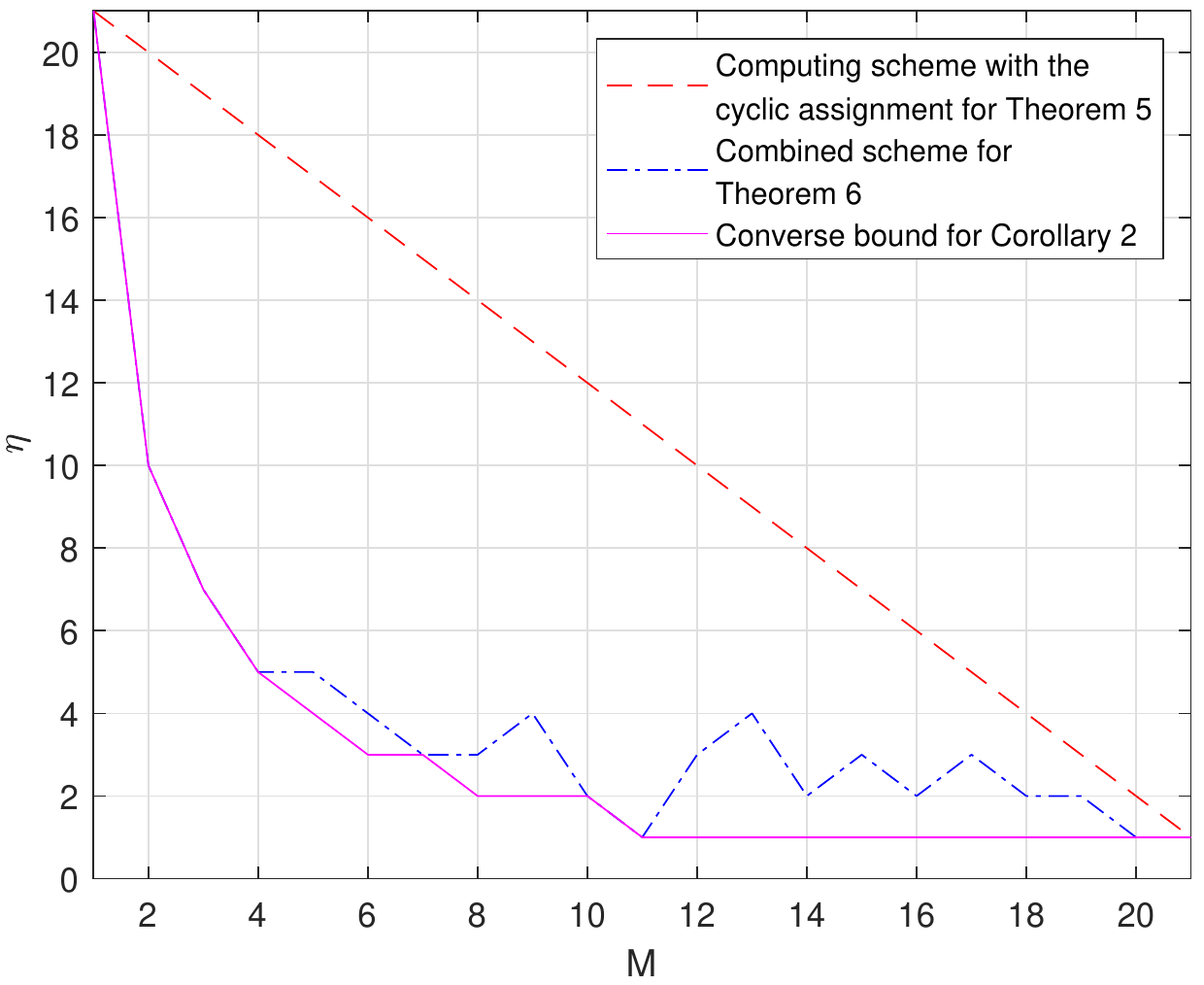}
        \caption{\small $(\Msf,\eta)$ tradeoff with $\Nsf=22$.}
        \label{fig:numerical 1a}
    \end{subfigure}%
    ~ 
    \begin{subfigure}[t]{0.5\textwidth}
        \centering
        \includegraphics[scale=0.6]{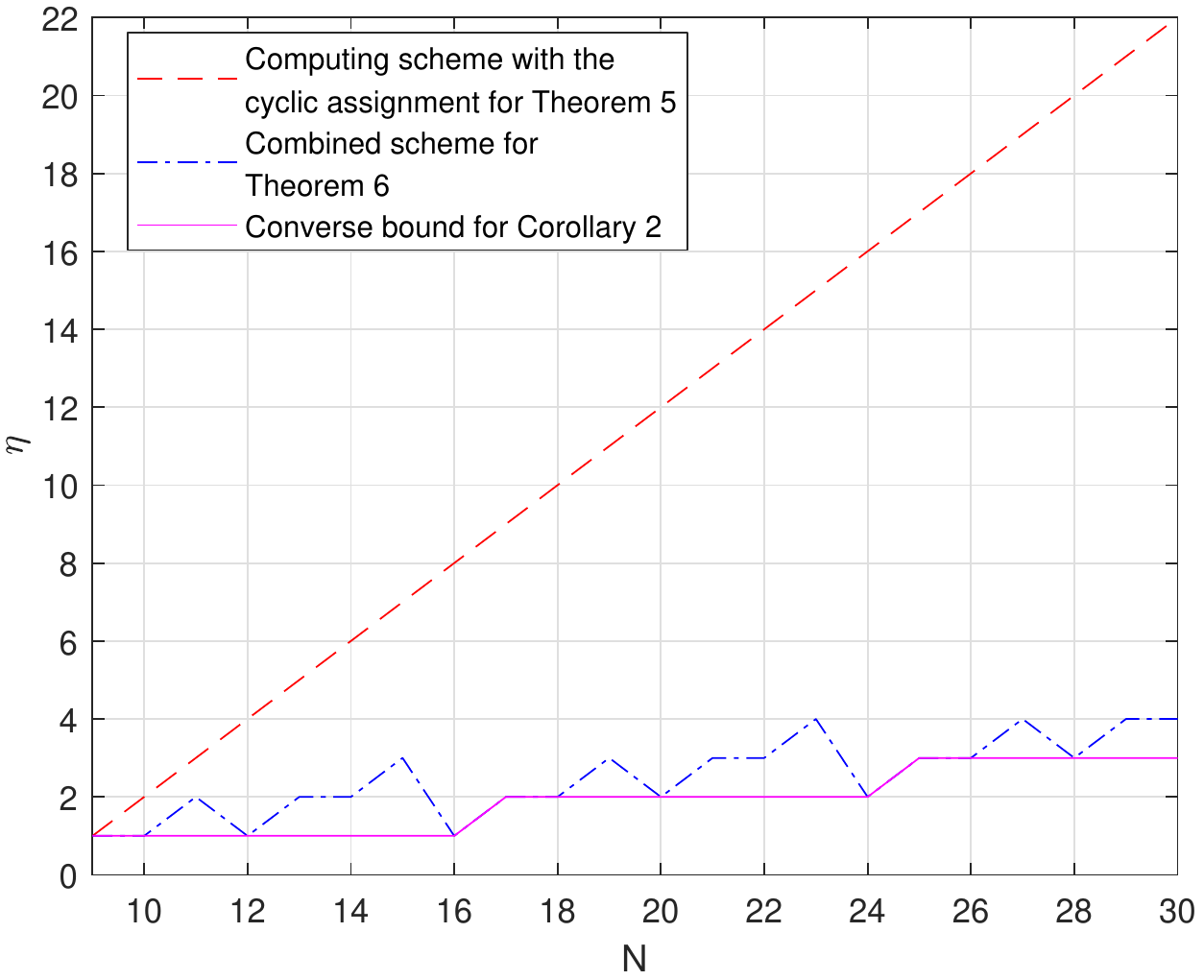}
        \caption{\small $(\Nsf,\eta)$ tradeoff with $\Msf=8$.}
        \label{fig:numerical 1b}
    \end{subfigure}\\
     \begin{subfigure}[t]{0.5\textwidth}
        \centering
        \includegraphics[scale=0.6]{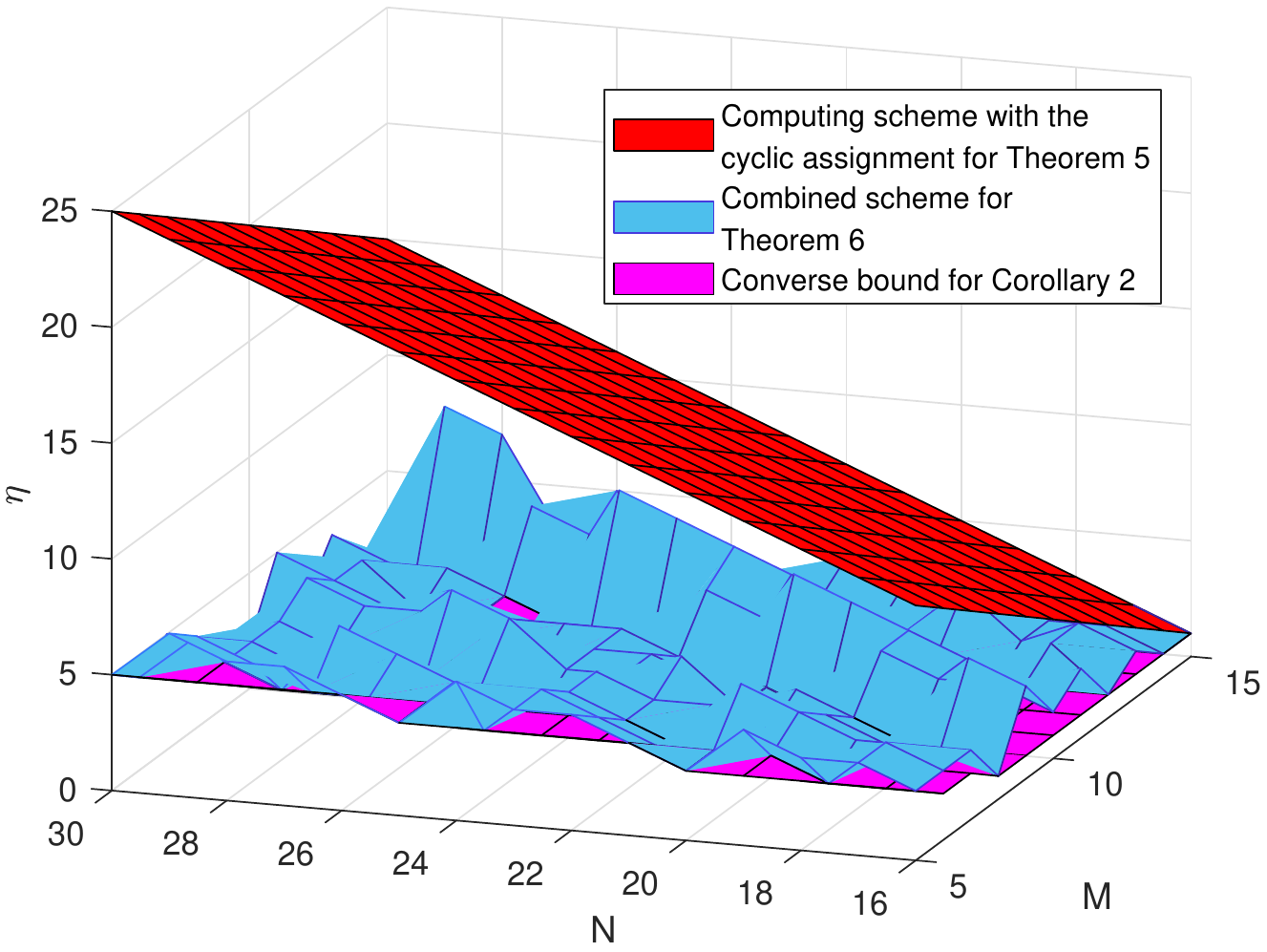}
        \caption{\small $(\Nsf,\Msf,\eta)$ tradeoff.}
        \label{fig:numerical 1c}
    \end{subfigure}
    \caption{\small Numerical evaluations for the considered secure distributed linearly separable computation problem.}
    \label{fig:numerical 1}
\end{figure}

At the end of this section, we provide some numerical evaluations to compare the needed randomness sizes of the computing scheme with the cyclic assignment for Theorem~\ref{thm:cyclic} (equal to $\Nsf_{\rm r}-1$) and the combined scheme for Theorem~\ref{thm:combined scheme} (equal to $h(\Nsf,\Msf^{\prime})-1$), while achieving the optimal communication cost. We consider the $(\Ksf,\Nsf,\Nsf_{\rm r},\Ksf_{\rm c},\Msf) $ secure distributed linearly separable computation problem with   $\Ksf=\Nsf$, $\Msf= \frac{\Ksf}{\Nsf} \Msf^{\prime}$, and $\Ksf_{\rm c}=1$.
In Fig.~\ref{fig:numerical 1a}, we fix $\Nsf=22$ and plot the tradeoffs between $\Msf$ and $\eta$. 
In Fig.~\ref{fig:numerical 1b}, we fix $\Msf=8$ and plot the tradeoffs between $\Nsf$ and $\eta$. 
 In Fig.~\ref{fig:numerical 1c}, we plot the $(\Nsf,\Msf,\eta)$ tradeoffs for the case where $16\leq \Nsf \leq 30$ and $5\leq \Msf \leq 15$.
 From all figures, it can be seen that the combined scheme for Theorem~\ref{thm:combined scheme} needs a much lower randomness size than that in Theorem~\ref{thm:cyclic}.

\section{Novel Achievable Schemes for Theorem~\ref{thm:combined scheme}}
\label{sec:achievable}
As explained in Remark~\ref{rem:highlight combined scheme}, by a grouping strategy, 
we treat the   $(\Ksf,\Nsf,\Nsf_{\rm r},1,\Msf)$ secure distributed linearly separable computation problem as the $(\Nsf,\Nsf,\Nsf_{\rm r},1,\Msf^{\prime})$ secure distributed linearly separable computation problem.  
For the ease of notation, in this section we directly consider the case where $\Ksf=\Nsf$;  thus we also have $\Msf=\Msf^{\prime}$.
 
The proposed schemes for Theorem~\ref{thm:combined scheme} contain two stages, where in the first stage we consider a 
$(\Nsf,\Nsf,\Nsf_{\rm r},1,\Msf)$ non-secure distributed linearly separable computation problem (a.k.a., $(\Nsf,\Msf)$ non-secure  problem), 
  and aim to minimize the number of totally transmitted linear combinations of messages $\lambda$ while achieving the optimal communication cost $\Nsf_{\rm r}$; then the second stage can be 
  immediately obtained by introducing $\lambda-1$ independent randomness    variables such that the security is guaranteed. 
 Because the second stage is unified for each proposed scheme, we only present the first stage (i.e., the $(\Nsf,\Msf)$ non-secure  problem) in the rest of this section.

\subsection{Scheme~1 for~\eqref{eq:from GCD}}
\label{sub:GCD}
 We consider the $(\Nsf,\Msf)$ non-secure problem where $\text{GCD}(\Nsf,\Msf)>1$, and aim to construct a scheme (Scheme~1) to prove~\eqref{eq:from GCD}. Intuitively,  we want to consider a set of $\text{GCD}(\Nsf,\Msf)$ messages as a single message and a set of $\text{GCD}(\Nsf,\Msf)$ servers as a single server. Thus
  Scheme~1 is a recursive scheme which is based on  the proposed scheme 
 for the  $\left( \frac{\Nsf}{\text{GCD}(\Nsf,\Msf)}  , \frac{ \Msf}{\text{GCD}(\Nsf,\Msf)}  \right )$ non-secure problem. We assume that the latter scheme has been designed before, whose number of totally transmitted linearly independent combinations of messages is $h\left( \frac{\Nsf}{\text{GCD}(\Nsf,\Msf)}  , \frac{ \Msf}{\text{GCD}(\Nsf,\Msf)}  \right )$.

 We first partition the $\Nsf$ datasets into $\frac{\Nsf}{\text{GCD}(\Nsf,\Msf)}$ groups, where the $i^{\text{th}}$ group is 
 $$
 \Kc_{i}= \left[(i-1) \text{GCD}(\Nsf,\Msf)+1 : i \ \text{GCD}(\Nsf,\Msf)\right]
 $$
  for each $i\in \left[\frac{\Nsf}{\text{GCD}(\Nsf,\Msf)} \right]$.
 In addition, we let $M_i= \sum_{k\in \Kc_{i} } W_{k}$; thus the task function could be expressed as 
 $
 W_{1}+\cdots+W_{\Nsf}= M_1+\cdots +M_{\frac{\Nsf}{\text{GCD}(\Nsf,\Msf)}}.
 $
  
We also partition the $\Nsf$ servers into $\frac{\Nsf}{\text{GCD}(\Nsf,\Msf)}$ groups, where the   $i^{\text{th}}$ group of servers is $$
\Uc_{i}= \left[(i-1) \text{GCD}(\Nsf,\Msf)+1 : i \  \text{GCD}(\Nsf,\Msf)\right]
$$
 for each $i\in \left[\frac{\Nsf}{\text{GCD}(\Nsf,\Msf)} \right]$.

We now prove that the  the proposed scheme for the  $\left( \frac{\Nsf}{\text{GCD}(\Nsf,\Msf)}  , \frac{ \Msf}{\text{GCD}(\Nsf,\Msf)}  \right )$ non-secure problem can be directly applied to  the $(\Nsf,\Msf)$ non-secure problem. 

In the proposed scheme for the  $\left( \frac{\Nsf}{\text{GCD}(\Nsf,\Msf)}  , \frac{ \Msf}{\text{GCD}(\Nsf,\Msf)}  \right )$ non-secure problem, we assume that
the set of assigned datasets to each server $n^{\prime} \in \left[\frac{\Nsf}{\text{GCD}(\Nsf,\Msf)}  \right] $ is  $\Zc^{\prime}_{n^{\prime}} \subseteq \left[\frac{\Nsf}{\text{GCD}(\Nsf,\Msf)}\right]$; obviously, $|\Zc^{\prime}_{n^{\prime}}|=  \frac{ \Msf}{\text{GCD}(\Nsf,\Msf)} $. 
In the computing phase, 
server $n^{\prime}$ computes a linear combination of the $\frac{\Nsf}{\text{GCD}(\Nsf,\Msf)}$ messages, where the coefficients of the messages with indices in $\left[\frac{\Nsf}{\text{GCD}(\Nsf,\Msf)}\right] \setminus  \Zc^{\prime}_{n^{\prime}} $ are $0$. We assume that the vector  of the coefficients in this linear combination  is $\vv_{n^{\prime}}$, containing $\frac{\Nsf}{\text{GCD}(\Nsf,\Msf)}$ elements.
From the answers of any $\frac{\Nsf}{\text{GCD}(\Nsf,\Msf)}-\frac{ \Msf}{\text{GCD}(\Nsf,\Msf)}+1$ servers, the user can recover the sum of the $\frac{\Nsf}{\text{GCD}(\Nsf,\Msf)}$ messages.

We then apply the above scheme to  the $(\Nsf,\Msf)$ non-secure problem.  

{\it Assignment phase.}
For each $i\in \left[ \frac{\Nsf}{\text{GCD}(\Nsf,\Msf)}\right]$, we assign all datasets in group $\Kc_{i}$ to each server in group $\Uc_{j}$ where $j\in \left[\frac{\Nsf}{\text{GCD}(\Nsf,\Msf)}  \right]$ and  $i\in \Zc^{\prime}_j$. 
As each group of servers contains $\text{GCD}(\Nsf,\Msf)$ servers,  each dataset is assigned 
to 
$
\text{GCD}(\Nsf,\Msf) \frac{ \Msf}{\text{GCD}(\Nsf,\Msf)} =\Msf
$
servers;
as  each group of datasets contains $\text{GCD}(\Nsf,\Msf)$, the number of datasets assigned to each server is
$
\text{GCD}(\Nsf,\Msf)  \frac{ \Msf}{\text{GCD}(\Nsf,\Msf)} =\Msf
$.
Thus the  assignment constraints are satisfied.

{\it Computing phase.}
For each $i\in \left[ \frac{\Nsf}{\text{GCD}(\Nsf,\Msf)}\right]$, we let each server in group $\Uc_i$ compute 
 $$
  \vv_n \ \left[ M_1;\ldots;M_{\frac{\Nsf}{\text{GCD}(\Nsf,\Msf)}} \right],
 $$
 where $\vv_n$ represents the   vector  of the coefficients in the linear combination sent by server $n$ in the $\left( \frac{\Nsf}{\text{GCD}(\Nsf,\Msf)}  , \frac{ \Msf}{\text{GCD}(\Nsf,\Msf)}  \right )$ non-secure problem.

{\it Decoding phase.}
Following the original  scheme for the  $\left( \frac{\Nsf}{\text{GCD}(\Nsf,\Msf)}  , \frac{ \Msf}{\text{GCD}(\Nsf,\Msf)}  \right )$ non-secure problem,
for any set $\Ac \subseteq \left[  \frac{\Nsf}{\text{GCD}(\Nsf,\Msf)}\right]$ where $|\Ac|=\frac{\Nsf}{\text{GCD}(\Nsf,\Msf)}-\frac{ \Msf}{\text{GCD}(\Nsf,\Msf)}+1$,
 if the user receives the answers of the servers in $\Ac$, it can recover the task function.

Let us go back to the   $(\Nsf,\Msf)$ non-secure problem. The user can receive the answers of $\Nsf-\Msf+1$ servers. 
As each group of servers contains $\text{GCD}(\Nsf,\Msf)$ servers,
it can be seen that these $\Nsf-\Msf+1$ servers  are from at least 
$\left\lceil  \frac{\Nsf-\Msf+1}{\text{GCD}(\Nsf,\Msf)} \right\rceil = \frac{\Nsf}{\text{GCD}(\Nsf,\Msf)}-\frac{ \Msf}{\text{GCD}(\Nsf,\Msf)}+1$ groups. Hence, the user recovers the task function.  

 In conclusion, we proved $h(\Nsf,\Msf)=h   \left( \frac{\Nsf}{\text{GCD}(\Nsf,\Msf)}  , \frac{ \Msf}{\text{GCD}(\Nsf,\Msf)}  \right ) $, coinciding with~\eqref{eq:from GCD}.

\subsection{Scheme~2 for~\eqref{eq:from partial rep}}
\label{sub:partial rep}
We  will start with an example to illustrate the main idea.
\begin{example}
\label{ex:two part example}
We consider the $(\Nsf,\Msf)=(5,2)$ non-secure  problem. It can be seen that in this example $\Nsf_{\rm r}= \Nsf-\Msf +1=4$. For the sake of simplicity, while illustrating the proposed schemes through examples, we assume that the field is a large enough prime field. It will be proved that in general this assumption is not necessary  in our proposed schemes. 

{\it Assignment phase.}  
We assign the datasets as follows.  
\begin{align*}
\begin{array}{rl|c|c|c|c|c|}\cline{3-3}\cline{4-4}\cline{5-5}\cline{6-6}\cline{7-7}
&&\rule{0pt}{1.2em}\mbox{Server 1} & \rule{0pt}{1.2em}\mbox{Server 2} &  \rule{0pt}{1.2em}\mbox{Server 3}&  \rule{0pt}{1.2em}\mbox{Server 4} &  \rule{0pt}{1.2em}\mbox{Server 5}  \\ \cline{3-3}\cline{4-4}\cline{5-5}\cline{6-6}\cline{7-7}
&& D_1&  D_1  &  D_3&  D_4 &  D_5\\
&& D_2&  D_2 &  D_4 & D_5 &  D_3\\ \cline{3-3}\cline{4-4}\cline{5-5}\cline{6-6}\cline{7-7}
\end{array}
\end{align*}

{\it Computing phase.}  
We let  servers $1$ and $2$ compute 
$
W_1 +W_2.
$
 
 We then focus on servers $3,4,5$. It can be seen that datasets $D_4, D_5, D_6$ are assigned to servers $3,4,5$ in a cyclic way. Hence, as the computing scheme illustrated in the Introduction,  
 we let  server $3$ compute $ 2W_3 +W_4$; let server $4$ compute 
 $
  W_4 + 2W_5 
 $;
 let server $5$ compute  $W_3-W_5$.

 {\it Decoding phase.}
 Among the answers of any $\Nsf_{\rm r}=4$ servers,
 there must exist $W_1+W_2$ and two answers of servers $3,4,5$.
 From any two answers of  servers $3,4,5$, the user can recover $W_3+W_4+W_5$. Together with $W_1+W_2$, the user can recover $W_1+\cdots+W_5$.
 
 It can be seen that the number of linearly independent combinations transmitted by servers $3,4,5$ is two. 
  Hence,    the number of totally transmitted linearly independent combinations is $h(5,2)=3$, which is equal to $h(3,2)+1$ coinciding with~\eqref{eq:from partial rep}.
   \hfill $\square$ 
\end{example}

We now consider  the $(\Nsf,\Msf)$ non-secure problem where $\Nsf > 2\Msf$, and   aim to construct a scheme (Scheme~2) to prove~\eqref{eq:from partial rep}. Scheme~2 is a recursive scheme which is based on  the proposed scheme 
 for the  $\left(\Nsf-\left\lfloor  \Nsf/\Msf -1 \right\rfloor\Msf, \Msf  \right )$ non-secure problem. We assume that the latter scheme has been designed before, whose number of totally transmitted linearly independent combinations of messages is $h\left(\Nsf-\left\lfloor  \Nsf/\Msf -1 \right\rfloor\Msf, \Msf  \right )$.

{\it Assignment phase.}
We divide the whole system into $\left\lfloor  \Nsf/\Msf \right\rfloor $ blocks. For each $i\in [\left\lfloor  \Nsf/\Msf \right\rfloor]$, 
  the $i^{\text{th}}$ block contains datasets $\{ D_{k}: k\in \Bc_i \}$ and servers in $ \Bc_i $, where 
\begin{align}
\Bc_{i}=
 \begin{cases}  \left[(i-1)\Msf+1: i \Msf \right]  , & \text{ if } i\in \left[\left\lfloor  \Nsf/\Msf-1 \right\rfloor \right]; \\  \left[\left\lfloor  \Nsf/\Msf -1 \right\rfloor\Msf+1 : \Nsf \right],  & \text{ if } i= \left\lfloor  \Nsf/\Msf \right\rfloor. \end{cases} 
\end{align}
The datasets in one block are only assigned to the servers in the same block. More precisely, for each $i\in [\left\lfloor  \Nsf/\Msf \right\rfloor]$,
\begin{itemize}
\item if $i\in \left[\left\lfloor  \Nsf/\Msf-1 \right\rfloor \right]$, we assign all datasets in $ \{ D_{k}: k\in \Bc_i \}$ to each server in  $\Bc_i$.
\item if $i= \left\lfloor  \Nsf/\Msf \right\rfloor$, the block contains $\Nsf-\left\lfloor  \Nsf/\Msf -1 \right\rfloor\Msf$ servers and $\Nsf-\left\lfloor  \Nsf/\Msf -1 \right\rfloor\Msf$ datasets, where each dataset should be assigned to $\Msf$ servers and each server should obtain $\Msf$ datasets.  Hence, we can apply the assignment phase of  the proposed scheme for the  $\left(\Nsf-\left\lfloor  \Nsf/\Msf -1 \right\rfloor\Msf, \Msf  \right )$ non-secure problem, to assign the datasets   $ \{ D_{k}: k\in \Bc_i \}$ to the servers in $\Bc_i $.
\end{itemize} 

{\it Computing phase.}  
For each $i\in \left[\left\lfloor  \Nsf/\Msf-1 \right\rfloor \right]$, we let the servers in the $i^{\text{th}}$ block compute
$$
 \sum_{k\in \Bc_i} W_k.
$$

We then focus on the $i^{\text{th}}$ block where  $i=\left\lfloor  \Nsf/\Msf \right\rfloor$ (i.e., the last block), to which we apply the computing phase of the proposed scheme for the    $\left(\Nsf-\left\lfloor  \Nsf/\Msf -1 \right\rfloor\Msf, \Msf  \right )$ non-secure problem. In the proposed scheme for the $\left(\Nsf-\left\lfloor  \Nsf/\Msf -1 \right\rfloor\Msf, \Msf  \right )$ non-secure problem,  
server $n^{\prime}\in \left[ \Nsf-\left\lfloor  \Nsf/\Msf -1 \right\rfloor\Msf \right]$ computes a linear combination of the $ \Nsf-\left\lfloor  \Nsf/\Msf -1 \right\rfloor\Msf$ messages, where the coefficients of the messages 
it cannot compute are $0$. We assume that the vector  of the coefficients in this linear combination  is $\vv_{n^{\prime}}$, containing $\Nsf-\left\lfloor  \Nsf/\Msf -1 \right\rfloor\Msf$ elements.

 Go back to  the  $i^{\text{th}}$ block where  $i=\left\lfloor  \Nsf/\Msf \right\rfloor$ of the $(\Nsf,\Msf)$ non-secure problem. For each $j\in [|\Bc_i|]$,  we let server $ \Bc_{i}(j)$ compute\footnote{\label{foot:Bcij}Recall that $\Bc_i(j)$ represents the $j^{\text{th}}$ element in $\Bc(i)$.}  
 $$
 \vv_j \ \left[ W_{\left\lfloor  \Nsf/\Msf -1 \right\rfloor\Msf+1}; W_{\left\lfloor  \Nsf/\Msf -1 \right\rfloor\Msf+2}; \ldots;W_{\Nsf} \right], 
 $$
where $\vv_j$ represents the vector  of the coefficients in the linear combination  sent by server $j$ in the  $\left(\Nsf-\left\lfloor  \Nsf/\Msf -1 \right\rfloor\Msf, \Msf  \right )$ non-secure problem.

{\it Decoding phase.}  
The user receives the answers of $\Nsf_{\rm r}=\Nsf-\Msf+1$ servers. In other words, the user does not receive the answers of  $\Msf-1$ servers.  Recall that in each of the  first $\left\lfloor  \Nsf/\Msf -1 \right\rfloor$ blocks there are $\Msf$ servers; in the last block there are $\Nsf-\left\lfloor  \Nsf/\Msf -1 \right\rfloor\Msf$ servers.
Hence, among these  $\Nsf_{\rm r}=\Nsf-\Msf+1$ responding servers, there must be at least one server in each of the  first $\left\lfloor  \Nsf/\Msf -1 \right\rfloor$ blocks, and at least $\Nsf-\left\lfloor  \Nsf/\Msf -1 \right\rfloor\Msf -\Msf +1 $ servers in the last block.
By construction, from the answers of any $\Nsf-\left\lfloor  \Nsf/\Msf -1 \right\rfloor\Msf -\Msf +1 $ servers in the last block, the user can recover 
$
\sum_{k\in \left[ \left\lfloor  \Nsf/\Msf -1 \right\rfloor\Msf+1: \Nsf \right]} W_k.
$
Together with the transmissions of the first $\left\lfloor  \Nsf/\Msf -1 \right\rfloor$ blocks, the user can recover $W_1 + \cdots +W_{\Nsf}$.

 In conclusion, we proved $h(\Nsf,\Msf)= \left\lfloor  \Nsf/\Msf -1 \right\rfloor + h   \left(  \Nsf-\left\lfloor  \Nsf/\Msf -1 \right\rfloor\Msf,\Msf   \right ) $, coinciding with~\eqref{eq:from partial rep}.
In addition, it can be seen  that $ \Msf <\Nsf-\left\lfloor  \Nsf/\Msf -1 \right\rfloor\Msf < 2\Msf$ if $\Nsf>2\Msf$ and $\Msf$ does not divide $\Nsf$.

\subsection{Scheme~3 for~\eqref{eq:M is even}}
\label{sub:M is even}
We first provide an example to illustrate the main idea.
\begin{example}
\label{ex:scheme 3 example}
We consider the $(\Nsf,\Msf)=(7,4)$ non-secure  problem. Notice that $\Nsf_{\rm r}= \Nsf-\Msf +1=4$.

{\it Assignment phase.}
We assign the datasets as follows.  
\begin{align*}
\begin{array}{rl|c|c|c|c|c|c|c|}\cline{3-3}\cline{4-4}\cline{5-5}\cline{6-6}\cline{7-7}\cline{8-8}\cline{9-9}
&&\rule{0pt}{1.2em}\mbox{Server 1}  &\rule{0pt}{1.2em}\mbox{Server 2}  & \rule{0pt}{1.2em}\mbox{Server 3} & \rule{0pt}{1.2em}\mbox{Server 4} & \rule{0pt}{1.2em}\mbox{Server 5} & \rule{0pt}{1.2em}\mbox{Server 6}    & \rule{0pt}{1.2em}\mbox{Server 7} \\ \cline{3-3}\cline{4-4}\cline{5-5}\cline{6-6}\cline{7-7}\cline{8-8}\cline{9-9}
&& D_1& D_1    & D_1 & D_1   & D_2  & D_3  & D_4 \\
&& D_2&  D_2   & D_5  & D_5   & D_5   & D_6 & D_7 \\ 
&& D_3&  D_3    &D_6  & D_6  & D_3  & D_4  & D_2 \\
&& D_4& D_4  & D_7   & D_7  & D_6   & D_7  & D_5 \\ 
\cline{3-3}\cline{4-4}\cline{5-5}\cline{6-6}\cline{7-7}\cline{8-8}\cline{9-9}
\end{array}
\end{align*}

{\it Computing phase.}  
We let servers $1,2$ compute a same linear combination of messages, assumed to be $A_1$. Similarly, we let servers $3,4$ compute a same linear combination of messages, assumed to be $A_2$.
Recall that $\Nsf_{\rm r}=4$. Thus from $A_1$ and $A_2$, the user should recover the task function.
We construct $A_1$ and $A_2$ such that from $A_1$ and $A_2$, we can recover the following two linear combinations,
 \begin{align*}
& F_1= W_1 + \cdots +W_7 ;\\
& F_2= W_2+W_3+W_4 +2(W_5+W_6+W_7).
\end{align*}
This can be done by letting 
$
A_1=  2 F_1 -F_2= W_1+W_2+W_3+W_4,
$
which can be computed by servers $1,2$, and letting 
$
A_2= F_2-F_1= -W_1+W_5+W_6+W_7,
$
which can be computed by servers $3,4$.

We then focus on servers $5,6,7$. The assignment for servers $5,6,7$ can be expressed as follows. We divide the datasets in $[2:7]$ into three pairs, $\Pc_1=\{2,5 \}$, $\Pc_2=\{3,6\}$, $\Pc_3 =\{4,7 \} $. The three pairs of datasets are assigned to servers $5,6,7$ in a cyclic way. We also let 
$P_1= W_2 +2W_5$, $P_2=W_3+2W_6$, $P_3=W_4+2W_7$. 
Hence, we can treat servers $5,6,7$ and $P_1,P_2,P_3$ as a  $(3,2)$ non-secure problem, where from the answers of any two servers we can recover $F_2=P_1+P_2+P_3$.
We construct the answers of servers $5,6,7$ (denoted by $A_3,A_4,A_5$, respectively) as 
$A_3= 2P_1+P_2$, $A_4=P_2+2P_3$, and $A_5=P_1- P_3$.

{\it Decoding phase.}
As shown before, if the set of $\Nsf_{\rm r}=4$ responding servers contains one server in $[2]$ and one server in $\{3,4\}$, the user can recover the task function from $A_1$ and $A_2$. 

We then consider the case where from the answers of the responding servers, the user can only receive one of $A_1$ and $A_2$. In this case, the set of responding servers must contain at least two servers in $[5:7]$. By construction, from the answers of any two servers in $[5:7]$, the user can recover $F_2$. 
Together with $A_1=2F_1-F_2$ or with $A_2=F_2-F_1$, the user can recover $F_1$, which is the task function.

  The number of totally transmitted linearly independent combinations is $h(7,4)=3$, which is equal to $h(3,2)+1$ coinciding with~\eqref{eq:M is even}.
 \hfill $\square$ 
\end{example}

We now consider  the $(\Nsf,\Msf)$ non-secure problem where $1.5 \Msf \leq  \Nsf < 2\Msf$ and $\Msf$ is even, and   aim to construct a scheme (Scheme~3) to prove~\eqref{eq:M is even}. Scheme~3 is a recursive scheme which is based on  the proposed scheme 
 for the  $\left(\Nsf-\Msf, \frac{\Msf}{2}\right)$ non-secure problem. We assume that the latter scheme has been designed before, whose number of totally transmitted linearly independent combinations of messages is $h\left(\Nsf-\Msf, \frac{\Msf}{2}\right)$.
 
We define that $\Nsf=2\Msf-\ysf$. In this case, we have  $\ysf\leq \Msf/2$ and
$
\Nsf_{\rm r}= \Nsf- \Msf+1  
= \Msf-\ysf +1 \leq \Msf.
$

{\it Assignment phase.}
We first focus on the assignment for the servers in $[\Msf]$, which is as follows. 
\begin{align*}
\begin{array}{rl|c|c|c|c|c|c|}\cline{3-3}\cline{4-4}\cline{5-5}\cline{6-6}\cline{7-7}\cline{8-8}
&&\rule{0pt}{1.2em}\mbox{Server 1} &\rule{0pt}{1.2em}\mbox{$\cdots$ } &  \rule{0pt}{1.2em}\mbox{Server $\frac{\Msf}{2}$ }&  \rule{0pt}{1.2em}\mbox{Server $\frac{\Msf}{2}  +1$}&  \rule{0pt}{1.2em}\mbox{$\cdots$ } &  \rule{0pt}{1.2em}\mbox{Server $\Msf$ } \\ 
\cline{3-3}\cline{4-4}\cline{5-5}\cline{6-6}\cline{7-7}\cline{8-8}
&& D_1&  \cdots       &   D_1 & D_{1}& \cdots &  D_{1} \\
&& \cdots &   \cdots &  \cdots &   \cdots &  \cdots &  \cdots  \\ 
&& D_{\ysf} &    \cdots    &  D_{\ysf} &  D_{\ysf}&  \cdots &  D_{\ysf}  \\
&& D_{\ysf+1}&  \cdots &  D_{\ysf+1} &  D_{\Msf+1}  & \cdots   & D_{\Msf+1} \\ 
&& \cdots & \cdots      &  \cdots &  \cdots& \cdots &  \cdots \\
&& D_{\Msf} &  \cdots  & D_{\Msf} & D_{\Nsf}&  \cdots &  D_{\Nsf} \\  
\cline{3-3}\cline{4-4}\cline{5-5}\cline{6-6}\cline{7-7}\cline{8-8}
\end{array} 
\end{align*}

It can be seen that we assign $D_{1},\ldots, D_{\ysf}$ to all servers in $[\Msf]$, and assign each dataset $D_k$ where $k\in [\ysf+1 : \Nsf]$ to $\frac{\Msf}{2}$  servers in $[\Msf]$. 

We then focus on the assignment for the servers in $[\Nsf-\Msf]$. 
We need to assign  $\Nsf-\ysf=2(\Nsf-\Msf)$ datasets (which are in $[\ysf+1 : \Nsf]$) to totally $\Nsf-\Msf$ servers, where each dataset is assigned to $\frac{\Msf}{2}$ servers and  each server obtains $\Msf$ datasets.
 We divide datasets  in $[\ysf+1 : \Nsf]$ into $\frac{\Nsf-\ysf}{2}=\Nsf-\Msf$ pairs,  where the $i^{\text{th}}$ pair is 
 $\Pc_i= \{\ysf+i,\Msf+i\}$ for each $i\in [\Nsf-\Msf]$.  
  Hence, we can apply the assignment phase of the proposed scheme for the $\left(\Nsf-\Msf, \frac{\Msf}{2}\right)$ non-secure problem, to assign $\frac{\Nsf-\ysf}{2}=\Nsf-\Msf$ pairs to $\Nsf-\Msf$ servers where each pair is assigned $\frac{\Msf}{2}$ servers and each server obtains $\frac{\Msf}{2}$ pairs.

{\it Computing phase.}  
 We first focus on the servers in $[\Msf]$. We let the  servers in $\left[\frac{\Msf}{2} \right]$ with the same datasets compute a same linear combination of messages, which 
 is denoted by $A_1$. Similarly, we let the  servers in $\left[\frac{\Msf}{2}+1 : \Msf \right]$ with the same datasets compute a same linear combination of messages, which 
 is denoted by $A_2$.
We construct $A_1$ and $A_2$ such that from $A_1$ and $A_2$, we can recover the following two linear combinations 
      \begin{subequations}
 \begin{align}
& F_1= W_1 + \cdots +W_{\Nsf} ;\label{eq: scheme 3 F1}\\
& F_2= W_{\ysf+1}+\cdots+ W_{\Msf}  + 2(W_{\Msf+1}+ \cdots +W_{\Nsf}). \label{eq: scheme 3 F2}
\end{align}
      \end{subequations}
 This can be done by letting 
 $$
 A_1 = 2F_1-F_2 =2 (W_1 + \cdots +   W_{\ysf}) + W_{\ysf+1} + \cdots +W_{\Msf}   
 $$
 which can be computed by servers in $\left[\frac{\Msf}{2} \right]$, and letting 
 $$
 A_2= F_2-F_1= -(W_1+ \cdots +W_{\ysf}) +W_{\Msf+1}+ \cdots +W_{\Nsf}
 $$
 which can be computed by servers in $\left[\frac{\Msf}{2}+1 : \Msf \right]$.
 
 We then focus on the servers in $[\Msf+1: \Nsf]$. 
  For each pair of datasets $\Pc_i=\{\ysf+i,\Msf+i\}$ where $i\in [\Nsf-\Msf]$, we let $P_i = W_{\ysf+i}+ 2 W_{\Msf+i}$. Hence, we can  express $F_2$ in~\eqref{eq: scheme 3 F2} as $P_1+\cdots + P_{\Nsf-\Msf}$.
 Next  we apply the computing phase of the proposed scheme for the    $\left(\Nsf-\Msf, \frac{\Msf}{2}\right)$   non-secure problem.
  In the proposed scheme for the $\left(\Nsf-\Msf, \frac{\Msf}{2}\right)$   non-secure problem,  
server $n^{\prime}\in \left[\Nsf-\Msf \right]$ computes a linear combination of the $\Nsf-\Msf$ messages,  
  where the coefficients of the messages   that server $n^{\prime}$ cannot compute are $0$. We assume that the vector  of the coefficients in this linear combination  is $\vv_{n^{\prime}}$, containing $\Nsf-\Msf$ elements.

 Go back to  the    $(\Nsf,\Msf)$ non-secure problem.   We let each server $n\in [\Msf+1: \Nsf]$ compute
 $$
A_{n-\Msf+2}= \vv_{n-\Msf} \ \left[ P_1 ;   \ldots; P_{\Nsf-\Msf} \right], 
 $$
  where $\vv_{n-\Msf}$ represents the   vector  of the coefficients in the linear combination sent by server $n-\Msf$ in the  $\left(\Nsf-\Msf, \frac{\Msf}{2}\right)$   non-secure problem.
 
  {\it Decoding phase.}  
  If the set of $\Nsf_{\rm r}=\Nsf-\Msf+1$ responding servers contains one server in $\left[\frac{\Msf}{2} \right]$ and one server in $\left[\frac{\Msf}{2}+1 :\Msf \right]$, from $A_1$ and $A_2$ the user can recover the task function. 
  
  We then consider the case where from the answers of the responding servers, the user can only receive one of $A_1$ and $A_2$. In this case, the set of $\Nsf_{\rm r}$ responding servers contains at least 
  $$
  \Nsf_{\rm r}-\frac{\Msf}{2}=\Nsf- \frac{3\Msf}{2}+1
  $$
   servers in $[\Msf+1 : \Nsf]$. Notice that in the    $\left(\Nsf-\Msf, \frac{\Msf}{2}\right)$   non-secure problem, the answers of any $\Nsf-\Msf-\frac{\Msf}{2}+1= \Nsf- \frac{3\Msf}{2}+1$ servers can re-construct the task function. Hence, in the  $\left(\Nsf,\Msf\right)$   non-secure problem, the answers of any  $\Nsf- \frac{3\Msf}{2}+1$ servers in $[\Msf+1:\Nsf]$ can re-construct $P_1+\cdots+P_{\Nsf-\Msf}=F_2$.
   Together with $A_1= 2F_1-F_2$ or with $A_2 =F_2-F_1$, the user can recover the task function $F_1$.

  It can be seen that the number of linearly independent combinations transmitted by servers in $[\Msf+1:\Nsf]$ is $h\left(\Nsf-\Msf, \frac{\Msf}{2}\right)$, the linear space of which contains $F_2$. 
  In addition, the number   of linearly independent combinations transmitted by servers in $[\Msf]$ is   two, the linear space of which also contains $F_2$.
  Hence,  the number of totally transmitted linearly independent combinations is $h(\Nsf,\Msf)= 2+h\left(\Nsf-\Msf, \frac{\Msf}{2}\right)-1=h\left(\Nsf-\Msf, \frac{\Msf}{2}\right)+1$, coinciding with~\eqref{eq:M is even}.

\subsection{Scheme~4 for~\eqref{eq:M is odd}}
\label{sub:M is odd}
We first provide an example to illustrate the main idea.
\begin{example}
\label{ex:scheme 4 example}
We consider the $(\Nsf,\Msf)=(8,5)$ non-secure  problem. It can be seen that in this example $\Nsf_{\rm r}= \Nsf-\Msf +1=4$.

{\it Assignment phase.}
We assign the datasets as follows. 
\begin{align*}
\begin{array}{rl|c|c|c|c|c|c|c|c|}\cline{3-3}\cline{4-4}\cline{5-5}\cline{6-6}\cline{7-7}\cline{8-8}\cline{9-9}\cline{10-10}
&&\rule{0pt}{1.2em}\mbox{Server 1} &\rule{0pt}{1.2em}\mbox{Server 2} & \rule{0pt}{1.2em}\mbox{Server 3}& \rule{0pt}{1.2em}\mbox{Server 4} & \rule{0pt}{1.2em}\mbox{Server 5} & \rule{0pt}{1.2em}\mbox{Server 6}   & \rule{0pt}{1.2em}\mbox{Server 7} & \rule{0pt}{1.2em}\mbox{Server 8}\\ \cline{3-3}\cline{4-4}\cline{5-5}\cline{6-6}\cline{7-7}\cline{8-8}\cline{9-9}\cline{10-10}
&& D_1& D_1  & D_1& D_1 & D_1 & D_3 & D_3  & D_3\\
&& D_2& D_2 & D_2 & D_2 & D_2 & D_4 & D_4 & D_4\\ 
&& D_3& D_3  &D_6& D_6 & D_6 & D_5 & D_5 &D_5\\
&& D_4& D_4 & D_7 & D_7 & D_7 & D_6 & D_7 & D_8\\ 
&& D_5& D_5 & D_8 & D_8 & D_8 &D_7 &D_8 &D_6\\ 
\cline{3-3}\cline{4-4}\cline{5-5}\cline{6-6}\cline{7-7}\cline{8-8}\cline{9-9}\cline{10-10}
\end{array}
\end{align*}

{\it Computing phase.}
We let each user  send one linear combination  of   messages, such that the user can recover ${\bf F} \ [W_{1}; \ldots; W_{\Nsf}]$ from the answers of  any $\Nsf_{\rm r}$ responding servers, where
\begin{equation}\setstretch{1.25}
 {\bf F} = \begin{bmatrix}  
 \fv_1 \\
\fv_2 \\
 \fv_{3} 
 \end{bmatrix}
 =
  \begin{bmatrix}  
  1 & 1    &   1 &1 & 1  &     1 &1 & 1     \\
  0 & 0    &  2 & 2 & 2    &  1 &1 & 1     \\
  0  & 0   &   0&0& 0   &  * &* & *    
\end{bmatrix} =
   \begin{bmatrix} 
  1 & 1    &   1 &1 & 1  &     1 &1 & 1     \\
  0 & 0    &  2 & 2 & 2    &  1 &1 & 1     \\
  0  & 0   &   0&0& 0    &   1 & 2 & 3    
\end{bmatrix},
 \label{eq:example4 division of S}
\end{equation}
 and each `$*$'  is uniformly i.i.d. over $\mathbb{F}_{\qsf}$ and in this example we assume that the last three `$*$' in $\fv_3$ are $(1,2,3)$.
 We also define that $[F_1;F_2;F_3]=  {\bf F} \  [W_1;\ldots;W_{\Nsf}] $.

We let servers $1,2$ with datasets $D_1,\ldots, D_5$ compute 
$$
X_1=X_2=F_1-F_2  = W_{1}+W_2-W_3-W_4-W_5.
$$ 

For servers in $[3:5]$   with  datasets $D_1,D_2, D_6,D_7,D_8$, we construct their transmissions such that from the answers of any two of them we can recover 
\begin{align*}
& 2F_1- F_2 =2 W_1+2W_2+W_6+W_7+W_8  ; \\
&F_3 = 2 W_6+W_7.
\end{align*}
Notice that both of $ 2F_1- F_2$ and $F_3$ can be computed by each server in $[3:5]$. Hence,
we each server in $[3:5]$ compute a random linear combination of $(2F_1-F_2)$ and $F_3$.
For example, we let servers $3,4,5$ compute $X_3,X_4,X_5$, respectively, where
\begin{align*}
& X_3= (2F_1- F_2 )+ F_3 ;\\
& X_4= (2F_1-F_2)+2F_3;\\
& X_5= (2F_1-F_2)+4F_3.\\
\end{align*} 
 
 For servers in $[6:8]$, we construct their transmissions such that from the answers of any two of them we can recover $F_2$ and $F_3$. This can be done by letting servers $6,7,8$ compute $X_6,X_7,X_8$, respectively, where 
 \begin{align*}
& X_6=  3 F_2 - F_3 = 6W_3+6W_4+6W_5+2W_6+W_7  ;\\
& X_7= F_2-F_3 = 2W_3+2W_4+2W_5 -W_7- 2W_8;\\
& X_8= 2F_2-F_3 = 4W_3+4W_4+4W_5 +W_6-W_8.
\end{align*} 

{\it Decoding phase.}
For any set of $\Nsf_{\rm r}=4$ servers, denoted by $\Ac$, we are in one of the following three cases:
\begin{itemize}
\item {\it Case 1: $\Ac$ contains at least two servers in $[3:5]$.} From the answers of any two servers in $[3:5]$, the user can recover $2F_1-F_2$ and $F_3$. Besides, $\Ac$ contains at least either one server in $[2]$ or one server in $[6:8]$. It can be seen that each of $X_1,X_2,X_6,X_7,X_8$ is linearly independent of $2F_1-F_2$ and $F_3$. Hence, the user then recovers $F_1$.
\item {\it Case 2:  $\Ac$ contains at least two servers in $[6:8]$.} From the   answers of any two servers in $[6:8]$, the user can recover $F_2$ and $F_3$. Besides, $\Ac$ contains at least   one server in $[5]$. It can be seen that in the transmitted linear combination of each server in $[5] $ contains $F_1$. Hence, the user then recovers $F_1$.
\item {\it Case 3: $\Ac$ contains servers $1,2$, one server in $[3:5]$, and one server in $[6:8]$.}  In this case, we can also check that the user receives three independent linear combinations in $F_1, F_2, F_3$, such that it can recover $F_1$.
\end{itemize}
It can be seen that 
the number of totally transmitted linearly independent combinations is $h(8,5)=3$, coinciding with~\eqref{eq:M is odd}.
 \hfill $\square$ 
\end{example}

We now consider  the $(\Nsf,\Msf)$ non-secure problem where $1.5 \Msf \leq  \Nsf < 2\Msf$ and $\Msf$ is odd, and   aim to construct a scheme (Scheme~4) to prove~\eqref{eq:M is odd}.
We also define that $\Nsf=2\Msf-\ysf$. 

{\it Assignment phase.}
The assignment is as follows.
\begin{align*}
\begin{array}{rl|c|c|c|c|c|c|c|c|c|c|c|}\cline{3-3}\cline{4-4}\cline{5-5}\cline{6-6}\cline{7-7}\cline{8-8}\cline{9-9}\cline{10-10}\cline{11-11}\cline{12-12}\cline{13-13}
&&\rule{0pt}{1.2em}\mbox{\small \negmedspace\negmedspace Server 1 \negmedspace\negmedspace} & \mbox{\small \negmedspace\negmedspace $\cdots$ \negmedspace\negmedspace} &  \rule{0pt}{1.2em}\mbox{\small \negmedspace\negmedspace Server $\ysf$ \negmedspace\negmedspace}& \rule{0pt}{1.2em}\mbox{\small \negmedspace\negmedspace Server $\ysf +1$ \negmedspace\negmedspace}&  \rule{0pt}{1.2em}\mbox{\small \negmedspace\negmedspace Server $\ysf+2$ \negmedspace\negmedspace}&  \rule{0pt}{1.2em}\mbox{\small \negmedspace\negmedspace $\cdots$ \negmedspace\negmedspace} &  \rule{0pt}{1.2em}\mbox{\small \negmedspace\negmedspace Server $\Msf$ \negmedspace\negmedspace} &\rule{0pt}{1.2em}\mbox{\small \negmedspace\negmedspace Server $\Msf$+1 \negmedspace\negmedspace} & \rule{0pt}{1.2em}\mbox{\small \negmedspace\negmedspace Server $\Msf+2$ \negmedspace\negmedspace} & \rule{0pt}{1.2em}\mbox{\small \negmedspace\negmedspace $\cdots$\negmedspace\negmedspace }& \rule{0pt}{1.2em}\mbox{\small \negmedspace\negmedspace Server $\Nsf$\negmedspace\negmedspace} \\ 
\cline{3-3}\cline{4-4}\cline{5-5}\cline{6-6}\cline{7-7}\cline{8-8}\cline{9-9}\cline{10-10}\cline{11-11}\cline{12-12}\cline{13-13}
&& D_1& \cdots       &   D_1 &  D_{1} &  D_{1} & \cdots &  D_{1}   &  D_{\frac{\Msf-1}{2}+1}   &  D_{\frac{\Msf-1}{2}+1} &  \cdots  &  D_{\frac{\Msf-1}{2}+1}  \\
&& \cdots & \cdots & \cdots &  \cdots & \cdots  & \cdots   & \cdots  & \cdots & \cdots & \cdots & \cdots  \\ 
&& D_{\frac{\Msf-1}{2}} &    \cdots    & D_{\frac{\Msf-1}{2}}  & D_{\frac{\Msf-1}{2}} & D_{\frac{\Msf-1}{2}}    & \cdots  & D_{\frac{\Msf-1}{2}} &   D_{\Msf}  &   D_{\Msf}  &  \cdots  &  D_{\Msf}  \\
&& D_{\frac{\Msf-1}{2}+1}&  \cdots  & D_{\frac{\Msf-1}{2}+1}  & D_{\Msf+1} & D_{\Msf+2} & \cdots & D_{\Nsf}& D_{\Msf+1}  &  D_{\Msf+2} &   \cdots  &  D_{\Nsf} \\ 
&& \cdots &  \cdots       & \cdots & \cdots & \cdots   & \cdots & \cdots & \cdots & \cdots   & \cdots & \cdots \\
&& D_{\Msf} &   \cdots  &  D_{\Msf} & D_{\frac{3\Msf+1}{2}} & D_{\frac{3\Msf+1}{2} +1}  & \cdots & D_{\frac{3\Msf+1}{2}-1} &  D_{\frac{3\Msf-1}{2}}  &  D_{\frac{3\Msf-1}{2}+1}   &  \cdots  &  D_{\frac{3\Msf-1}{2}-1}\\  
\cline{3-3}\cline{4-4}\cline{5-5}\cline{6-6}\cline{7-7}\cline{8-8}\cline{9-9}\cline{10-10}\cline{11-11}\cline{12-12}\cline{13-13}
\end{array} 
\end{align*}
In the  assignment,   we divide the $\Nsf$ datasets into three parts, where the first part contains $D_1,\ldots, D_{t}$ (later we will explain the reason to choose $t=\frac{\Msf-1}{2}$) which are all assigned 
to servers in $[\Msf]$; the second part contains  $D_{t+1},\ldots, D_{\Msf}$ which are all assigned to servers in $[\ysf] \cup [\Msf+1:\Nsf]$; the third part contains  $D_{\Msf+1},\ldots, D_{\Nsf}$, which are assigned to servers in $[\ysf+1:\Msf]$ in a cyclic way where each server   obtains $\Msf-t$ neighbouring datasets in $\left[{\frac{\Msf-1}{2}+1}: \Nsf \right]$. The datasets  $D_{\Msf+1},\ldots, D_{\Nsf}$ are also assigned to servers in  $[\Msf+1 : \Nsf]$  in a cyclic way where 
each server obtains $t$ neighbouring datasets in $\left[{\frac{\Msf-1}{2}+1}: \Nsf \right]$.

As we assign the datasets in $[\Msf+1:\Nsf]$ to  the servers in $[\ysf+1:\Msf]$ in a cyclic way where each server obtains $\Msf-t$ datasets,  we can choose $\Nsf-\Msf-(\Msf-t)+1=\Nsf-2\Msf+t+1$ neighbouring servers in $[\ysf+1:\Msf]$ satisfying the   constraint   in~\eqref{eq:vector constraint}; in addition, server $1$ has $D_{t+1},\ldots, D_{\Msf}$, which are not assigned to the servers in $[\ysf+1:\Msf]$. Hence, the ordered set of the above $\Nsf-2\Msf+t+2$ servers satisfies the   constraint in~\eqref{eq:vector constraint}. 

Similarly, we assign the datasets in $[\Msf+1:\Nsf]$ to  the servers in $[\Msf+1:\Nsf]$ in a cyclic way where each server obtains $ t$ datasets,  we can choose $\Nsf-\Msf-t+1$ neighbouring servers in $[\Msf+1:\Nsf]$ satisfying the   constraint   in~\eqref{eq:vector constraint}; in addition, server $1$ has $D_{1},\ldots, D_{t}$, which are not assigned to the servers in $[\Msf+1:\Nsf]$. Hence, the ordered set of the above $\Nsf-\Msf-t+2$ servers satisfies the   constraint in~\eqref{eq:vector constraint}. 

Similar to the derivation of~\eqref{eq:lemma converse all term}, by the chain rule of entropy, under the above assignment  we have  
$$
H(X_{[\Nsf]})/\Lsf \geq  
\max\{\Nsf-2\Msf+t+2,  \Nsf-\Msf-t+2\} \stackrel{t=\frac{\Msf-1}{2}}{=} \Nsf-\Msf- \frac{\Msf-1}{2}+2= \frac{\Msf+5}{2}-y.
$$ 
Hence, we let $t=\frac{\Msf-1}{2}$.

  {\it Computing phase.}
We design the computing phase such that the total number of independent transmitted linear combinations of messages by all servers is $\frac{\Msf+5}{2}-\ysf $.
These linear combinations are in ${\bf F} \ [W_1;\ldots;W_{\Nsf}]$ where
\begin{equation}\setstretch{1.25}
 {\bf F} = \begin{bmatrix}\ 
 \fv_1 \\
 \vdots \\
 \fv_{\frac{\Msf+5}{2}-\ysf} 
 \end{bmatrix}
 =
  \begin{bmatrix}\ 
\tikzmark{left4} \textcolor{white}{0} 1 ,\ldots , 1  \textcolor{white}{0} & \tikzmark{left5}  \textcolor{white}{0} 1 ,\ldots , 1\textcolor{white}{0} &   \tikzmark{left6}  \textcolor{white}{0}  1 ,\ldots , 1 \textcolor{white}{0}  \    \\
 \ \textcolor{white}{0} 0 ,\ldots , 0  \textcolor{white}{0} &\textcolor{white}{0} a ,\ldots , a  \textcolor{white}{0} &\textcolor{white}{0} 1 ,\ldots , 1  \textcolor{white}{0}   \   \\
\ \textcolor{white}{0} 0 ,\ldots , 0  \textcolor{white}{0}& \textcolor{white}{0} 0  ,\ldots , 0   \textcolor{white}{0}& \textcolor{white}{0} * ,\ldots , *   \textcolor{white}{0}   \    \\
\ \textcolor{white}{0} \ddots  \textcolor{white}{0}  &\textcolor{white}{0} \ddots  \textcolor{white}{0} &\textcolor{white}{0} \ddots  \textcolor{white}{0}   \   \\ 
\ \textcolor{white}{0} 0 ,\ldots , 0  \textcolor{white}{0} \tikzmark{right4} & \textcolor{white}{0} 0,\ldots , 0\textcolor{white}{0} \tikzmark{right5}  & \textcolor{white}{0} * ,\ldots , *  \textcolor{white}{0} \tikzmark{right6}  \  
\end{bmatrix}. \label{eq:example division of S}
\end{equation}
 \DrawBox[thick, black,  dashed ]{left4}{right4}{\textcolor{black}{\footnotesize$ {\bf F}_1 $}}
\DrawBox[thick, red, dashed]{left5}{right5}{\textcolor{red}{\footnotesize${\bf F}_2 $}}
\DrawBox[thick, blue, dashed]{left6}{right6}{\textcolor{blue}{\footnotesize${\bf F}_3 $}}

Notice that $a$  represents a  symbol uniformly  over $\mathbb{F}_{\qsf}\setminus \{0,1\}$, and  `$*$'  represents a symbol uniformly i.i.d. over $\mathbb{F}_{\qsf}$. 
We divide matrix $ {\bf F} $ into three column-wise sub-matrices, ${\bf F}_1$ with dimension $\left(\frac{\Msf+5}{2}-\ysf \right) \times \frac{\Msf-1}{2}$ which corresponds to the messages in $\left[\frac{\Msf-1}{2} \right]$,
 ${\bf F}_2$ with dimension $\left(\frac{\Msf+5}{2}-\ysf \right) \times \frac{\Msf+1}{2}$ which corresponds to the messages in $\left[\frac{\Msf+1}{2} :\Msf \right]$, and ${\bf F}_3$ with dimension $\left(\frac{\Msf+5}{2}-\ysf \right) \times (\Nsf-\Msf)$ which corresponds to the messages in $[\Msf+1 : \Nsf]$.
 We also define that $F_i= \fv_i [W_1;\ldots;W_{\Nsf}]$ for each $i\in \left[ \frac{\Msf+5}{2}-\ysf\right]$.
 Thus the transmission of each server could be expressed as a linear combination of $\left[F_1;\ldots;F_{ \frac{\Msf+5}{2}-\ysf} \right]$.
 
 As each server   $n\in \left[ \ysf \right]$  cannot compute  $W_{\Msf+1} , \ldots , W_{\Nsf} $, we let it compute 
      \begin{subequations}
\begin{align}
&{\bf s}_n  \  {\bf F} \  [W_1;\ldots;W_{\Nsf}] = [1,-1,0,\ldots,0]  \  {\bf F} \  [W_1;\ldots;W_{\Nsf}]\\
&= W_1+\cdots + W_{\frac{\Msf-1}{2}} + (1-a)(W_{\frac{\Msf+1}{2}}+\cdots+W_{\Msf}), \label{eq:scheme 4 transmission of class 1}
\end{align}  
     \end{subequations}
 such that the coefficients of $W_{\Msf+1},\ldots,W_{\Nsf}$ which it cannot compute are $0$.
 
 For the servers in $[\ysf+1 : \Msf]$, we construct their transmissions  such that from   the answers of any $ \frac{\Msf+3}{2}-\ysf$ servers in $[\ysf+1: \Msf]$, the user can recover 
 $
 aF_1 - F_2, F_3, \ldots, F_{\frac{\Msf+5}{2}-\ysf}.
 $
More precisely, we let server $n\in [\ysf+1 : \Msf]$ compute 
      \begin{subequations}
\begin{align}
&{\bf s}_{n}   \left[a \fv_1- \fv_2; \fv_3;\ldots;\fv_{ \frac{\Msf+5}{2}-\ysf} \right]   [W_1;\ldots;W_{\Nsf}], \label{eq:scheme 4 transmission of second class}\\
 &\text{where }  \left[a \fv_1- \fv_2; \fv_3;\ldots;\fv_{ \frac{\Msf+5}{2}-\ysf} \right]= \begin{bmatrix}  
 a & \ldots & a   &  0 & \ldots & 0   & \tikzmark{left7} a-1 & \ldots & a-1    \ \\
  0 & \ldots & 0  &   0   & \ldots & 0    &   * & \ldots &\textcolor{white}{00}  *     \textcolor{white}{0} \ \\
& \ddots &  & &\ddots&   & & \ddots &  \ \\ 
 0 & \ldots & 0   &  0 & \ldots & 0  &  * & \ldots &\textcolor{white}{00} * \textcolor{white}{0} \tikzmark{right7} \
\end{bmatrix}  .
\end{align}
     \end{subequations}
     \DrawBox[thick, blue, dashed]{left7}{right7}{\textcolor{blue}{\footnotesize$({\bf F}^{\prime}_3)_{\left(\frac{\Msf+3}{2}-\ysf \right) \times (\Nsf-\Msf)} $}}
     
     We design ${\bf s}_{n} $ as follows.
Notice that $W_1,\ldots,W_{\frac{\Msf-1}{2}}$ can be computed by server $n$; and that in the linear combination~\eqref{eq:scheme 4 transmission of second class} the coefficients of   $W_{\frac{\Msf-1}{2}+1}, \ldots, W_{\Msf}$ are $0$. 
Hence, in order to guarantee that in~\eqref{eq:scheme 4 transmission of second class} the coefficients of the messages which server $n$ cannot compute are $0$, we only need to consider the messages in $W_{\Msf+1}, \ldots, W_{\Nsf}$, whose related columns are in ${\bf F}^{\prime}_3 $.
Server $n$ cannot compute  $\Nsf-\Msf-\frac{\Msf+1}{2}=\frac{\Msf-1}{2}-\ysf$ messages in $W_{\Msf+1}, \ldots, W_{\Nsf}$; thus the column-wise 
sub-matrix of ${\bf F}^{\prime}_3 $ corresponding to these $\frac{\Msf-1}{2}-\ysf$ messages has the dimension $\left(\frac{\Msf+3}{2}-\ysf \right) \times \left(\frac{\Msf-1}{2}-\ysf \right)$. In addition, each `$*$'  is uniformly i.i.d. over $\mathbb{F}_{\qsf}$. Hence, 
the left-hand side nullspace of this sub-matrix  contains $\frac{\Msf+3}{2}-\ysf-\left(\frac{\Msf-1}{2}-\ysf \right)=2$ vectors, each of which has $\frac{\Msf+3}{2}-\ysf$ elements. 
We let  ${\bf s}_{n} $ be a random linear combination of these two vectors, where each of the two coefficients is uniformly i.i.d over $\mathbb{F}_{\qsf}$.

The following lemma will be proved in Appendix~\ref{sub:proof of lemma 1 scheme 2}.
\begin{lem}
\label{lem:lemma 1 scheme 2}
From any $ \frac{\Msf+3}{2}-\ysf$ answers of servers in $[\ysf +1: \Msf]$, the user can recover $a F_1-F_2, F_3,\ldots, F_{\frac{\Msf+3}{2}-\ysf}$ with high probability. 
\hfill $\square$ 
\end{lem}

Finally, we focus on the servers in $[\Msf+1 : \Nsf]$. 
We construct their transmissions  such that from any the answers of any $ \frac{\Msf+3}{2}-\ysf$ servers in $[\Msf+1 : \Nsf]$, the user can recover 
 $
  F_2, F_3, \ldots, F_{\frac{\Msf+5}{2}-\ysf}.
 $
More precisely, we let server $n\in [\Msf+1 : \Nsf]$ compute 
\begin{align}
{\bf s}_{n}   \left[  \fv_2; \fv_3;\ldots;\fv_{ \frac{\Msf+5}{2}-\ysf} \right]   [W_1;\ldots;W_{\Nsf}]. \label{eq:scheme 4 transmission of second class last group}
\end{align}
 We design ${\bf s}_{n} $ as follows.
Notice that  in the linear combination~\eqref{eq:scheme 4 transmission of second class last group} the coefficients of   $W_{1}, \ldots, W_{\frac{\Msf-1}{2}}$ are $0$; and that $W_{\frac{\Msf+1}{2}},\ldots,W_{\Msf}$ can be computed by server $n$. 
 Hence, in order to guarantee that in~\eqref{eq:scheme 4 transmission of second class last group} the coefficients of the messages which server $n$ cannot compute are $0$, we only need to consider the messages in $W_{\Msf+1}, \ldots, W_{\Nsf}$, whose related columns are in ${\bf F}_3^{([2:\frac{\Msf+5}{2}-y])_{\rm r}} $.\footnote{\label{foot:recall column wise} Recall that $\mathbf{M}^{(\Sc)_{\rm r}}$ represents the sub-matrix of $\mathbf{M}$ which is composed of the rows  of $\mathbf{M}$ with indices in $\Sc$.}
Server $n$ cannot compute  $\Nsf-\Msf-\frac{\Msf-1}{2}=\frac{\Msf+1}{2}-\ysf$ messages in $W_{\Msf+1}, \ldots, W_{\Nsf}$; thus the column-wise 
sub-matrix of ${\bf F}_3^{([2:\frac{\Msf+5}{2}-y])_{\rm r}} $ corresponding to these $\frac{\Msf+1}{2}-\ysf$ messages has the dimension $\left(\frac{\Msf+3}{2}-\ysf \right) \times \left(\frac{\Msf+1}{2}-\ysf \right)$. In addition, each `$*$'  is uniformly i.i.d. over $\mathbb{F}_{\qsf}$. Hence, 
the left-hand side nullspace of this sub-matrix  contains $\frac{\Msf+3}{2}-\ysf-\left(\frac{\Msf+1}{2}-\ysf \right)=1$ vector,  which has $\frac{\Msf+3}{2}-\ysf$ elements. 
We let  ${\bf s}_{n} $ be this vector.
 
It can be seen that the choice of ${\bf s}_{n} $ for the servers in $[\Msf+1 :\Nsf]$ is from the computing scheme with the cyclic assignment in~\cite[Section IV-B]{linearcomput2020wan} for the $\left(\Nsf-\Msf, \frac{\Msf-1}{2} \right)$ non-secure problem. Hence, as proved in~\cite[Section IV-B]{linearcomput2020wan},  from any $\Nsf-\Msf-\frac{\Msf-1}{2}+1= \frac{\Msf+3}{2}-\ysf$ answers of servers in $[\Msf+1: \Nsf]$, the user can recover  $F_2, \ldots, F_{\frac{\Msf+5}{2}-\ysf}$ with high probability.
 
 {\it Decoding phase.}
 Now we analyse each possible set  of $\Nsf_{\rm r}= \Nsf- \Msf+1= \Msf- \ysf+1$ responding servers,  assumed to be $\Ac$.
 \begin{itemize}
 \item {\it First case}: $\Ac$ contains at least $\frac{\Msf+3}{2}-\ysf$ servers in  $[\ysf+1 : \Msf]$. Here we consider the worst case, where $\Ac$  contains all servers in $[\ysf+1 : \Msf]$.
From Lemma~\ref{lem:lemma 1 scheme 2}, the user can recover $aF_1-F_2, F_3,\ldots,F_{\frac{\Msf+5}{2}-\ysf}$ with high probability. 
 
In $\Ac$, there remains $\Nsf_{\rm r} - (\Msf-\ysf)= 1$ server outside   $[\ysf+1 : \Msf]$. 
If this server is in $[\ysf]$, it computes $ F_1-F_2 $. The user can recover $F_1$ from $F_1-F_2$ and $a F_1-F_2$ because $a\notin \{0,1\}$.
If this server is in $[\Msf+1 : \Nsf]$,    it can be seen from~\eqref{eq:example division of S} that  each `$*$' in ${\bf F}_3^{([2:\frac{\Msf+5}{2}-y])_{\rm r}} $ is generated uniformly i.i.d over $\mathbb{F}_{\qsf}$, and thus
the transmission of this server can be expressed as $ a_1 F_2+ a_2 F_3 +\cdots + a_{\frac{\Msf+3}{2}-\ysf} F_{\frac{\Msf+5}{2}-\ysf} $, where $a_1$ is not zero
  with high probability  (from the proof in~\cite[Appendix C]{linearcomput2020wan}). Hence, the user then recovers $F_1$ with high probability.
  \item  {\it Second case}: $\Ac$ contains at least $\frac{\Msf+3}{2}-\ysf$ servers in  $[\Msf+1 : \Nsf]$. Here we consider the worst case, where $\Ac$  contains all servers in $[\Msf+1 : \Nsf]$. By construction, the user can recover $ F_2, \ldots,F_{\frac{\Msf+5}{2}-\ysf}$ with high probability.  
  
In $\Ac$, there remains $\Nsf_{\rm r} - (\Nsf-\Msf)= 1$ server outside   $[\Msf+1 : \Nsf]$. Similar to Case 1 described above,  with the answer from any other server  outside   $[\Msf+1 : \Nsf]$, the user then recovers $F_1$ with high probability. 
\item {\it Third case}: $\Ac$ contains  $\frac{\Msf+1}{2}-\ysf$ servers in $[\ysf+1 : \Msf]$,  $\frac{\Msf+1}{2}-\ysf$ servers in $[\Msf+1:\Nsf]$, and $\ysf$ servers in $[\ysf]$. 
Notice that the servers in $[\ysf]$ compute $F_1 -F_2$, and that the servers in $[\Msf+1:\Nsf]$ compute linear combinations of $F_2,\ldots,F_{\frac{\Msf+5}{2}-\ysf}$. Hence,
 the union of the answers of servers in $[\ysf]$ and the $\frac{\Msf+1}{2}-\ysf$ servers in $[\Msf+1:\Nsf]$, contains $\frac{\Msf+1}{2}-\ysf+1$ linearly independent combinations of $F_1,\ldots, F_{\frac{\Msf+5}{2}-\ysf}$.
 We denote the set of these linear combinations by $\Lc_1$.  
Moreover, it can be seen 
the coefficients in the linear combinations in  $\Lc_1$ are independent of the value of $a$. This is because the answer of the servers in $[\ysf]$ is $F_1-F_2$, and the answer  of each server in $[\Msf+1:\Nsf]$ is a linear combination of $F_2,\ldots,  F_{\frac{\Msf+5}{2}-\ysf}$ whose coefficients are determined by  ${\bf F}_3^{([2:\frac{\Msf+5}{2}-y])_{\rm r}} $, independent of $a$.

We then introduce the following lemma which will be proved in Appendix~\ref{sub:proof lemma 2 scheme 2}.
\begin{lem}
\label{lem:lemma 2 scheme 2}
Among the answers    of  any $ \frac{\Msf+1}{2}-\ysf$ servers in $[\ysf+1:\Msf]$,   with high probability there exists some linear combination which is independent of the linear combinations in $\Lc_1$. 
  \hfill $\square$ 
\end{lem}
By Lemma~\ref{lem:lemma 2 scheme 2},  the user can totally obtain $\frac{\Msf+1}{2}-\ysf+1+1=\frac{\Msf+5}{2}-\ysf$ linearly independent combinations of 
$F_1,\ldots,  F_{\frac{\Msf+5}{2}-\ysf}$ with high probability, and thus it can recover its desired task function $F_1$. 
 \end{itemize}
 We have proved that  if  $a$ is generated uniformly  over $\mathbb{F}_{\qsf} \setminus \{0,1\}$ and each   $*$  in ${\bf F}_3^{([3:\frac{\Msf+5}{2}-y])_{\rm r}} $ is generated uniformly i.i.d. over $\mathbb{F}_{\qsf}$, the user can   recover the task function with high probability.  Hence, we only need to pick one realization of ${\bf F}_3^{([3:\frac{\Msf+5}{2}-y])_{\rm r}} $ and $a$, such that we can guarantee the successful decoding.

 By the above scheme,  the number of linearly independent transmissions by all servers is equal to the number of rows in ${\bf F}$, i.e.,
$\frac{\Msf+5}{2}-\ysf =\Nsf- \frac{3\Msf-5}{2} $, coinciding with~\eqref{eq:M is odd}.

 \subsection{Scheme~5 for~\eqref{eq:less than 1.5M}}
\label{sub:less than 1.5M}
 Finally, we consider the case where $\Msf< \Nsf <  1.5 \Msf$, and   aim to construct a scheme (Scheme~5) to prove~\eqref{eq:less than 1.5M}. Scheme~5 is a recursive scheme which is based on  the proposed scheme 
 for the  $(\Msf, 2\Msf-\Nsf)$ non-secure problem. We assume that the latter scheme has been designed before, whose number of totally transmitted linearly independent combinations of messages is $h(\Msf, 2\Msf-\Nsf)$.
 
  \paragraph*{Assignment phase.}
We first assign datasets $D_{1},\ldots, D_{\Nsf-\Msf}$ to each server in $[\Msf]$. Then we assign datasets $D_{\Nsf-\Msf+1},\ldots,D_{\Nsf}$ to each server in $[\Msf+1:\Nsf]$. So far, each server in $[\Msf+1:\Nsf]$ has obtained $\Msf$ datasets, while each server in $[\Msf]$ has obtained $\Nsf-\Msf<\Msf$ datasets. In addition, each dataset in $[\Nsf-\Msf+1 : \Nsf]$ has been assigned to $\Nsf-\Msf<\Msf$ servers. 
Hence, in the next step we should assign each dataset $D_k$ where $k\in [\Nsf-\Msf+1 : \Nsf]$ to $\Msf-(\Nsf-\Msf)=2\Msf-\Nsf$ servers in $[\Msf]$, such that each server in $[\Msf]$ obtains $\Msf-(\Nsf-\Msf)=2\Msf-\Nsf$ datasets in $[\Nsf-\Msf+1 : \Nsf]$. 
Thus we can apply the assignment phase of the proposed scheme  for the $(\Msf, 2\Msf-\Nsf)$ non-secure problem, to assign datasets  $D_{\Nsf-\Msf+1},\ldots,D_{\Nsf}$ to servers in $[\Msf]$.

   \paragraph*{Computing phase.}
Let us first focus on the    $(\Msf, 2\Msf-\Nsf)$ non-secure problem, where the $\Msf$ messages are assumed to be $W^{\prime\prime}_1,\ldots, W^{\prime\prime}_{\Msf}$. 
In the proposed scheme for the $(\Msf, 2\Msf-\Nsf)$  non-secure problem,  
each server  computes a linear combination of the $ \Msf$ messages. 
 Considering   the transmitted linear  combinations by  all servers, the number of linearly independent combinations is denoted by $h(\Msf, 2\Msf-\Nsf)$ and these $h(\Msf, 2\Msf-\Nsf)$ linear combinations can be expressed as  
\begin{align}
{\bf F}_4 \  [W^{\prime\prime}_1;\ldots;W^{\prime\prime}_{\Msf}].\label{eq:F_1}
\end{align} 
 The transmission of each server $n^{\prime}\in \left[\Msf \right]$ can be expressed as 
$$
 \sv_{n^{\prime}} \ {\bf F}_4  \  [W^{\prime\prime}_1;\ldots;W^{\prime\prime}_{\Msf}].
 $$

Let us then go back to the $(\Nsf,\Msf)$ non-secure problem. We construct the answer of the $\Nsf$ servers, such that the transmissions of all servers totally contain 
$h(\Msf, 2\Msf-\Nsf)$ linearly independent combinations and these $h(\Msf, 2\Msf-\Nsf)$ linear combinations can be expressed as $ {\bf F} [W_{1};\ldots;W_{\Nsf}]$, where   (each $a_i$ where $i\in [h(\Msf, 2\Msf-\Nsf)-1]$ represents a     symbol uniformly i.i.d. over $\mathbb{F}_{\qsf}$)
\begin{equation}\setstretch{1.25}
 {\bf F} =  
  \begin{bmatrix}\ 
\tikzmark{left9}  \textcolor{white}{00000} 1 &\ldots & 1  \textcolor{white}{0} & \tikzmark{left10}  \textcolor{white}{0} 1 &\ldots & 1\textcolor{white}{0}    \\
\textcolor{white}{00000} a_1 &\ldots &a_1  \textcolor{white}{0} &\textcolor{white}{0} + &\ldots & +  \textcolor{white}{0}     \\
\textcolor{white}{00000} & \ddots & \textcolor{white}{0}  &\textcolor{white}{0} &\ddots&  \textcolor{white}{0}     \\ 
\textcolor{white}{00000} a_{h(\Msf, 2\Msf-\Nsf)-1}&\ldots &  a_{h(\Msf, 2\Msf-\Nsf)-1} \textcolor{white}{0}  \tikzmark{right9}  & \textcolor{white}{0} + &\ldots & + \textcolor{white}{0} \tikzmark{right10}    \ \
\end{bmatrix}. \label{eq:example division of S1}
\end{equation} 
\DrawBox[thick, blue, dashed]{left9}{right9}{\textcolor{blue}{\footnotesize$  {\bf F}_5  $}}
\DrawBox[thick, red, dashed]{left10}{right10}{\textcolor{red}{\footnotesize $  {\bf F}_4  $ in~\eqref{eq:F_1}}}

Each `$+$' represents  an   element of $  {\bf F}_4  $ in~\eqref{eq:F_1}.
Notice that the dimension of ${\bf F}_5 $ is $\big( h(\Msf, 2\Msf-\Nsf)-1 \big) \times (\Nsf-\Msf) $ and the dimension of ${\bf F}_4 $ is $\big( h(\Msf, 2\Msf-\Nsf)-1 \big) \times \Msf $.

For each server $n\in [\Msf]$,  by the construction of the assignment phase,  datasets $D_1,\ldots,D_{\Nsf-\Msf}$ are assigned to it and the assignment on the datasets $D_{\Nsf-\Msf+1},\ldots,D_{\Nsf}$ is from the assignment phase of the proposed scheme for the $(\Msf, 2\Msf-\Nsf)$  non-secure problem. Hence, we let server $n $ compute
$ 
 \sv_{n }   {\bf F }   [W_1;\ldots;W_{\Nsf}],
$ 
where $\sv_n$ is the same as the transmission vector of server $n$ in the   $(\Msf, 2\Msf-\Nsf)$  non-secure problem.

For each server  $n \in [\Msf+1 : \Nsf]$, it cannot compute $W_{1}, \ldots, W_{\Nsf-\Msf}$, which correspond to the column-wise sub-matrix  ${\bf F}_5$, whose rank is $1$. So the left-hand side null space of ${\bf F}_5$ contains $h(\Msf, 2\Msf-\Nsf)-1$ linearly independent vectors, each of which has $h(\Msf, 2\Msf-\Nsf)$ elements. 
 We let the transmission vector of server $n$, denoted by $\sv_n$, be a random linear combinations of these $h(\Msf, 2\Msf-\Nsf)-1$ linearly independent vectors, where the $h(\Msf, 2\Msf-\Nsf)-1$ coefficients are uniformly i.i.d. over $\mathbb{F}_{\qsf}$; in other words, server $n$ computes  $\sv_n   {\bf F}   [W_{1};\ldots; W_{\Nsf}]$.  

     \paragraph*{Decoding phase.}
  Assume the set of responding servers is $\Ac$, where $\Ac \subseteq [\Nsf]$ and $|\Ac|=\Nsf_{\rm r}$. We also define that 
$\Ac_1 = \Ac \cap [\Msf]$ and $\Ac_2= \Ac \setminus [\Msf]$. By definition,  $|\Ac_2| \leq \Nsf-\Msf =\Nsf_{\rm r}-1$. 

As $\Msf\geq \Nsf-\Msf+1=\Nsf_{\rm r}$, it can be seen that $|\Ac_2|$ could be $0$. 
If $|\Ac_2|=0$,  
from the answers of the servers in $\Ac_1$, the user can recover $\fv_1  [W_1;\ldots; W_{\Nsf}]$, where $\fv_1$ is the first row of ${\bf F}$. This is   from the decodability of the proposed scheme for the  $(\Msf, 2\Msf-\Nsf)$  non-secure problem.

We then consider the case where  $0<|\Ac_2| \leq \Nsf-\Msf$ and from the answers of $\Ac_1$ the user cannot recover the task function.
 Assume the number of linearly independent combinations of messages transmitted by the servers in $\Ac_1 $ is $\lambda_1$. 
Besides the answers of the servers in $\Ac_1$, if the user obtains any $h(\Msf, 2\Msf-\Nsf)- \lambda_1$ linear combinations in    ${\bf F} [W_{1}; \ldots; W_{\Nsf}]$, which are all independent of the answers   of the servers in $\Ac_1$, the user can recover ${\bf F} [W_{1}; \ldots; W_{\Nsf}]$.\footnote{\label{foot:scheme 5 deco} The reason is as follows. In each of the proposed schemes (except a special case in Scheme~3), from the answers of any $\Nsf_{\rm r}$ responding servers, the user can recover   the transmissions  by all servers. The only special case is in Scheme~3 when the set of responding servers contains one server in $\left[\frac{\Msf}{2} \right]$ and one server in $\left[ \frac{\Msf}{2}+1 :\Msf \right]$. However, if $\Ac_1$ contains such two servers, the user can directly recover the task function; otherwise, we can   find $\Nsf_{\rm r}-|\Ac_1|$ servers in $[\Msf+1:\Nsf]$, such that from the answers of the total $\Nsf_{\rm r}$ servers, the user can recover the   transmissions  by all servers.  } 
In Appendix~\ref{sec:proof of lemma scheme 3}, we will prove the following lemma.
\begin{lem}
\label{lem:lem scheme 3}
  The  answers in servers $\Ac_2$ contains $h(\Msf, 2\Msf-\Nsf)- \lambda_1$ linear combinations,   independent of  the answers of the servers in $\Ac_1$   with high probability. 
\end{lem}
 By Lemma~\ref{lem:lem scheme 3}, from the answers of servers in $\Ac$, the user can recover ${\bf F} [W_{1}; \ldots; W_{\Nsf}]$.

 In conclusion, we have $h(\Nsf,\Msf)=h(\Msf, 2\Msf-\Nsf)$, which coincides with~\eqref{eq:less than 1.5M}.
We have proved that  if  $a_j$ where $j\in [h(\Msf, 2\Msf-\Nsf)-1]$ is generated uniformly i.i.d. over $\mathbb{F}_{\qsf}$,  the user can   recover the task function with high probability.  Hence, we only need to pick one realization of them, such that we can guarantee the successful decoding.

  \section{Conclusions}
  \label{sec:con}
  In this paper, we formulated the secure  distributed linearly separable computation problem, where the user should only recover the desired task  function without retrieving any other information about the datasets. It is interesting to see that to preserve this security, we need not to increase the communication cost if the computing scheme is based on linear coding. We then focused on the problem where the computation cost is minimum. In this case, 
   while achieving the optimal communication cost, we aim to   minimize the size of the randomness variable which is independent of the datasets and is introduced in the system to preserve the security. 
   For this purpose, we proposed   an information theoretic   converse bound on the randomness size for each  possible assignment. We then proposed a   secure computing scheme with novel assignment strategies, which outperforms the optimal computing schemes with the well-know fractional repetition assignment and cyclic assignment in terms of the randomness size. Exact optimality results have been obtained from the proposed computing scheme under some system parameters. 
    Ongoing work includes   deriving tighter converse  bounds over all possible assignments and minimizing the needed randomness size for more general case where the computation cost is not minimum and the user requests multiple linear combinations of messages.
   
    \appendices
    
\section{Proof of Theorem~\ref{thm:extension of linear coding}}
\label{sec:extension of linear coding}
We  consider the distributed linearly separable computation problem in~\cite{cctradeoff2020wan} where $\Msf=\frac{\Ksf}{\Nsf}(\Nsf-\Nsf_{\rm r}+\msf)$ for $\msf\in [\Nsf_{\rm r}]$ and the user requests $\Ksf_{\rm c} \in [\Ksf]$ linearly independent combinations of messages. 
We now describe on a general linear coding computing scheme.  In this scheme, we divide each message $W_k$ where $k\in [\Ksf]$ into $\ell$ non-overlapping and equal-length sub-messages, $W_{k}=\{W_{k,i}: i\in [\ell]\}$. Server $n\in [\Nsf]$ sends $\frac{\ell T_n }{\Lsf}$ linearly independent  combinations of $W_{1,1},W_{1,2},\ldots,W_{\Ksf,\ell}$.

Considering   the transmitted linear  combinations by  all servers, the number of linearly independent combinations is denoted by $\lambda$ and these $\lambda$ linear combinations can be expressed as  ${\bf F}   [W_{1,1};W_{1,2};\ldots;W_{\Ksf,\ell}]$, where
\begin{align}
{\bf F} = \begin{bmatrix}      
\fv_1 \\
\vdots\\
\fv_{\lambda}
\end{bmatrix}
\ =   \begin{bmatrix}  
f_{1,1}& \cdots & f_{1,\ell\Ksf} \\
\vdots & \ddots & \vdots\\
f_{\lambda,1}& \cdots & f_{\lambda,\ell\Ksf}
\end{bmatrix}.  \label{eq:all linear}
\end{align}    
Notice that any linear scheme can be transformed in the above manner.
    Among the linear combinations in~\eqref{eq:all linear}, $\fv_i [W_{1,1};W_{1,2};\ldots;W_{\Ksf,\ell}]$ where $i\in [\ell \Ksf_{\rm c}]$ represent the desired task  function of the user.
   The transmission of each server $n \in [\Nsf]$ can be express as 
   \begin{align}
   \mathbf{S}_n \ {\bf F} \  [W_{1,1};W_{1,2};\ldots;W_{\Ksf,\ell}], \label{eq:app transmission of server n}
   \end{align}
   where the dimension of $ \mathbf{S}_n$ is $\frac{\ell T_n }{\Lsf} \times \lambda$. 
   
   For any set of $\Nsf_{\rm r}$ responding servers (denoted by $\Ac=\{\Ac(1),\ldots,\Ac(\Nsf_{\rm r})\}$), the user receives 
   $$
\mathbf{S}_{\Ac} \ {\bf F} \  [W_{1,1};W_{1,2};\ldots;W_{\Ksf,\ell}]
   $$
where $\mathbf{S}_{\Ac}$ represents the row-wise sub-matrix of   $\left[\mathbf{S}_{\Ac(1)};  \ldots; \mathbf{S}_{\Ac(\Nsf_{\rm r})}\right]$ which has the same rank as $\left[\mathbf{S}_{\Ac(1)};  \ldots; \mathbf{S}_{\Ac(\Nsf_{\rm r})}\right]$. Assume $\mathbf{S}_{\Ac}$ contains $r_{\Ac}$ rows.
In the decoding phase, 
the user multiply $\mathbf{S}_{\Ac}   {\bf F}    [W_{1,1};W_{1,2};\ldots;W_{\Ksf,\ell}] $ by $\mathbf{D}_{\Ac}$, where
the dimension of $\mathbf{D}_{\Ac}$ is $\ell\Ksf_{\rm c} \times r_{\Ac}$ and 
   \begin{align}
\mathbf{D}_{\Ac} \ \mathbf{S}_{\Ac}  = [\mathbf{I}_{\ell\Ksf_{\rm c}} ,  {\bf 0}_{\ell\Ksf_{\rm c} \times (r_{\Ac}-\ell\Ksf_{\rm c} )}], \label{eq:decoding linear}
   \end{align}
  where     $\mathbf{I}_n$ represents the identity matrix with dimension $n \times n$ and
${\bf 0}_{m \times n}$ represents the zero  matrix with dimension $m\times n$.
    Hence, the user can recover the desired task function from  $\mathbf{D}_{\Ac}   \mathbf{S}_{\Ac}   {\bf F}   [W_{1,1};W_{1,2};\ldots;W_{\Ksf,\ell}] $.
    
    Now we take the security constraint~\eqref{eq:security} into consideration, and  
      extend the above general linear coding scheme. 
      We introduce $\lambda-\ell \Ksf_{\rm c}$ independent randomness    variables $Q_{1}, \ldots, Q_{\lambda-\ell \Ksf_{\rm c}}$, where $Q_i, i\in [\lambda-\ell \Ksf_{\rm c}]$ is uniformly i.i.d. over $[\mathbb{F}_{\qsf}]^{\frac{\Lsf}{\ell}}$. 
      We then let  $
      {\bf F}^{\prime}= [{\bf F}, \mathbf{S}],
      $
    where 
$
    \mathbf{S}= [{\bf 0}_{\ell\Ksf_{\rm c} \times (\lambda-\ell\Ksf_{\rm c} )}; \mathbf{S}^{\prime} ]
$
    and 
    $
    \mathbf{S}^{\prime}
    $
    is full-rank with dimension $(\lambda-\ell\Ksf_{\rm c} ) \times (\lambda-\ell\Ksf_{\rm c} )$.
   
   We let each server $n\in [\Nsf]$ transmit 
\begin{align}
 \mathbf{S}_n \ {\bf F}^{\prime} \  [W_{1,1};W_{1,2};\ldots;W_{\Ksf,\ell};Q_1;\ldots;Q_{\lambda-\ell \Ksf_{\rm c}}], \label{eq:new transmission}
\end{align}   
  where $\mathbf{S}_n $ is the same as that in~\eqref{eq:app transmission of server n}.
It can be seen that in the transmitted linear combinations~\eqref{eq:new transmission}, the coefficients of the sub-messages which server $n$ cannot compute are still $0$ as the original non-secure scheme.

    For any set of $\Nsf_{\rm r}$ responding servers $\Ac$, the user receives 
   $$
\mathbf{S}_{\Ac} \ {\bf F}^{\prime} \  [W_{1,1};W_{1,2};\ldots;W_{\Ksf,\ell}],
   $$
and    recovers its desired task function from 
  $
  \mathbf{D}_{\Ac}  \mathbf{S}_{\Ac}   {\bf F}^{\prime}    [W_{1,1};W_{1,2};\ldots;W_{\Ksf,\ell};Q_1;\ldots;Q_{\lambda-\ell \Ksf_{\rm c}}],
  $ 
  since~\eqref{eq:decoding linear} holds.
  
  Finally, we will prove that the above scheme is secure, i.e., the security constraint~\eqref{eq:security} holds. 
From the answers of all servers, the user  can only recover   totally 
$\lambda$ linearly independent combinations, which are 
\begin{align}
{\bf F}^{\prime} \ [W_{1,1};W_{1,2};\ldots;W_{\Ksf,\ell};Q_1;\ldots;Q_{\lambda-\ell \Ksf_{\rm c}}].\label{eq:total lambda}
\end{align} 
In addition, 
$    \mathbf{S}^{\prime}$ is full-rank (with rank equal to $\lambda-\ell\Ksf_{\rm c} $).
  Hence, the user can only recover the desired task function (i.e., the first $\ell\Ksf_{\rm c} $ linear combinations in~\eqref{eq:total lambda}) without $Q_1,\ldots,Q_{\lambda-\ell \Ksf_{\rm c}}$, and thus the above scheme is secure.

    \section{Proof of Theorem~\ref{thm:converse lemma}}
    \label{sec:compound}
 By the security constraint in~\eqref{eq:security}, the user can only obtain $W_{1}+\cdots+W_{\Ksf}$ without knowing any other information about the messages after receiving   the answers of all servers. 
Recall that $X_{\Sc}=\{X_n: n\in \Sc \}$.
Intuitively by~\cite{shannonsecurity},  we need a key with length at least $H(X_{[\Nsf]}) - H(W_{1}+\cdots+W_{\Ksf})$ such that except $W_{1}+\cdots+W_{\Ksf}$, the other information about $W_1,\ldots, W_{\Ksf}$ transmitted in $X_{[\Nsf]}$ is hidden. More precisely, 
 from~\eqref{eq:security} we have 
         \begin{subequations}
 \begin{align}
0&= I\big(W_{1},\ldots, W_{\Ksf};   X_{[\Nsf]} | W_1+W_2+\cdots+W_{\Ksf}  \big) \\
&= H(X_{[\Nsf]}| W_1+W_2+\cdots+W_{\Ksf} ) - H( X_{[\Nsf]}|W_{1},\ldots, W_{\Ksf} ) \\
& \geq  H(X_{[\Nsf]}| W_1+W_2+\cdots+W_{\Ksf} ) - H( Q, W_{1},\ldots, W_{\Ksf} |W_{1},\ldots, W_{\Ksf} ) \label{eq:function of QW}\\
& =H(X_{[\Nsf]}| W_1+W_2+\cdots+W_{\Ksf} ) - H( Q  ) \label{eq:independent noise}\\
& =H(X_{[\Nsf]}) -I(X_{[\Nsf]}; W_1+W_2+\cdots+W_{\Ksf}) - H( Q  )\\
&\geq H(X_{[\Nsf]})-H(W_1+W_2+\cdots+W_{\Ksf}) - H( Q  ) \label{eq:shannon}
 \end{align}
       \end{subequations}
       where~\eqref{eq:function of QW} comes from that the $X_{[\Nsf]}$ is a function of $Q, W_{1},\ldots, W_{\Ksf}$,~\eqref{eq:independent noise} comes from that $Q$ is independent of $ W_{1},\ldots, W_{\Ksf}$. 
  Hence, from~\eqref{eq:shannon} and  we have 
        \begin{subequations}
\begin{align}
&   \eta \Lsf  \geq  H(Q) \geq H(X_{[\Nsf]}) -  H(W_{1}+\cdots+W_{\Ksf})  \\
& \geq H(X_{[\Nsf]}) -  \Lsf \label{eq:W1 inde} \\
&\geq  H(X_{s_1},\ldots, X_{s_v} )-  \Lsf \\
& =H(X_{s_1})+ H(X_{s_2}|X_{s_1}) + \cdots+ H(X_{s_v}|X_{s_1},\ldots, X_{s_{v-1}} ) -\Lsf, \label{eq:lemma converse all term}  
\end{align}
      \end{subequations}
      where~\eqref{eq:W1 inde} comes from that the $\Ksf$ messages are independent and each message is uniformly i.i.d. over $[\mathbb{F}_{\qsf}]^{\Lsf}$,~\eqref{eq:lemma converse all term} comes from the chain rule of entropy.

Let us then focus on each entropy term in~\eqref{eq:lemma converse all term}, $H(X_{s_i}| X_{s_1},\ldots, X_{s_{i-1}} )$ where
 $i\in [v]$. Recall that server $s_{i}$ can compute some message (assumed to be message $W_j$) which cannot be computed by servers $s_1,\ldots, s_{i-1}$, and that each message cannot computed by $\Nsf_{\rm r}-1$ servers. We assume that the set of servers which cannot compute $W_j$ is $\overline{\Ac_j}$. Obviously, $\{s_1,\ldots,s_{i-1} \} \subseteq \overline{\Ac_j}$.
Now consider  that the set of responding servers is $\overline{\Ac_j} \cup \{s_i\}$, totally containing  $\Nsf_{\rm r}$ servers.
 As   $W_j$ can only computed by server $s_i$ among the servers in $\overline{\Ac_j} \cup \{s_i\}$, and
from the answers of servers in    $\overline{\Ac_j} \cup \{s_i\}$ the user should recover $W_{1}+\cdots+W_{\Ksf}$,
  we have 
\begin{align}
H(X_{s_i}| X_{s_1},\ldots,X_{s_{i-1}}) \geq  H(X_{s_i}|X_{k}: k \in \overline{\Ac_j} ) \geq \Lsf. \label{eq:i term converse}
\end{align}
Hence, we take~\eqref{eq:i term converse} into~\eqref{eq:lemma converse all term} to obtain,
\begin{align}
\eta \Lsf \geq  v \Lsf -\Lsf,
\end{align} 
which proves~\eqref{eq:converse lemma}.

\section{Proof of Theorem~\ref{thm:exact optimality Mleq4}}
\label{sec:exact optimality Mleq4}
We first introduce the following lemma which will be proved in Appendix~\ref{sub:2M-1lemma} and will be used in the proof of Theorem~\ref{thm:exact optimality Mleq4}.
\begin{lem}
\label{lem:2M-1lemma}
For the $(\Ksf,\Nsf,\Nsf_{\rm r},\Ksf_{\rm c},\Msf)$ secure distributed linearly separable computation problem with  $\Msf= \frac{\Ksf}{\Nsf} \Msf^{\prime}$, $\Ksf_{\rm c}=1$, $\frac{\Msf^{\prime}}{\text{GCD}(\Nsf, \Msf^{\prime})} \geq 3$, and $ \text{Mod} \left( \frac{\Nsf}{\text{GCD}(\Nsf,\Msf^{\prime})}, \frac{\Msf^{\prime}}{\text{GCD}(\Nsf,\Msf^{\prime})} \right) = \frac{\Msf^{\prime}}{\text{GCD}(\Nsf,\Msf^{\prime})} -1$,
 to achieve the optimal communication cost, it must hold that
 \begin{align}
 \eta^{\star} \geq \left\lceil \Nsf/\Msf^{\prime} \right\rceil  . \label{eq:2M-1lemma}
 \end{align}
  \hfill $\square$ 
\end{lem}

We then start to prove Theorem~\ref{thm:exact optimality Mleq4} and focus on the $(\Ksf,\Nsf,\Nsf_{\rm r},\Ksf_{\rm c},\Msf)$ secure distributed linearly separable computation problem where  $\Msf= \frac{\Ksf}{\Nsf}\Msf^{\prime}$, $\Ksf_{\rm c}=1$, and $\frac{\Msf^{\prime}}{\text{GCD}(\Nsf, \Msf^{\prime})} \leq 4$. Notice that when $\frac{\Msf^{\prime}}{\text{GCD}(\Nsf, \Msf^{\prime})} =1$ (i.e., $\Msf^{\prime}$ divides $\Nsf$), the optimality directly comes from Theorem~\ref{thm:division}. Hence, in the following, we consider the case where $\frac{\Msf^{\prime}}{\text{GCD}(\Nsf, \Msf^{\prime})} \in [2:4]$.

\subsection{$\frac{\Msf^{\prime}}{\text{GCD}(\Nsf, \Msf^{\prime})}=2$}
\label{sub:proof M2} 
 When $\frac{\Msf^{\prime}}{\text{GCD}(\Nsf, \Msf^{\prime})}=2$,  it can be seen that $ \text{Mod} \left( \frac{\Nsf}{\text{GCD}(\Nsf,\Msf^{\prime})}, \frac{\Msf^{\prime}}{\text{GCD}(\Nsf,\Msf^{\prime})} \right)=1$. Hence,
 by the proposed scheme for Theorem~\ref{thm:combined scheme}, the needed randomness size is 
         \begin{subequations}
\begin{align}
h(\Nsf, \Msf^{\prime})-1 & =  h\left(\frac{\Nsf}{\text{GCD}(\Nsf, \Msf^{\prime})},\frac{\Msf^{\prime}}{\text{GCD}(\Nsf,\Msf^{\prime})} \right)- 1 \label{eq:case 2from GCD} \\
&=\left\lfloor \Nsf/\Msf^{\prime} -1\right\rfloor + h(3,2) -1  \label{eq:case 2 first step}\\
& =\left\lfloor \Nsf/\Msf^{\prime} -1\right\rfloor + 2 -1= \left\lceil \Nsf/\Msf^{\prime} \right\rceil -1 ,\label{eq:case 2 sec step}
\end{align}
        \end{subequations}
where~\eqref{eq:case 2from GCD} comes from~\eqref{eq:from GCD},~\eqref{eq:case 2 first step} comes from~\eqref{eq:from partial rep},~\eqref{eq:case 2 sec step}  comes from that
$h(3,2)=2$ and that $ \left\lfloor \Nsf/\Msf^{\prime} +1\right\rfloor =\left\lceil \Nsf/\Msf^{\prime} \right\rceil $ since $\Msf^{\prime}$ does not divide $\Nsf$.
It can be seen that the needed randomness size~\eqref{eq:case 2 sec step} coincides with the converse bound in Corollary~\ref{cor:converse cor}, and thus is optimal. 
 
\subsection{$\frac{\Msf^{\prime}}{\text{GCD}(\Nsf, \Msf^{\prime})}=3$}
\label{sub:proof M3} 
When  $\frac{\Msf^{\prime}}{\text{GCD}(\Nsf, \Msf^{\prime})}=3$,  it can be seen that $ \text{Mod} \left( \frac{\Nsf}{\text{GCD}(\Nsf,\Msf^{\prime})}, \frac{\Msf^{\prime}}{\text{GCD}(\Nsf,\Msf^{\prime})} \right) \in [2]$.

For the case where  $ \text{Mod} \left( \frac{\Nsf}{\text{GCD}(\Nsf,\Msf^{\prime})}, \frac{\Msf^{\prime}}{\text{GCD}(\Nsf,\Msf^{\prime})} \right)=1$, by the proposed scheme for Theorem~\ref{thm:combined scheme}, the needed randomness size is 
         \begin{subequations}
\begin{align}
h(\Nsf, \Msf^{\prime})-1& =   h\left(\frac{\Nsf}{\text{GCD}(\Nsf, \Msf^{\prime})},\frac{\Msf^{\prime}}{\text{GCD}(\Nsf,\Msf^{\prime})} \right)- 1 \label{eq:case 31from GCD} \\
& =\left\lfloor \Nsf/\Msf^{\prime} -1\right\rfloor + h(4,3) -1  \label{eq:case 31 first step}\\
& =\left\lfloor \Nsf/\Msf^{\prime} -1\right\rfloor + 2 -1 = \left\lceil \Nsf/\Msf^{\prime} \right\rceil -1,\label{eq:case 31 sec step}
\end{align}
        \end{subequations}
where~\eqref{eq:case 31from GCD} comes from~\eqref{eq:from GCD},~\eqref{eq:case 31 first step} comes from~\eqref{eq:from partial rep}, and~\eqref{eq:case 31 sec step}  comes from that
$h(4,3)=2$.
The needed randomness size in~\eqref{eq:case 31 sec step} coincides with the converse bound in Corollary~\ref{cor:converse cor}, and thus is optimal.

For the case where  $ \text{Mod} \left( \frac{\Nsf}{\text{GCD}(\Nsf,\Msf^{\prime})}, \frac{\Msf^{\prime}}{\text{GCD}(\Nsf,\Msf^{\prime})} \right)=2$, by the proposed scheme for Theorem~\ref{thm:combined scheme}, the needed randomness size is 
         \begin{subequations}
\begin{align}
h(\Nsf, \Msf^{\prime}) -1 & =    h\left(\frac{\Nsf}{\text{GCD}(\Nsf, \Msf^{\prime})},\frac{\Msf^{\prime}}{\text{GCD}(\Nsf,\Msf^{\prime})} \right)- 1 \label{eq:case 32from GCD} \\
&=\left\lfloor \Nsf/\Msf^{\prime} -1\right\rfloor + h(5,3) -1  \label{eq:case 32 first step}\\
& =\left\lfloor \Nsf/\Msf^{\prime} -1\right\rfloor + 3 -1 = \left\lceil \Nsf/\Msf^{\prime} \right\rceil  ,\label{eq:case 32 sec step}
\end{align}
        \end{subequations}
    where~\eqref{eq:case 32from GCD} comes from~\eqref{eq:from GCD},~\eqref{eq:case 32 first step} comes from~\eqref{eq:from partial rep}, and~\eqref{eq:case 32 sec step}  comes from that
$h(5,3)=3$.    
The needed randomness size in~\eqref{eq:case 32 sec step} coincides with the converse bound in Lemma~\ref{lem:2M-1lemma}, and thus is optimal. 
        
\subsection{$\frac{\Msf^{\prime}}{\text{GCD}(\Nsf, \Msf^{\prime})}=4$}
\label{sub:proof M4}         
    When  $\frac{\Msf^{\prime}}{\text{GCD}(\Nsf, \Msf^{\prime})}=4$,  it can be seen that $ \text{Mod} \left( \frac{\Nsf}{\text{GCD}(\Nsf,\Msf^{\prime})}, \frac{\Msf^{\prime}}{\text{GCD}(\Nsf,\Msf^{\prime})} \right) \in \{1,3\}$.    
        
        For the case where  $ \text{Mod} \left( \frac{\Nsf}{\text{GCD}(\Nsf,\Msf^{\prime})}, \frac{\Msf^{\prime}}{\text{GCD}(\Nsf,\Msf^{\prime})} \right)=1$, by the proposed scheme for Theorem~\ref{thm:combined scheme}, the needed randomness size is 
         \begin{subequations}
\begin{align}
h(\Nsf, \Msf^{\prime})  -1  & =    h\left(\frac{\Nsf}{\text{GCD}(\Nsf, \Msf^{\prime})},\frac{\Msf^{\prime}}{\text{GCD}(\Nsf,\Msf^{\prime})} \right)- 1 \label{eq:case 41from GCD} \\
&=\left\lfloor \Nsf/\Msf^{\prime} -1\right\rfloor + h(5,4) -1 \label{eq:case 41 first step}\\
& =\left\lfloor \Nsf/\Msf^{\prime} -1\right\rfloor + 2 -1 = \left\lceil \Nsf/\Msf^{\prime} \right\rceil -1 ,\label{eq:case 41 sec step}
\end{align}
        \end{subequations}
where~\eqref{eq:case 41from GCD} comes from~\eqref{eq:from GCD},~\eqref{eq:case 41 first step} comes from~\eqref{eq:from partial rep}, and~\eqref{eq:case 41 sec step}  comes from that
$h(5,4)=2$.
The needed randomness size in~\eqref{eq:case 41 sec step} coincides with the converse bound in Corollary~\ref{cor:converse cor}, and thus is optimal. 
        
For the case where  $ \text{Mod} \left( \frac{\Nsf}{\text{GCD}(\Nsf,\Msf^{\prime})}, \frac{\Msf^{\prime}}{\text{GCD}(\Nsf,\Msf^{\prime})} \right)=3$, by the proposed scheme for Theorem~\ref{thm:combined scheme}, the needed randomness size is 
         \begin{subequations}
\begin{align}
h(\Nsf, \Msf^{\prime})-1 &  =    h\left(\frac{\Nsf}{\text{GCD}(\Nsf, \Msf^{\prime})},\frac{\Msf^{\prime}}{\text{GCD}(\Nsf,\Msf^{\prime})} \right)- 1 \label{eq:case 42from GCD} \\
& =\left\lfloor \Nsf/\Msf^{\prime} -1\right\rfloor + h(7,4) -1  \label{eq:case 42 first step}\\
& =\left\lfloor \Nsf/\Msf^{\prime} -1\right\rfloor + 3 -1 = \left\lceil \Nsf/\Msf^{\prime} \right\rceil  ,\label{eq:case 42 sec step}
\end{align}
        \end{subequations}
    where~\eqref{eq:case 42from GCD} comes from~\eqref{eq:from GCD},~\eqref{eq:case 42 first step} comes from~\eqref{eq:from partial rep}, and~\eqref{eq:case 42 sec step}  comes from that
$h(7,4)=3$.    
The needed randomness size in~\eqref{eq:case 42 sec step} coincides with the converse bound in Lemma~\ref{lem:2M-1lemma}, and thus is optimal.         
        
\subsection{Proof of Lemma~\ref{lem:2M-1lemma}}
\label{sub:2M-1lemma}
Recall that $\Msf=\frac{\Ksf}{\Nsf}\Msf^{\prime}$, and that $ \text{Mod} \left( \frac{\Nsf}{\text{GCD}(\Nsf,\Msf^{\prime})}, \frac{\Msf^{\prime}}{\text{GCD}(\Nsf,\Msf^{\prime})} \right) = \frac{\Msf^{\prime}}{\text{GCD}(\Nsf,\Msf^{\prime})} -1$.  
For the ease of notation, we let $\asf :=\frac{\Msf^{\prime}}{\text{GCD}(\Nsf,\Msf^{\prime})}$. By the constraints in Lemma~\ref{lem:2M-1lemma}, we have $\asf \geq 3$.

 We assume that   $\Nsf=(2\asf-1+\asf\bsf)\text{GCD}(\Nsf,\Msf^{\prime})$, where $\bsf $ is a non-negative integer. To prove~\eqref{eq:2M-1lemma}, it is equivalent to prove
$$
\eta \geq \left\lceil \Nsf/\Msf^{\prime} \right\rceil  = \left\lceil 2+\bsf-\frac{1}{\asf}\right\rceil = \bsf+2.
$$
{\bf In the following we use the induction method to prove that for any assignment, there must exist some ordered set of $\bsf+3$ servers  satisfying the constraint in~\eqref{eq:vector constraint}.  }

{\it Proof step 1.}
We first consider the case where   $\bsf=0$,  and
  aim to prove that for any possible assignment,   we can always find an ordered set of three servers $\sv=(s_1,s_2,s_3)$ such that server $s_2$ has some dataset not assigned to  server $s_1$ and server $s_3$ has some dataset not assigned to servers $s_1, s_2$.

In this case, $\Nsf=2\Msf^{\prime}-\text{GCD}(\Nsf, \Msf^{\prime})$. 
For any possible assignment, 
  each dataset is assigned to $\Msf^{\prime}$ servers, and each server has (recall that $\asf :=\frac{\Msf^{\prime}}{\text{GCD}(\Nsf,\Msf^{\prime})} \geq 3$)
$$
\Msf =\frac{\Ksf}{\Nsf} \Msf^{\prime} =  \frac{\Msf^{\prime} \Ksf}{2\Msf^{\prime}-\text{GCD}(\Nsf, \Msf^{\prime})}= \frac{\asf \Ksf}{2\asf-1}
$$
 datasets. It can be seen that $\frac{\Ksf}{2}< \Msf  < \frac{2}{3} \Ksf $.
We can prove that   there exist two servers (assumed to be server $s_1 ,s_2$) such that $\Zc_{s_1}\neq \Zc_{s_2}$ and $\Zc_{s_1} \cup \Zc_{s_2} \neq [\Ksf]$.\footnote{\label{foot:3servers}    As $\frac{\Ksf}{2}< \Msf  < \frac{2}{3} \Ksf $, there must be three servers with three different sets of obtained datasets. We denote these three servers by servers $n_1,n_2,n_3$. 
We then  prove by contradiction that   there exist two servers in $\{n_1,n_2,n_3 \}$, where some dataset is not assigned to any of them.  
Assume that $\Zc_{n_1} \cup \Zc_{n_2}= [\Ksf]$, $\Zc_{n_1} \cup \Zc_{n_3}=[\Ksf]$, and $\Zc_{n_2} \cup \Zc_{n_3}= [\Ksf]$. Hence, each dataset must be assigned to at least two servers in  $\{n_1,n_2,n_3 \}$, and thus there are at least $2 \Ksf/3$ datasets assigned to each server, which contradicts $\Msf  < \frac{2}{3} \Ksf $. 
} 
 Assumed   that dataset $D_k$ is not assigned to  server $s_1 ,s_2$. We then pick one server which has dataset $D_k$ (assumed to be server $s_3$). It can be see that 
 the ordered set of servers $(s_1,s_2,s_3)$ satisfies the constraint in~\eqref{eq:vector constraint}. 

{\it Proof step 2.}
We then focus on the case   where  $\bsf=1$.  In this case, 
\begin{align*}
&\Nsf=(3\asf-1 )\text{GCD}(\Nsf,\Msf^{\prime})=3\Msf^{\prime}-\text{GCD}(\Nsf,\Msf^{\prime}), \\
& \Ksf= \frac{\Ksf}{\Nsf} \Nsf=3 \frac{\Ksf}{\Nsf} \Msf^{\prime} - \frac{\Ksf}{\Nsf}\text{GCD}(\Nsf,\Msf^{\prime})= 3\Msf -\frac{\Ksf}{\Nsf}\text{GCD}(\Nsf,\Msf^{\prime}).
\end{align*}
Hence, $\Ksf-\Msf=2\Msf -\frac{\Ksf}{\Nsf}\text{GCD}(\Nsf,\Msf^{\prime}) > 2\Msf -\frac{\Ksf}{\Nsf} \Msf^{\prime}= \Msf$.
WLOG, we assume that server $1$ has datasets $D_{1},\ldots, D_{\Msf}$. Let us then focus on the datasets $D_{\Msf+1},\ldots,D_{\Ksf}$, and consider the following two cases:
\begin{itemize}
\item {\it Case 1: among $\Zc_{2} \cap [\Msf+1:\Ksf],\ldots,\Zc_{\Nsf}\cap [\Msf+1:\Ksf]$, there are at least three different non-empty sets.} As shown before, there must exist an ordered set of three servers in $[2:\Nsf]$, which is assumed to be $(s_1,s_2,s_3)$, satisfying the constraint in~\eqref{eq:vector constraint}. In addition, server $1$ does not have any datasets in $[\Msf+1:\Ksf]$. Hence, the ordered set of servers $(1,s_1,s_2,s_3)$ also satisfies the constraint in~\eqref{eq:vector constraint}.
\item {\it Case 2: among $\Zc_{2} \cap [\Msf+1:\Ksf],\ldots,\Zc_{\Nsf}\cap [\Msf+1:\Ksf]$, there are only two different non-empty sets.} 
In this case, there are $\Msf^{\prime}$ servers (assumed to be $\Vc_1$) whose obtained datasets in $[\Msf+1:\Ksf]$ are non-empty and   the same; there are other   
 $\Msf^{\prime}$ servers (assumed to be $\Vc_2$) whose obtained datasets in $[\Msf+1:\Ksf]$ are non-empty  and   the same. 
We then consider the sets $\Zc_{n}$ where $n\in (\Vc_1 \cup \Vc_2)$. It is not possible that there are only two different sets among them, because the datasets in $[\Msf]$ are only assigned to the 
servers in  
$[\Nsf]\setminus (\Vc_1 \cup \Vc_2)$ where $|[\Nsf]\setminus (\Vc_1 \cup \Vc_2)|< \Msf^{\prime}$. Hence, there must exist at least three different sets among $\Zc_{n}$ where $n\in (\Vc_1 \cup \Vc_2)$. 
Thus there must be an ordered set of three servers in $(\Vc_1 \cup \Vc_2)$ satisfying the constraint in~\eqref{eq:vector constraint}, which is assumed to be $(s_1,s_2,s_3)$. In addition,
 in the union set of the assigned datasets to these three servers, there are at most $2\Msf- (\Ksf-\Msf) < \Msf$ datasets in $[\Msf]$.
Hence, the ordered set of servers $(s_1,s_2,s_3,1)$ also satisfies the  constraint in Theorem~\ref{thm:converse lemma}
 \end{itemize}

{\it Proof step 3.}
 Finally, we assume when $\bsf \in [x]$, there   exists some ordered set of servers $(s_1,\ldots,s_{\bsf+3})$ satisfying the  constraint   in Theorem~\ref{thm:converse lemma}.  
We will show that when $\bsf=x+1$, there   exists some ordered set of servers $(s_1,\ldots,s_{x+4})$ satisfying the  constraint in Theorem~\ref{thm:converse lemma}.

In this case, 
\begin{align*}
&\Nsf=((3+x)\asf-1 )\text{GCD}(\Nsf,\Msf^{\prime})=(3+x)\Msf^{\prime}-\text{GCD}(\Nsf,\Msf^{\prime}), \\
& \Ksf= \frac{\Ksf}{\Nsf} \Nsf= (3+x) \frac{\Ksf}{\Nsf}\Msf^{\prime}-\frac{\Ksf}{\Nsf}\text{GCD}(\Nsf,\Msf^{\prime})=(3+x)\Msf-\frac{\Ksf}{\Nsf}\text{GCD}(\Nsf,\Msf^{\prime}).
\end{align*}
Hence, we have $(x+2)\Msf < \Ksf < (x+3)\Msf$.
As there are $\Ksf$ datasets, each of which is assigned to $\Msf^{\prime}$ servers, and the number of datasets assigned to each server is $\Msf$, we can   
 find $x+1$ servers where each server has some dataset  not assigned to other $x$ servers. 
 
 If there exists some server  $j\in [x+1]$ where at least $\frac{\Ksf}{\Nsf}\text{GCD}(\Nsf,\Msf^{\prime})+1$ datasets assigned to server $j$ have also been  assigned to some other server in $[x+1]\setminus \{j\}$, 
it can be seen that there are at least  
$$
\Ksf- (x+1)\Msf +  \frac{\Ksf}{\Nsf}\text{GCD}(\Nsf,\Msf^{\prime})+1= 2\Msf+1
$$
 datasets not assigned to $[x+1]$; thus we can   find three servers which has some dataset not assigned to the servers in $[x+1]$ nor the other two servers.  Hence, we can find an ordered set of $x+4$ servers satisfying the constraint in~\eqref{eq:vector constraint}. 
 
 Hence, in the following we consider that at most 
  $\frac{\Ksf}{\Nsf}\text{GCD}(\Nsf,\Msf^{\prime}) $ datasets assigned to each server in $[x+1]$ have also  been  assigned to some other server in $[x+1]$. 
 There are totally at most $(x+1)\Msf$ different datasets to these $x+1$ servers, and thus there remains at least 
$
\Ksf-(x+1)\Msf= 2\Msf - \frac{\Ksf}{\Nsf}\text{GCD}(\Nsf,\Msf^{\prime})
$
datasets which are not assigned to these $x+1$ servers. WLOG, we assume that 
the $x+1$ servers are in $[x+1]$ and
these $2\Msf - \frac{\Ksf}{\Nsf}\text{GCD}(\Nsf,\Msf^{\prime})$ datasets are $D_{(x+1)\Msf+1},\ldots, D_{\Ksf}$.

We then consider two cases:
\begin{itemize}
\item {\it Case 1: among $\Zc_{x+2} \cap [(x+1)\Msf+1:\Ksf],\ldots,\Zc_{\Nsf} \cap [(x+1)\Msf+1:\Ksf]$, there are at least three different non-empty sets.} As shown before, there must exist an  ordered set of three servers in $[x+2:\Nsf]$, which is assumed to be $(s_1,s_2,s_3)$, satisfying the constraint in~\eqref{eq:vector constraint}. In addition, servers in $[x+1]$ do not have any datasets in $[\Msf+1:\Ksf]$. Hence, the ordered set of servers $(1,2,\ldots, x+1, s_1,s_2,s_3)$ also satisfies the constraint in~\eqref{eq:vector constraint}.
\item {\it Case 2: among $\Zc_{x+2} \cap [(x+1)\Msf+1:\Ksf],\ldots,\Zc_{\Nsf} \cap [(x+1)\Msf+1:\Ksf]$, there are only two different non-empty sets.} 
In this case, there are $\Msf^{\prime}$ servers (assumed to be $\Vc_1$) whose obtained datasets in $[(x+1)\Msf+1:\Ksf]$ are non-empty and   the same; there are other   
 $\Msf^{\prime}$ servers (assumed to be $\Vc_2$) whose obtained datasets in $[(x+1)\Msf+1:\Ksf]$ are non-empty and   the same. 
We then consider the sets $\Zc_{n}$ where $n\in (\Vc_1 \cup \Vc_2)$. 
\begin{itemize}
\item If there are two different sets among them, we have completely assigned $2\Msf$ datasets to $2\Msf^{\prime}$ servers, each of  which  has $\Msf$ datasets. Hence,  if we focus on the assignment for the servers in $[\Nsf] \setminus (\Vc_1 \cup \Vc_2)$, the assignment is equivalent to the problem where
  we assign $\Ksf_1=\Ksf- 2\Msf$ datasets to $\Nsf_1=\Nsf-2\Msf^{\prime}=(2\asf-1+(x-1)\asf)\text{GCD}(\Nsf,\Msf^{\prime})$ servers, where each dataset is assigned to $\Msf^{\prime}$ servers and the number of assigned datasets to each server is $\Msf$. Therefore, we can use the induction assumption to find  an ordered set of $(x-1)+3=x+2$ servers satisfying the constraint in~\eqref{eq:vector constraint}, which are assumed to be servers $(s_1,\ldots,s_{x+2})$. In addition, we can pick two server in  $\Vc_1 \cup \Vc_2$  with different sets of datasets, which are assumed to be servers $s_{x+3}, s_{x+4}$. In summary, the ordered set of servers $(s_1,\ldots,s_{x+4})$ also satisfies the constraint in~\eqref{eq:vector constraint}.
  \item If there are at least three different sets among them, we can   find  an ordered set of three servers in $(\Vc_1 \cup \Vc_2)$ satisfying the constraint in~\eqref{eq:vector constraint}, which is assumed to be $(s_1,s_2,s_3)$. In addition,
 in the union set of the assigned datasets to these three servers, there are at most $2\Msf- (\Ksf-(x+1)\Msf ) = \frac{\Ksf}{\Nsf}\text{GCD}(\Nsf,\Msf^{\prime}) $ datasets in $[(x+1)\Msf]$.
 In addition, each server in $[x+1]$ has $\Msf \geq 3  \frac{\Ksf}{\Nsf}\text{GCD}(\Nsf,\Msf^{\prime})  $ datasets and at most 
  $\frac{\Ksf}{\Nsf}\text{GCD}(\Nsf,\Msf^{\prime}) $ datasets assigned to each server in $[x+1]$ have also been  assigned to some other server in $[x+1]$. 
  As a result, the ordered set of servers $(s_1,s_2,s_3,1,2,\ldots,x+1)$ satisfies the constraint in~\eqref{eq:vector constraint}. 
\end{itemize}
\end{itemize}

In conclusion, by the induction method, we proved Lemma~\ref{lem:2M-1lemma}.

  \section{Proof of Decodability of Scheme~4}
\label{sec:scheme 2 deco}

\subsection{Proof of Lemma~\ref{lem:lemma 1 scheme 2}}
\label{sub:proof of lemma 1 scheme 2}
Recall that for each server $n\in [\ysf+1:\Msf]$, it computes a linear combination ${\bf s}_{n} \left[a\fv_1-\fv_2; \fv_3;\ldots ;  \fv_{\frac{\Msf+5}{2}-\ysf} \right]$, where ${\bf s}_n$ is   a random linear combination of the  two linearly independent vectors in the left-hand side nullspace of  $\frac{\Msf-1}{2}-\ysf$ neighbouring columns in
 ${\bf F}^{\prime}_3 $.  
 We aim to prove that  for any set $\Vc \subseteq [\ysf+1:\Msf]$ where $|\Vc|= \frac{\Msf+3}{2}-\ysf$, 
  the vectors $\sv_n$ where $n\in \Vc$ are linearly independent.

As the field size $\qsf$ is large enough, following the decodability proof in~\cite[Appendix C]{linearcomput2020wan} (which also proves that the transmission vectors by a set of servers are linearly independent with high probability) based on the Schwartz-Zippel lemma~\cite{Schwartz,Zippel,Demillo_Lipton},  we only need to find out one specific realization of 
$$
 {\bf F}^{\prime}_3 =
\begin{bmatrix}\ 
  a & a& \ldots & a \\
  * & * &\ldots & *\\
  \vdots & \vdots & \vdots& \vdots\\
  * & * &\ldots & *
  \end{bmatrix},
$$
  such that  the vectors $\sv_n$ where $n\in \Vc$ are linearly independent.

We sort the servers in $\Vc$ in an increasing order, $\Vc=\{ v_1, \ldots, v_{\frac{\Msf+3}{2}-\ysf} \}$, where $v_i < v_j$  if $i<j$.
For each $i\in [\frac{\Msf+1}{2}-\ysf]$, we let the $(i+1)^{\text{th}}$ row of  $ {\bf F}^{\prime}_3$ be 
\begin{align*}
 [ *, *, \cdots,*, 0, 0, \cdots, 0, *,*,\cdots ,*],  
\end{align*}
 where each $0$ corresponds to one distinct message which server $v_i$ cannot compute.

By the construction of $ {\bf F}^{\prime}_3$, we let the transmission vector $\sv_{v_i}$ of each server $v_i$ where $i\in [\frac{\Msf+1}{2}-\ysf]$ be as follows,\footnote{\label{foot:2dimen} Notice that $ {\bf F}^{\prime}_3$ contains $\frac{\Msf+3}{2}-\ysf$ rows, and server $v_i$ cannot compute $ \frac{\Msf-1}{2}-\ysf$ messages in $W_{\Msf+1}, \ldots, W_{\Nsf}$; thus we need to fix two positions in  $\sv_{v_i}$   and then to determine the other elements $\sv_{v_i}$ by solving linear equations. More precisely, we let the first element is $0$ and  the $(i+1)^{th}$ element be $1$. With the above construction, by the proof in~\cite[Appendix D]{linearcomput2020wan}, the remaining elements in $\sv_{v_i}$ are obtained by solving linear equations, which are all $0$.} 
\begin{align*}
  \sv_{v_i}=[0,\cdots,0,1,0,\cdots,0], 
\end{align*}
where  $\sv_{v_i}$ has $\frac{\Msf+3}{2}-\ysf$ elements and  the $(i+1)^{th}$ element is $1$.
 
 Let us then design the transmission vector $\sv_{v_{\frac{\Msf+3}{2}-\ysf}}$ of server $v_{\frac{\Msf+3}{2}-\ysf}$ by letting its first element be $1$ and second element be $0$.
 By the proof in~\cite[Appendix D]{linearcomput2020wan}, the remaining elements in $\sv_{v_{\frac{\Msf+3}{2}-\ysf}}$ are obtained by solving linear equations (but not necessary be all $0$). 
Hence, by construction $\sv_{v_1},\ldots, \sv_{v_{\frac{\Msf+3}{2}-\ysf}}$ 
are linearly independent.

 \subsection{Proof of Lemma~\ref{lem:lemma 2 scheme 2}}
\label{sub:proof lemma 2 scheme 2}
Recall that  $\Lc_1$ contains $\frac{\Msf+1}{2}-\ysf$ linearly independent combinations of $F_2, \ldots, f_{\frac{\Msf+5}{2}-\ysf }$, and one linear combination $F_1- F_2$. 
 The coefficients in these linear combinations are independent of $a$.

Let us focus on the servers in $[\ysf+1:\Msf]$ and the servers in $[\Msf+1: \Nsf]$.
 Recall that the transmission of each server $n \in [\ysf+1:\Msf]$ is 
\begin{align*}
\sv_{n} \ \left[a \fv_1- \fv_2;  \fv_3; \ldots; \fv_{\frac{\Msf+5}{2}-\ysf} \right]  \ [W_1;\ldots; W_{\Nsf}].
\end{align*}
Server $n$ can compute $\frac{\Msf+1}{2}$ neighbouring messages in $W_{\Msf+1},\ldots, W_{\Nsf}$, which are 
\begin{align}
W_{\text{Mod}(n-\ysf,\Nsf-\Msf)+\Msf},W_{\text{Mod}(n-\ysf+1,\Nsf-\Msf)+\Msf},  \ldots, W_{\text{Mod}\left(n-\ysf+\frac{\Msf-1}{2},\Nsf-\Msf\right)+\Msf}. \label{eq:class 2 server knows}
\end{align} 
The transmission vector $\sv_{n}$ of server $n$ is in the left-hand side null space   of the column-wise sub-matrix of ${\bf F}^{\prime}_3 $, including the columns corresponding to    the  $\Nsf-\Msf-\frac{\Msf+1}{2}=\frac{\Msf-1}{2}-\ysf$ messages in $W_{\Msf+1},\ldots, W_{\Nsf}$  which server $n$ cannot compute.
Notice that this null space contains $\frac{\Msf+1}{2}-\ysf+1-\frac{\Msf-1}{2}-\ysf=2$ linearly independent vectors, and $\sv_{n}$  is a random combination of them.

It can be seen that the number of servers in $[\ysf +1: \Msf]$ is the same as the number of servers in $[\Msf+1 : \Nsf]$, which are both equal to $\Nsf-\Msf=\Msf-\ysf$. 
In addition, for each $n \in [\ysf+1: \Msf]$,  
 server $n + (\Msf-\ysf)$ (which is in $[\Msf+1:\Nsf]$) can compute the messages
$$
W_{\text{Mod}(n-\ysf,\Nsf-\Msf)+\Msf}, W_{\text{Mod}(n-\ysf+1,\Nsf-\Msf)+\Msf}, \ldots, W_{\text{Mod}\left(n-\ysf+\frac{\Msf-1}{2}-1,\Nsf-\Msf\right)+\Msf} 
$$
among the messages in $[\Msf+1: \Nsf]$. 
Hence, compared to the set of messages  in $[\Msf+1: \Nsf]$ that server  $n + (\Msf-\ysf)$ can compute, it can be seen from~\eqref{eq:class 2 server knows} that server $n$ additionally has $W_{\text{Mod}\left(n-\ysf+\frac{\Msf-1}{2},\Nsf-\Msf\right)+\Msf}$.

Let us then prove Lemma~\ref{lem:lemma 2 scheme 2} by the   Schwartz-Zippel lemma~\cite{Schwartz,Zippel,Demillo_Lipton}, as in~\cite[Appendix C]{linearcomput2020wan}.  More precisely, 
for any set of $\frac{\Msf+1}{2}-\ysf$ servers in $[\Msf+1:\Nsf]$ (denoted by $\Ac_1$) and any set of   $\frac{\Msf+1}{2}-\ysf$ servers in $[ \ysf+1:\Msf]$ (denoted by $\Ac_2$), 
we aim to find one specific realization of $a$ and   $
{\bf F}^{(\left[3:\frac{\Msf+5}{2}-\ysf \right])_{\rm r}}_3 $ 
 such that 
there exists one server in $\Ac_2$ whose transmission is linearly independent of the linear combinations in $\Lc_1$.

We sort the servers in $\Ac_1$ in an increasing order, $\Ac_1(1)<\cdots<\Ac_1\left(\frac{\Msf+1}{2}-\ysf \right)$. For each $i\in \left[ \frac{\Msf+1}{2}-\ysf \right]$,
we let the $i^{\text{th}}$ row of ${\bf F}^{(\left[3:\frac{\Msf+5}{2}-\ysf \right])_{\rm r}}_3 $  be
$$
  [ *, *, \cdots,*, 0, 0, \cdots, 0, *,*,\cdots ,*],  
$$
  where each $0$ corresponds to one distinct message which server $\Ac_1(i)$ cannot compute. 
 After determining such ${\bf F}^{(\left[3:\frac{\Msf+5}{2}-\ysf \right])_{\rm r}}_3 $, as shown in~\cite[Appendix D]{linearcomput2020wan},
  the transmission   of server $\Ac_1(i)$  is $ \sv_{\Ac_1(i)}  \left[\fv_2;\ldots;\fv_{\frac{\Msf+5}{2}-\ysf}\right] [W_1;\ldots;W_{\Nsf}]$, where 
  $$
 \sv_{\Ac_1(i)}= [0,\cdots,0,1,0,\cdots,0].
  $$
  Notice that  $\sv_{\Ac_1(i)}$ contains $\frac{\Msf+3}{2}-\ysf$ elements and $1$ is located at the $(i+1)^{\text{th}}$ position.

In other words,  it can be seen that the transmissions of the $\frac{\Msf+1}{2}-\ysf$ servers in $[\Msf+1:\Nsf]$  are 
$
F_{3}, \ldots, F_{\frac{\Msf+5}{2}-\ysf}.
$
In addition, the transmission of each server in $[\ysf]$ is $F_1-F_2$.

Let us then determine the transmissions of the servers in $\Ac_2 \subseteq [\ysf+1:\Msf]$. 
It can be easily proved that among the servers in $\Ac_2$, there must exist one server (assumed to be $n^{\prime}$) the set of whose available messages in $W_{\Msf+1},\ldots,W_{\Nsf}$, is a super set of the available messages in  $W_{\Msf+1},\ldots,W_{\Nsf} $ to some server (assumed to be $n^{\prime\prime}$) in $[\Msf+1:\Nsf]\setminus \Ac_1$.\footnote{\label{foot:super set} The proof is as follows.
For any $\Nsf-\Msf-|\Ac_1|=\frac{\Msf-1}{2}$ servers in  $[\Msf+1:\Nsf]$, there are at least $\frac{\Msf-1}{2}+2$ servers in $[\ysf+1:\Msf]$ the set of whose obtained datasets is a super set of the set of the obtained datasets  by some of these $\frac{\Msf-1}{2}$  servers. Hence,
  $\Ac_2$ contains $\frac{\Msf+1}{2}-\ysf$ servers in $[ \ysf+1:\Msf]$, which must contain some of these  $\frac{\Msf-1}{2}+2$  servers because $\frac{\Msf+1}{2}-\ysf+ \frac{\Msf-1}{2}+2=\Msf-\ysf+2 > \Msf-\ysf$.
}
The transmission of $n^{\prime\prime}$ is assumed to be 
\begin{align*}
\sv_{n^{\prime\prime}} \  [\fv_2;\ldots; \fv_{\frac{\Msf+5}{2}-\ysf }] \ [W_1;\ldots; W_{\Nsf}],
\end{align*}
where $\sv_{n^{\prime\prime}}$ is the vector in the    left-hand side null space (which contains two independent vectors) of the column-wise sub-matrix of ${\bf F}^{(\left[2: \frac{\Msf+5}{2}-\ysf\right])_{\rm r}}_3$  corresponding the unavailable messages in $[\Msf+1:\Nsf]$ to server $n^{\prime\prime}$.
By the proof in~\cite[Appendix C]{linearcomput2020wan},  the first element of $\sv_{n^{\prime\prime}}$ is not $0$ with high probability; thus $\sv_{n^{\prime\prime}}$ is linearly independent of the transmissions of the servers in $\Ac_1$ with high probability.  
Recall that the   transmission vector of server $n^{\prime}$ is a random vector in the left-hand side null space (which contains two independent vectors) of the column-wise sub-matrix of ${\bf F}^{\prime}_3$  corresponding the 
unavailable messages in $[\Msf+1:\Nsf]$ to server $n^{\prime}$. In addition,  the set of   available messages in $[\Msf+1:\Nsf]$ to server $n^{\prime}$ is a super set of that to server $n^{\prime\prime}$. Hence, $\sv_{n^{\prime\prime}}$ can be also the transmission vector of server $n^{\prime}$; that is, server $n^{\prime}$ transmits
\begin{align}
\sv_{n^{\prime}} \ \left[a \fv_1- \fv_2;  \fv_3; \ldots; \fv_{\frac{\Msf+5}{2}-\ysf} \right]  \ [W_1;\ldots; W_{\Nsf}], \label{eq:transmission K''}
\end{align}
where $\sv_{n^{\prime}}$ is obtained by replacing the first element of $\sv_{n^{\prime\prime}}$ (assumed to be $s$) by $\frac{s}{a-1}$.

Recall that   $\Lc_1$ contains $F_3,\ldots, F_{\frac{\Msf+5}{2}- \ysf}$  and $F_1-F_2$. As $a$ is uniformly  over $F_{\qsf} \setminus \{0,1\}$, it can be seen that the transmission in~\eqref{eq:transmission K''} is independent of the linear combinations  in $\Lc_1$ with high probability.  

\section{Proof of Lemma~\ref{lem:lem scheme 3}}
\label{sec:proof of lemma scheme 3}
We will prove Lemma~\ref{lem:lem scheme 3} by  the Schwartz-Zippel lemma~\cite{Schwartz,Zippel,Demillo_Lipton}. Hence, we need to find one specific realization of $a_1,\ldots, a_{h(\Msf, 2\Msf-\Nsf)-1}$, such that the answers of any $h(\Msf, 2\Msf-\Nsf)- \lambda_1$ servers in $\Ac_2$ are linearly independent of the answers of the servers in $\Ac_1$.\footnote{\label{foot:A2>h-lam} It holds that $|\Ac_2|\geq h(\Msf, 2\Msf-\Nsf)- \lambda_1$. This is because, as shown in Footnote~\ref{foot:scheme 5 deco}, if from $\Ac_1$ the user cannot recover the task function, then there must exist   $\Nsf_{\rm r}=|\Ac_1|+|\Ac_2| $ servers in $[\Msf]$  containing the servers in $\Ac_1$, such that   the answers of these $\Nsf_{\rm r}$ servers contain $h(\Msf, 2\Msf-\Nsf)$ linearly independent combinations. 
Recall that the number of linearly independent combinations transmitted by the servers in $\Ac_1$ is $\lambda$. Hence, we have $|\Ac_2|\geq h(\Msf, 2\Msf-\Nsf)- \lambda_1$. }

Without loss of generality, we assume a possible set of  $h(\Msf, 2\Msf-\Nsf)- \lambda_1$ linearly independent transmission vectors is, 
\begin{align}
\left[ ({\bf S}^{\prime}_1)_{\left( h(\Msf, 2\Msf-\Nsf)- \lambda_1 \right) \times \lambda_1}  ,\mathbf{I}_{h(\Msf, 2\Msf-\Nsf)- \lambda_1} \right], \label{eq:cont rows}
\end{align}  
 where the $h(\Msf, 2\Msf-\Nsf)- \lambda_1$ linear combinations in  $\left[  {\bf S}^{\prime}_1   ,\mathbf{I}_{h(\Msf, 2\Msf-\Nsf)- \lambda_1} \right] {\bf F} [W_1;\ldots;W_{\Nsf}]$  are linearly independent of the answers of the servers in $\Ac_1$.

We now prove that there exist a matrix $[1;a_1;\ldots; a_{h(\Msf, 2\Msf-\Nsf)-1}]$, whose left-hand side null space contains all row vectors in~\eqref{eq:cont rows}.
This can be proved by  randomly choosing the values of $a_1,\ldots,a_{\lambda_1-1}$, and then determining the   values in $[a_{\lambda_1};\ldots; a_{h(\Msf, 2\Msf-\Nsf)-1}]$ as follows,
\begin{align*}
\begin{bmatrix} \
a_{\lambda_1}\\
\vdots\\
a_{h(\Msf, 2\Msf-\Nsf)-1}
  \end{bmatrix}
  = 
  -  {\bf S}^{\prime}_1
  \begin{bmatrix} \
1\\
a_1\\
\vdots\\
a_{\lambda_1-1}
  \end{bmatrix}.
\end{align*}

\bibliographystyle{IEEEtran}
\bibliography{IEEEabrv,IEEEexample}

\begin{thebibliography}{10}
\providecommand{\url}[1]{#1}
\csname url@samestyle\endcsname
\providecommand{\newblock}{\relax}
\providecommand{\bibinfo}[2]{#2}
\providecommand{\BIBentrySTDinterwordspacing}{\spaceskip=0pt\relax}
\providecommand{\BIBentryALTinterwordstretchfactor}{4}
\providecommand{\BIBentryALTinterwordspacing}{\spaceskip=\fontdimen2\font plus
\BIBentryALTinterwordstretchfactor\fontdimen3\font minus
  \fontdimen4\font\relax}
\providecommand{\BIBforeignlanguage}[2]{{%
\expandafter\ifx\csname l@#1\endcsname\relax
\typeout{** WARNING: IEEEtran.bst: No hyphenation pattern has been}%
\typeout{** loaded for the language `#1'. Using the pattern for}%
\typeout{** the default language instead.}%
\else
\language=\csname l@#1\endcsname
\fi
#2}}
\providecommand{\BIBdecl}{\relax}
\BIBdecl

\bibitem{gradiencoding}
R.~Tandon, Q.~Lei, A.~G. Dimakis, and N.~Karampatziakis, ``Gradient coding:
  Avoiding stragglers in distributed learning,'' \emph{in Advances in Neural
  Information Processing Systems (NIPS)}, p. 3368–3376, 2017.

\bibitem{shortdot2016dutta}
S.~Dutta, V.~Cadambe, and P.~Grover, ``Short-dot: Computing large linear
  transforms distributedly using coded short dot products,'' \emph{in Advances
  in Neural Information Processing Systems (NIPS)}, pp. 2100--2108, 2016.

\bibitem{linearcomput2020wan}
K.~Wan, H.~Sun, M.~Ji, and G.~Caire, ``Distributed linearly separable
  computation,'' \emph{available at arXiv:2007.00345}, Jul. 2020.

\bibitem{efficientgradientcoding}
M.~Ye and E.~Abbe, ``Communication computation efficient gradient coding,''
  \emph{in Advances in Neural Information Processing Systems (NIPS)}, pp.
  5610--5619, 2018.

\bibitem{adaptiveGC2020}
H.~Cao, Q.~Yan, and X.~Tang, ``Adaptive gradient coding,''
  \emph{arXiv:2006.04845}, Jun. 2020.

\bibitem{cctradeoff2020wan}
K.~Wan, H.~Sun, M.~Ji, and G.~Caire, ``On the tradeoff between computation and
  communication costs for distributed linearly separable computation,''
  \emph{available at arXiv:2010.01633}, Oct. 2020.

\bibitem{BGW1988}
M.~Ben-Or, S.~Goldwasser, , and A.~Wigderson, ``Completeness theorems for
  non-cryptographic fault-tolerant distributed computation,'' \emph{in
  Proceedings of the twentieth annual ACM symposium on Theory of computing},
  pp. 1--10, 1988.

\bibitem{CCD1988}
D.~Chaum, C.~Cr{\'e}peau, and I.~Damgard, ``Multiparty unconditionally secure
  protocols,'' \emph{in Proceedings of the twentieth annual ACM symposium on
  Theory of computing}, pp. 11--19, 1988.

\bibitem{practicalsecure2016Bonawitz}
K.~Bonawitz, V.~Ivanov, B.~Kreuter, A.~Marcedone, H.~B. McMahan, S.~Patel,
  D.~Ramage, A.~Segal, and K.~Seth, ``Practical secure aggregation for
  federated learning on user-held data,'' \emph{available at arXiv:1611.04482},
  Nov. 2016.

\bibitem{privacy2017Bonawitz}
------, ``Practical secure aggregation for privacy-preserving machine
  learning,'' \emph{in Proceedings of the 2017 ACM SIGSAC Conference on
  Computer and Communications Security}, pp. 1175--1191, 2017.

\bibitem{lagrange2019yu}
Q.~Yu, S.~Li, N.~Raviv, S.~M.~M. Kalan, M.~Soltanolkotabi, and S.~A.
  Avestimehr, ``Lagrange coded computing: Optimal design for resiliency,
  security, and privacy,'' \emph{in Proceedings of Machine Learning Research
  (PMLR)}, pp. 1215--1225, Apr. 2019.

\bibitem{chang2019privatesecure}
W.~T. Chang and R.~Tandon, ``On the upload versus download cost for secure and
  private matrix multiplication,'' \emph{arXiv:1906.10684}, Jun. 2019.

\bibitem{aliasgari2019private}
M.~Aliasgari, O.~Simeone, and J.~Kliewer, ``Private and secure distributed
  matrix multiplication with flexible communication load,''
  \emph{arXiv:1909.00407}, Sep. 2019.

\bibitem{secure2019jia}
Z.~Jia and S.~A. Jafar, ``On the capacity of secure distributed matrix
  multiplication,'' \emph{arXiv:1908.06957}, Aug. 2019.

\bibitem{updowlink2019kakar}
J.~Kakar, A.~Khristoforov, S.~Ebadifar, and A.~Sezgin, ``Uplink-downlink
  tradeoff in secure distributed matrix multiplication,'' \emph{available at
  arXiv:1910.13849}, Oct. 2019.

\bibitem{yu2020entangle}
Q.~Yu and A.~S. Avestimehr, ``Entangled polynomial codes for secure, private,
  and batch distributed matrix multiplication: Breaking the cubic barrier,''
  \emph{arXiv:2001.05101}, Jan. 2020.

\bibitem{GCSAsecure2020chen}
Z.~Chen, Z.~Jia, Z.~Wang, , and S.~A. Jafar, ``{GCSA} codes with noise
  alignment for secure coded multi-party batch matrix multiplication,''
  \emph{arXiv:2002.07750}, Feb. 2020.

\bibitem{securezhu2021}
J.~Zhu and X.~Tang, ``Secure batch matrix multiplication from grouping lagrange
  encoding,'' \emph{IEEE Communications Letters}, Dec. 2020.

\bibitem{improvedGC2017halbawi}
W.~Halbawi, N.~Azizan-Ruhi, F.~Salehi, and B.~Hassibi, ``Improving distributed
  gradient descent using reed-solomon codes,'' \emph{available at
  arXiv:1706.05436}, Jun. 2017.

\bibitem{MDSGC2018raviv}
N.~Raviv, R.~Tandon, A.~Dimakis, and I.~Tamo, ``Gradient coding from cyclic
  {MDS} codes and expander graphs,'' \emph{in Proc. Int. Conf. on Machine
  Learning (ICML)}, pp. 4302--4310, Jul. 2018.

\bibitem{yangelastic2019}
Y.~Yang, M.~Interlandi, P.~Grover, S.~Kar, S.~Amizadeh, and M.~Weimer, ``Coded
  elastic computing,'' \emph{in IEEE International Symposium on Information
  Theory (ISIT)}, pp. 2654--2658, Jul. 2019.

\bibitem{replicationcode2020}
A.~Behrouzi-Far and E.~Soljanin, ``Efficient replication for straggler
  mitigation in distributed computing,'' \emph{available at arXiv:2006.02318},
  Jun. 2020.

\bibitem{shannonsecurity}
C.~E. Shannon, ``Communication theory of secrecy systems,'' \emph{in The Bell
  System Technical Journal}, vol.~28, no.~4, pp. 656--715, Oct. 1949.

\bibitem{Schwartz}
J.~T. Schwartz, ``Fast probabilistic algorithms for verification of polynomial
  identities,'' \emph{Journal of the ACM (JACM)}, vol.~27, no.~4, pp. 701--717,
  1980.

\bibitem{Zippel}
R.~Zippel, ``Probabilistic algorithms for sparse polynomials,'' in
  \emph{International symposium on symbolic and algebraic manipulation}.\hskip
  1em plus 0.5em minus 0.4em\relax Springer, 1979, pp. 216--226.

\bibitem{Demillo_Lipton}
R.~A. Demillo and R.~J. Lipton, ``A probabilistic remark on algebraic program
  testing,'' \emph{Information Processing Letters}, vol.~7, no.~4, pp.
  193--195, 1978.

\end{thebibliography}

\end{document}